\documentclass[12pt]{article}

\usepackage{amsfonts}
\usepackage{amsmath}
\usepackage{amscd}
\usepackage{amsopn}
\usepackage{eucal}

\begin{document}

\setlength{\baselineskip}{12pt}
\setlength{\normalbaselineskip}{24pt}



\newcommand\chapter[2]{{\newpage\refstepcounter{section}%
\section*{Chapter \Roman{section}. \quad #1}\label{#2}}\bigskip}


%

\newtheorem{theorem}{Theorem}
\renewcommand{\thetheorem}{\arabic{section}.\arabic{theorem}}
\newtheorem{calculation}[theorem]{Calculation}
\newtheorem{observation}[theorem]{Observation}
\newtheorem{teo}[theorem]{Theorem}
\newtheorem{acknowledgement}[theorem]{Acknowledgement}
\newtheorem{algorithm}[theorem]{Algorithm}
\newtheorem{axiom}[theorem]{Axiom}
\newtheorem{case}[theorem]{Case}
\newtheorem{claim}[theorem]{Claim}
\newtheorem{conclusion}[theorem]{Conclusion}
\newtheorem{condition}[theorem]{Condition}
\newtheorem{conjecture}[theorem]{Conjecture}
\newtheorem{corollary}[theorem]{Corollary}
\newtheorem{criterion}[theorem]{Criterion}
\newtheorem{definition}[theorem]{Definition}
\newtheorem{example}[theorem]{Example}
\newtheorem{exercise}[theorem]{Exercise}
\newtheorem{lemma}[theorem]{Lemma}
\newtheorem{notation}[theorem]{Notation}
\newtheorem{problem}[theorem]{Problem}
\newtheorem{proposition}[theorem]{Proposition}
\newtheorem{remark}[theorem]{Remark}
\newtheorem{solution}[theorem]{Solution}
\newtheorem{summary}[theorem]{Summary}
\newtheorem{diagram}[theorem]{Diagram}

\newcommand\proof[1]{\noindent{Proof}:\ #1\hfill $\square$}



\def\tomath#1{{\ifmmode{#1}\else${#1}$\fi}}


\def\Nset{\tomath{\mathbb N}}
\def\Qset{\tomath{\mathbb Q}}
\def\Zset{\tomath{\mathbb Z}}
\def\Cset{\tomath{\mathbb C}}



\def\a{\tomath{\mathcal A}} \def\b{\tomath{\mathcal B}}
\def\c{\tomath{\mathcal C}} \def\d{\tomath{\mathcal D}}
\def\e{\tomath{\mathcal E}} \def\f{\tomath{\mathcal F}}
\def\g{\tomath{\mathcal G}} \def\h{\tomath{\mathcal H}}
\def\i{\tomath{\mathcal I}} \def\j{\tomath{\mathcal J}}
\def\k{\tomath{\mathcal K}} \def\l{\tomath{\mathcal L}}
\def\m{\tomath{\mathcal M}} \def\n{\tomath{\mathcal N}}
\def\o{\tomath{\mathcal O}} 
\def\q{\tomath{\mathcal Q}} \def\r{\tomath{\mathcal R}}
\def\s{\tomath{\mathcal S}} \def\t{\tomath{\mathcal T}}
\def\u{\tomath{\mathcal U}} \def\v{\tomath{\mathcal V}}
\def\w{\tomath{\mathcal W}} \def\x{\tomath{\mathcal X}}
\def\y{\tomath{\mathcal Y}} \def\z{\tomath{\mathcal Z}}



\def\p{\tomath{\mathfrak P}}

\def\AA{\tomath{\mathfrak A}} \def\BB{\tomath{\mathfrak B}}
\def\CC{\tomath{\mathfrak C}} \def\DD{\tomath{\mathfrak D}}
\def\EE{\tomath{\mathfrak E}} \def\FF{\tomath{\mathfrak F}}
\def\GG{\tomath{\mathfrak G}} \def\HH{\tomath{\mathfrak H}}
\def\II{\tomath{\mathfrak I}} \def\JJ{\tomath{\mathfrak J}}
\def\KK{\tomath{\mathfrak K}} \def\LL{\tomath{\mathfrak L}}
\def\MM{\tomath{\mathfrak M}} \def\NN{\tomath{\mathfrak N}}
\def\OO{\tomath{\mathfrak O}} \def\PP{\tomath{\mathfrak P}}
\def\QQ{\tomath{\mathfrak Q}} \def\RR{\tomath{\mathfrak R}}
\def\SS{\tomath{\mathfrak S}} \def\TT{\tomath{\mathfrak T}}
\def\UU{\tomath{\mathfrak U}} \def\VV{\tomath{\mathfrak V}}
\def\WW{\tomath{\mathfrak W}} \def\XX{\tomath{\mathfrak X}}
\def\YY{\tomath{\mathfrak Y}} \def\ZZ{\tomath{\mathfrak Z}}


\def\aa{\tomath{\mathfrak a}} \def\bb{\tomath{\mathfrak b}}
\def\cc{\tomath{\mathfrak c}} \def\dd{\tomath{\mathfrak d}}
\def\ee{\tomath{\mathfrak e}} \def\ff{\tomath{\mathfrak f}}
\def\gg{\tomath{\mathfrak g}} \def\hh{\tomath{\mathfrak h}}
\def\ii{\tomath{\mathfrak i}} \def\jj{\tomath{\mathfrak j}}
\def\kk{\tomath{\mathfrak k}} \def\ll{\tomath{\mathfrak l}}
\def\mm{\tomath{\mathfrak m}} \def\nn{\tomath{\mathfrak n}}
\def\oo{\tomath{\mathfrak o}} \def\pp{\tomath{\mathfrak p}}
\def\qq{\tomath{\mathfrak q}} \def\rr{\tomath{\mathfrak r}}
\def\ss{\tomath{\mathfrak s}} \def\tt{\tomath{\mathfrak t}}
\def\uu{\tomath{\mathfrak u}} \def\vv{\tomath{\mathfrak v}}
\def\ww{\tomath{\mathfrak w}} \def\xx{\tomath{\mathfrak x}}
\def\yy{\tomath{\mathfrak y}} \def\zz{\tomath{\mathfrak z}}



\def\Epsilon{{\rm E}}
\newcommand\E{\tomath\Delta}
\newcommand\Del{{\rm E}}


\def\gf{$\g${\rm\ over\ }$\f$}
\def\gsg{\g\sigma\g}
\def\gsp{\g\langle\sigma'\rangle}
\def\fo{\f^{\circ}}
\def\ds{$\Delta$-}
\def\fp{\f'}
\def\fon{\f_1}
\def\ft{\f_2}
\def\gpsgp{\g'\sigma\g'}
\def\sg{\sigma\g}
\def\gspg{\g\sigma'\g}
\def\dsi{$\Del$-strong $(\Del,\E)$-isomorphism}
\def\ds{$\Del$-strong}
\def\ge{\g^{\E}}
\def\fe{\f^{\E}}
\def\fes{\f^{\E} \langle \sigma \rangle}
\def\ges{\g^{\E} \langle \sigma \rangle}
\def\ue{\u^{\E}}
\def\get{\g{^{\E}} \langle \tau \rangle}
\def\gest{\g{^{\E}} \langle \sigma \tau \rangle}
\def\gesi{\g{^{\E}} \langle \sigma^{-1} \rangle}
\def\gesp{\g^{\E} \langle \sigma' \rangle}
\def\gespp{\g^{\E} \langle \sigma'' \rangle}
\def\getp{\g^{\E} \langle \tau' \rangle}
\def\dsn{$\Del$-strongly normal }
\def\gp{\g'}
\def\cs{\c\langle\sigma\rangle}
\def\ct{\c\langle\tau\rangle}
\def\csp{\c\langle\sigma'\rangle}
\def\ctp{\c\langle\tau'\rangle}
\def\cp{\c'}
\def\dc{$\Del$-$\c$-}
\def\dcg{$\Del$-$\c$-group}
\def\dcsg{$\Del$-$\c$-subgroup}
\def\ggf{G(\g/\f)}
\def\hp{\h'}
\def\da{\d_a}
\def\de{$(\Del,\E)$-}
\newcommand\trde[2]{{\rm\ tr\ deg\,}_{#1}\ #2}
\newcommand\trdeg[2]{${\rm\ tr\ deg\,}_{#1}\ #2$}
\newcommand{\td[2]}{${\rm\ tr\ deg\,}_{#1}\ #2$}
\newcommand\deleted[1]{}
\newcommand\mkset[1]{{\{\, #1 \,\}}}
\def\spec{spec}
\def\diffspec{diffspec}
\def\varkappa{\kappa}
\def\nsubseteqq{\nsubseteq}
\def\Hom{Hom}
\def\Hom{{\rm Hom}}  
\def\Isom{Isom}
\def\Isom{{\rm Isom}}  
\def\ann{{\rm ann}}     
\def\spec{{\rm spec\,}}     
\def\diffspec{{\rm diffspec\,}} 
\def\wass{{\rm ass_f}}      
\def\fop{{\rm diffspec}~ \o_{f(\p)}}
\def\Mor{Mor}
\def\Mor{{\rm Mor}}
\def\Ep{{\mathrm E}}
\def\id{{\rm id}}

%
%
%
%
%

\makeatletter

\newbox\CDP@dotbox

\newdimen\CDP@dotdim

\def\math@axis{\fontdimen22\textfont\tw@}

\setboxz@h{$\m@th\cdot$}\CDP@dotdim\wd\z@
\setbox\@ne\hb@xt@\z@{\hss\unhbox\z@\hss}\setbox\CDP@dotbox\vbox
to\z@
 {\kern-\ht\@ne\kern\math@axis\box\@ne\vss}


\def\dotline(#1,#2)#3{\bgroup\@linelen#3\unitlength\@dotgetargs(#1,#2)
 \let\reserved@a\@badlinearg\ifdim\@linelen>\z@\ifnum\@xarg=\z@
 \let\reserved@a\@vdotline\else\ifnum\@yarg=\z@\let\reserved@a\@hdotline\else
 \let\reserved@a\@sdotline\fi\fi\fi\reserved@a\egroup}


\def\@dotgetargs(#1,#2){\m@th\@xarg#1\@yarg#2\@negargfalse
 \ifnum\@xarg=\z@\ifnum\@yarg<\z@\@negargtrue\fi
 \else\ifnum\@xarg<\z@\@negargtrue\@xarg-\@xarg\fi\fi
 \@yyarg\ifnum\@yarg<\z@-\fi\@yarg}


\def\@vdotline{\@clnwd\@linelen\@clnht\z@\@whiledim\@clnwd>\z@
 \do{\raise\@clnht\copy\CDP@dotbox\advance\@clnht
 \if@negarg-\fi\CDP@dotdim\advance\@clnwd-\CDP@dotdim}}


\def\@hdotline{\@clnwd\@linelen\@whiledim\@clnwd>\z@\do
 {\copy\CDP@dotbox\kern\if@negarg-\fi\CDP@dotdim
 \advance\@clnwd-\CDP@dotdim}\kern\if@negarg-\fi\@clnwd}


\def\@sdotline{\ifnum\@yyarg>\@xarg\@xdim\CDP@dotdim\multiply\@xdim\@xarg
 \divide\@xdim\@yyarg\@ydim\ifnum\@yarg<\z@-\fi\CDP@dotdim\else
 \@xdim\CDP@dotdim\@ydim\CDP@dotdim\multiply\@ydim\@yarg
 \divide\@ydim\@xarg\fi\@clnwd\@linelen\@clnht\z@
 \@whiledim\@clnwd>\z@\do{\raise\@clnht\copy\CDP@dotbox
 \advance\@clnht\@ydim\kern\if@negarg-\fi\@xdim
 \advance\@clnwd-\@xdim}\kern\if@negarg-\fi\@clnwd}


\def\dotvector(#1,#2)#3{\bgroup\@linelen#3\unitlength\@dotgetargs(#1,#2)
 \let\reserved@a\@badlinearg\ifdim\@linelen>\z@\ifnum\@xarg=\z@
 \let\reserved@a\@vdotvector\else\ifnum\@yarg=\z@
 \let\reserved@a\@hdotvector\else\ifnum\@xarg<5\ifnum\@yyarg<5
 \let\reserved@a\@sdotvector\fi\fi\fi\fi\fi\reserved@a\egroup}

\def\@hvectdots#1{\@vectdots{#1}\kern\if@negarg-.5\CDP@dotdim\else
 .5\CDP@dotdim\fi}

\def\@vectdots#1{\advance\@linelen-.5\CDP@dotdim{#1}\advance
 \@linelen.5\CDP@dotdim}


\def\@vdotvector{\@vectdots{\@vdotline}\setboxz@h to\z@{\@linefnt\char
 \if@negarg'77\else'66\fi\hss}\@ydim\@linelen\if@negarg\lower\else
 \advance\@ydim-\ht\z@\raise\fi\@ydim\boxz@}


\def\@hdotvector{\@hvectdots{\@hdotline}\hb@xt@\z@{\@linefnt
 \if@negarg\@getlarrow(1,0)\hss\else\hss\@getrarrow(1,0)\fi}}


\def\@sdotvector{\@hvectdots{\@sdotline}\setboxz@h to\z@{\@linefnt
 \if@negarg\@getlarrow(\@xarg,-\@yarg)\hss\else\hss\@getrarrow(\@xarg,\@yarg)
 \fi}\@ydim\@linelen\multiply\@ydim\@yarg\divide\@ydim\@xarg
 \ifnum\@yarg>\z@\advance\@ydim-\ht\z@\fi\raise\@ydim\boxz@}


\def\dashline(#1,#2)#3{\bgroup\@linelen#3\unitlength\@dotgetargs(#1,#2)
 \let\reserved@a\@badlinearg\ifdim\@linelen>\z@\ifnum\@xarg=\z@
 \let\reserved@a\@vdashline\else\ifnum\@yarg=\z@
 \let\reserved@a\@hdashline\else\ifnum\@xarg<7\ifnum\@yyarg<7
 \let\reserved@a\@sdashline\fi\fi\fi\fi\fi\reserved@a\egroup}


\def\@vdashline{\setbox\@linechar\hb@xt@\z@{\hss\vrule\@width\@wholewidth
 \@height10pt\hss}\@clnwd\@linelen\@clnht\if@negarg-10pt\else\z@\fi
 \@whiledim\@clnwd>9.9pt\do{\raise\@clnht\copy\@linechar\advance\@clnht
 \if@negarg-\fi12pt\advance\@clnwd-12pt\relax}}


\def\@hdashline{\setbox\@linechar\hb@xt@\z@{\if@negarg\hss\fi\vrule
 \@width10pt\@height\@halfwidth\@depth\@halfwidth\if@negarg\else\hss\fi}%
 \@clnwd\@linelen\@whiledim\@clnwd>9.9pt\do{\copy\@linechar
 \kern\if@negarg-\fi12pt\advance\@clnwd-12pt\relax}\kern\if@negarg-\fi\@clnwd}


\def\@sdashline{\setboxz@h{\@linefnt\@getlinechar(\@xarg,\if@negarg-\fi\@yarg)}%
 \@xdim1.2\wd\z@\@ydim\ifnum\@yarg<\z@-\fi1.2\ht\z@
 \@clnwd\@linelen\@clnht\ifnum\@yarg<\z@-\ht\z@\else\z@\fi\setbox\@linechar
 \hb@xt@\z@{\if@negarg\hss\fi\unhbox\z@\if@negarg\else\hss\fi}%
 \@whiledim\@clnwd>\@xdim\do{\raise\@clnht\copy\@linechar
 \advance\@clnht\@ydim\kern\if@negarg-\fi\@xdim
 \advance\@clnwd-\@xdim}\kern\if@negarg-\fi\@clnwd}


\def\dashvector(#1,#2)#3{\bgroup\@linelen#3\unitlength\@dotgetargs(#1,#2)
 \let\reserved@a\@badlinearg\ifdim\@linelen>\z@\ifnum\@xarg=\z@
 \let\reserved@a\@vdashvector\else\ifnum\@yarg=\z@
 \let\reserved@a\@hdashvector\else\ifnum\@xarg<5\ifnum\@yyarg<5
 \let\reserved@a\@sdashvector\fi\fi\fi\fi\fi\reserved@a\egroup}


\def\@vdashvector{\@vdashline\setboxz@h to\z@{\@linefnt\char
 \if@negarg'77\else'66\fi\hss}\@ydim\@linelen\if@negarg\lower\else
 \advance\@ydim-\ht\z@\raise\fi\@ydim\boxz@}


\def\@hdashvector{\@hdashline\hb@xt@\z@{\@linefnt
 \if@negarg\@getlarrow(1,0)\hss\else\hss\@getrarrow(1,0)\fi}}


\def\@sdashvector{\@sdashline\setboxz@h to\z@{\@linefnt
 \if@negarg\@getlarrow(\@xarg,-\@yarg)\hss\else\hss\@getrarrow(\@xarg,\@yarg)
 \fi}\@ydim\@linelen\multiply\@ydim\@yarg\divide\@ydim\@xarg
 \ifnum\@yarg>\z@\advance\@ydim-\ht\z@\fi\raise\@ydim\boxz@}


\newcount\CDP@arrow
\def\CDPnormal{\global\CDP@arrow\z@}
\def\CDPdot{\global\CDP@arrow\@ne}
\def\CDPset{\global\CDP@arrow\tw@}
\def\CDPequal{\global\CDP@arrow\thr@@}
\def\CDPdash{\global\CDP@arrow 4}


\def\CDP@def#1{\expandafter\def\csname CDP@\string#1\endcsname}
\atdef@ +#1{\expandafter\ifx\csname CDP@\string#1\endcsname\relax
 \DN@{\at@@@+}\else\DN@{\csname CDP@\string#1\endcsname}\fi\next@}

\CDP@def >#1>#2>{\CDP@@@horz<#1,#2>\iffalse} \CDP@def
<#1<#2<{\CDP@@@horz<#1,#2>\iftrue} \CDP@def
=#1=#2={\CDPequal\CDP@@@horz<#1,#2>\iffalse}


\def\CDP@@@horz<#1,#2>#3{\@ifstar{\bgroup\let\if@negarg#3\DN@
 {\egroup}\CDP@@horz<#1,#2>}{\ampersand@\bgroup\let\if@negarg#3\DN@
 {\egroup\ampersand@}\CDP@@horz<#1,#2>}}


\def\CDP@@horz<#1,#2>{\@testopt{\CDP@horz<#1,#2>}{30}}


\def\CDP@horz<#1,#2>[#3]{\enskip\@linelen#3\unitlength
 \CDP@hbox<#1,#2>{\ifcase\CDP@arrow\CDP@hnorm\or\CDP@hdot\or
 \CDP@hset\or\CDP@hequal\or\CDP@hdash\fi}\CDPnormal\enskip\next@}


\def\CDP@hbox<#1,#2>#3{\m@th\mathop{\setbox\@tempboxa
 \hb@xt@\@linelen{#3}\@tempdima\fontdimen6\textfont\tw@
 \ht\@tempboxa.366875\@tempdima\dp\@tempboxa.083333\@tempdima\box\@tempboxa}%
 \limits^{\CDP@hlabel<#1>}\@ifnotempty{#2}{_{\CDP@hlabel<#2>}}}

\def\CDP@hlabel<#1>{\hb@xt@\z@{$\hss\scriptstyle#1\hss$}}

\def\CDP@hnorm{\if@negarg\kern\@linelen\fi\raise\math@axis
 \hbox{\@yarg\z@\@xarg\if@negarg\m@ne\else\@ne\fi\@hvector}%
 \if@negarg\kern\@linelen\fi}

\def\CDP@hdot{\if@negarg\kern\@linelen\fi\raise\math@axis
 \hbox{\@hdotvector}\if@negarg\kern\@linelen\fi}


\def\CDP@hset{\setboxz@h{$\m@th\if@negarg\supset\else\subset\fi$}%
 \advance\@linelen-.8\wdz@\@xdim.2\wdz@\@ydim\math@axis
 \advance\@ydim1.5\@halfwidth\if@negarg\CDP@hnorm\kern-\@xdim\fi
 \raise\@ydim\boxz@\if@negarg\else\kern-\@xdim\CDP@hnorm\fi}


\def\CDP@hequal{\setboxz@h{\vrule\@width\@linelen\@height\@halfwidth
 \@depth\@halfwidth}\@tempdima\math@axis\advance\@tempdima\@ne\p@
 \raise\@tempdima\copy\z@\kern-\@linelen\advance\@tempdima-\tw@\p@
 \raise\@tempdima\box\z@}

\def\CDP@hdash{\if@negarg\kern\@linelen\fi\raise\math@axis
 \hbox{\@hdashvector}\if@negarg\kern\@linelen\fi}

\CDP@def .{\ampersand@\ampersand@}

\CDP@def A#1A#2A{\CDP@@@vert<#1,#2>\iffalse} \CDP@def
V#1V#2V{\CDP@@@vert<#1,#2>\iftrue} \CDP@def
|#1|#2|{\CDPequal\CDP@@@vert<#1,#2>\iffalse}


\def\CDP@@@vert<#1,#2>#3{\bgroup\let\if@negarg#3\@ifstar{\DN@
 {\egroup}\CDP@@vert<#1,#2>}{\DN@{\egroup\ampersand@\ampersand@}%
 \CDP@@vert<#1,#2>}}


\def\CDP@@vert<#1,#2>{\@testopt{\CDP@vert<#1,#2>}{30}}


\def\CDP@vert<#1,#2>[#3]{\@linelen#3\unitlength
 \CDP@vbox<#1,#2>{\ifcase\CDP@arrow\CDP@vnorm\or\CDP@vdot\or
 \CDP@vset\or\CDP@vequal\or\CDP@vdash\fi}\CDPnormal\next@}


\def\CDP@vbox<#1,#2>#3{\llap{\CDP@vlabel<#1>\;}\setbox\@tempboxa
 \hbox{#3}\@tempdima.5\@linelen\ifdim\ht\@tempboxa=\z@\advance\@tempdima
 \math@axis\raise\else\advance\@tempdima-\math@axis\lower\fi\@tempdima\box
 \@tempboxa\rlap{\;\CDP@vlabel<#2>}}

\def\CDP@vlabel<#1>{$\m@th\vcenter{\hbox{$\scriptstyle#1$}}$}

\def\CDP@vnorm{\@xarg\z@\@yarg\if@negarg\m@ne\else\@ne\fi\@vvector}

\def\CDP@vdot{\@vdotvector}


\def\CDP@vset{\setboxz@h to\z@{\hss$\m@th\if@negarg\cap\else\cup\fi
 \mkern-1.6mu$}\setbox\@ne\hb@xt@\z@{\advance\@linelen-\ht\z@
 \if@negarg\@downvector\else\@upvector\fi\hss}\if@negarg\lower
 \else\raise\fi\ht\z@\box\@ne\if@negarg\lower\ht\z@\fi\boxz@}


\def\CDP@vequal{\setboxz@h{\vrule\@width\@wholewidth
 \@height\@linelen\@depth\z@}\copy\z@\kern\tw@\p@\boxz@}

\def\CDP@vdash{\@vdashvector}

\CDP@def d#1d#2d#3{\CDP@@@diag<#1,#2>(1,1)}

\CDP@def n#1{\expandafter\ifx\csname
CDP@n\string#1\endcsname\relax
 \def\DN@{\at@@@+n}\else\DN@{\csname CDP@n\string#1\endcsname}\fi\next@}

\CDP@def s#1{\expandafter\ifx\csname
CDP@s\string#1\endcsname\relax
 \def\DN@{\at@@@+s}\else\DN@{\csname CDP@s\string#1\endcsname}\fi\next@}

\def\CDP@ne #1d#2d{\CDP@@@diag<#1,#2>(1,1)}
\def\CDP@nw #1d#2d{\CDP@@@diag<#1,#2>(-1,1)}
\def\CDP@se #1d#2d{\CDP@@@diag<#1,#2>(1,-1)}
\def\CDP@sw #1d#2d{\CDP@@@diag<#1,#2>(-1,-1)}

\def\CDP@@@diag<#1,#2>(#3,#4){\@ifstar{\bgroup\DN@{\egroup}%
 \CDP@@@diagA<#1,#2>(#3,#4)}{\ampersand@\bgroup
 \DN@{\egroup\ampersand@}\CDP@@@diagA<#1,#2>(#3,#4)}}

\def\CDP@@@diagA<#1,#2>(#3,#4){\@ifnextchar({\CDP@@@diagB<#1,#2>(#3,#4)}%
 {\CDP@@@diagC<#1,#2>(#3,#4)}}

\def\CDP@@@diagB<#1,#2>(#3,#4)(#5,#6){\@xarg#3\multiply\@xarg#5
 \@yarg#4\multiply\@yarg#6\CDP@@@diagC<#1,#2>(\@xarg,\@yarg)}


\def\CDP@@@diagC<#1,#2>(#3,#4){\@dotgetargs(#3,#4)\def\reserved@a{\@badlinearg}
 \ifnum\@xarg=\z@\def\reserved@a{\CDP@@vert<#1,#2>}\else\ifnum\@yarg=\z@
 \def\reserved@a{\CDP@@horz<#1,#2>}\else\ifnum\@xarg<5\ifnum\@yyarg<5
 \def\reserved@a{\CDP@@diag<#1,#2>}\fi\fi\fi\fi\reserved@a}

\def\CDP@@diag<#1,#2>{\@testopt{\CDP@diag<#1,#2>}{30}}


\def\CDP@diag<#1,#2>[#3]{\enskip\@linelen#3\unitlength
 \ifnum\@linelen>\z@\CDP@dbox<#1,#2>{\ifcase\CDP@arrow\CDP@dnorm\or
 \CDP@ddot\or\@badlinearg\or\CDP@dequal\or\CDP@dash\fi}\else\@badlinearg\fi
 \CDPnormal\enskip\next@}


\def\CDP@dbox<#1,#2>#3{\hb@xt@\z@{\kern.5\@linelen
 \@tempdima\if@negarg\ifnum\@yarg>\z@-\fi\else
 \ifnum\@yarg<\z@-\fi\fi1ex
 \raise\@tempdima\llap{\CDP@vlabel<#1>}\lower\@tempdima
 \rlap{\CDP@vlabel<#2>}\hss}
 \setbox\@tempboxa\hb@xt@\@linelen{\hss#3\hss}\ifdim\ht\@tempboxa=\z@
 \@tempdima.5\dp\@tempboxa\advance\@tempdima\math@axis\raise\else\@tempdima.5
 \ht\@tempboxa\advance\@tempdima-\math@axis\lower\fi\@tempdima\box\@tempboxa}

\def\CDP@dnorm{\if@negarg\kern\@linelen\@xarg-\@xarg\fi
 \@svector\if@negarg\kern-\@ne em\kern\@linelen\fi}

\def\CDP@ddot{\if@negarg\kern\@linelen\fi\@sdotvector
 \if@negarg\kern\@linelen\hss\fi}

\def\CDP@dequal{\setboxz@h to\z@{\if@negarg\kern\@linelen\@xarg-\@xarg\fi
 \@sline\hss}\copy\z@\ifnum\@yarg<\z@\lower\else\raise\fi\thr@@\p@
 \boxz@\kern\@linelen}

\def\CDP@ddash{\if@negarg\kern\@linelen\fi\@sdashvector
 \if@negarg\kern\@linelen\hss\fi}

\makeatother

\thispagestyle{empty}
  \begin{center}
    {\LARGE Generalized Differential Galois Theory}
    \vskip 3em%
    {\large
     \lineskip .75em%
      \begin{tabular}[t]{c}%
        by
      \end{tabular}\par}%
      \vskip 1.0em%
    {\large
     \lineskip .75em%
      \begin{tabular}[t]{c}%
       Peter Landesman
      \end{tabular}\par}%

  \end{center}\par

{\bf A Galois theory of differential fields with parameters is
developed in a manner that generalizes Kolchin's theory. It is
shown that all connected differential algebraic groups are Galois
groups of some appropriate differential field extension. }\\

\noindent{\bf Introduction} This paper may be viewed as the next
step in E. R. Kolchin's work on the foundations of differential
Galois theory. In \cite[1948]{kol48}, Kolchin was the first to
formulate the Galois theory of differential fields in the current
standard of mathematical rigor. In \cite[1953]{kol53}, he defined
strongly normal differential field extensions, generalizing
Picard-Vessiot extensions, so as to include the non-linear
algebraic groups as Galois groups. In his first book
\cite[1973]{kol73}, the properties of these Galois groups are
axiomatized as the category of $\c$-groups for a field of
constants $\c$ and are shown to be the Galois groups of strongly
normal differential field extensions. In his second book
\cite[1985]{kol85}, Kolchin develops more general axioms to define
the category of differential algebraic groups. This paper defines
a generalization of strongly normal differential field extensions
and shows that these extensions have a good Galois theory for
which the Galois groups are differential algebraic groups.

A {\it differential algebraic group} or $\Ep$-$\c$-{\it group}
(Definition \ref{Definition of E-group}), where
$\Ep=\{\epsilon_1,\ldots,\epsilon_m\}$ is a commuting set of
derivations acting on a field $\c$, is a group that may be thought
of a set of zeros of a system of differential equations in
$\Ep$-derivatives over $\c$. It is endowed with the
$\Ep$-$\c$-Zariski topology for which the closed sets are the
zeros of a system of differential equations and the operations of
the group structure are defined by differential rational
functions. In Cassidy's treatment of affine differential algebraic
groups \cite{cass72}, this is how they are defined. However, in
Kolchin's exposition \cite{kol85}, this definition is a
consequence of the extensive development of Kolchin's differential
algebraic group axioms.

To see how Kolchin's theory of strongly normal extensions can be
enriched, consider two sets of mutually commuting derivations
$\Ep$ and $\E$ acting on a field $\f$. Let $\u$ be a universal
differential extension field of $\f$ with respect to both $\Ep$
and $\E$, and let $\f$ contain the $\E$-constants $\c$ of $\u$.
Let $\g$ be a subfield of $\u$ containing $\f$ which is closed
under the operation of $\E$. If $\g$ over $\f$ is a strongly
normal extension of $\E$-fields, in the sense of Kolchin, the set
of $\E$-isomorphisms of $\g$ into $\u$ over $\f$, when $\u$ is
viewed as a universal differential extension of $\f$ with respect
to $\E$, has the structure of an algebraic group defined over
$\c$. All the $\E$-isomorphisms are obtained by sending the
$\E$-generators of $\g$ to rational expressions in the
$\E$-generators of $\g$ and their $\E$-derivatives, with
$\E$-constants in $\u$ as coefficients. These constants are not
necessarily constants with respect to $\Ep$. The generators of
$\g$ may satisfy differential equations in $\Ep$ as well as $\E$.
A $\E$-isomorphism of $\g$ into $\u$ will extend to an $\Ep$ and
$\E$ isomorphism of the $\Ep$ and $\E$ field $\h$ generated by
$\Ep$-derivatives of $\g$ if and only if it maps solutions of the
system of differential equations in $\Ep$ and $\E$ to other
solutions of the same system. The $\E$-isomorphisms of $\g$ into
$\u$ which extend to $\Ep$ and $\E$ isomorphisms of $\h$ form a
differential subgroup $H$ of $G$ defined by differential equations
with respect to $\Ep$. If the field  $\c$ of $\E$-constants of
$\f$ is equal to the field of $\E$-constants of $\h$, then it will
be shown that $H$ is a Galois group for $\h$ over $\f$ (Corollary
\ref{Corollary on embeddability}). That is: subfields of $\h$
closed under both $\Ep$ and $\E$ are in bijection with subgroups
of $H$ defined over $\c$ by $\Ep$-equations.

In addition to proving the fundamental theorems of a Galois
theory, this paper will show that each differential algebraic
group is the Galois group of some generalized strongly normal
differential field extension (Theorem \ref{Theorem on the
construction of an $E$-strongly normal extension from any
$E$-group}). Then there is a short section on the generalized
strongly normal extensions that are induced from strongly normal
extensions (Section \ref{Section of extension of Galois groups by
Delta and Galois theory}). At the end of the paper, examples of
generalized strongly normal field extensions are constructed for
each differential algebraic subgroup of $G_a$ and $G_m$. This
section is dependent on ideas of Johnson, Reinhart and Rubel
\cite{johnsonrr}, which are developed in an appendix. Examples
with non-linear Galois groups will appear in another paper, and
the geometric consequences of this Galois theory is a work in
progress.

Several other people have developed Galois theories of
differential fields: Drach \cite{drach}, Vessiot \cite{vessiot},
Pommerat \cite{pomm}, Umemura \cite{umem1} \cite{umem2}, Pillay
\cite{pillay1} \cite{pillay2} \cite{pillay3} and Kovacic
\cite{kov3}. Although the Galois groups of Pillay's theory are
differential groups, they are only algebraically finite
dimensional. Since a differential algebraic group may have
infinite algebraic dimension (even though its differential
dimension is finite), the Galois theory developed here includes
infinite dimensional groups.

One may speculate as to why this generalized Galois theory was not
previously realized.  One reason may be that since the simplest
new examples are of infinite algebraic dimension the symmetries
are difficult intuit. Also, because the finite dimension examples
all necessitate two commuting derivation, those working in
Picard-Vessiot theory do not usually work with two derivations
since Kolchin showed that the Picard-Vessiot theory with several
derivations is subsumed in that with one derivation \cite{kol52}.

I wish I could thank Professor Kolchin for teaching me his special
field of expertise. I also wish to acknowledge the assistance and
encouragement of Professors Phyllis Cassidy, Richard C. Churchill,
Jerold Kovacic and William Sit, who sat through a series lectures
in 2001-2004 during which I explained the theory presented in this
paper. Cassidy and Singer used these ideas to write an exposition
of the linear case \cite{cass04} where they cite this work as my
forthcoming thesis.

\bigskip
\noindent \textbf{A Simple Example}\\

The following finite dimensional example will serve to further
elucidate the nature of the generalization herein and to exhibit
its relationship with the standard Picard-Vessiot. Let
$\mathbb{C}$ be the complex numbers, and let $\mathbb{C}[t,x]$ be
a polynomial ring in two variables with standard derivations $D_t$
and $ D_x$. Consider \nolinebreak$\f = \mathbb{C}(t,x,\cos t,\sin
t)$ and $\g = \f(\log x \sin t ) =\f (\log x )$ as differential
fields with respect to $D_t$ and $ D_x$.  Let $\c = \mathbb{C}(t,
\cos t, \sin t)$ be the field of $D_x$-constants of $\f$, and let
$\u^{D_x}$ be the same of $\u$. Note that in this example the
$D_t$-field generated by $\g$ is $\g$, and the field of
$D_x$-constants of $\g$ equals that of $\f$.

Let $\eta=\log x \sin t$, and $\zeta =\sin t /x \in \f$.  Then
$\eta$ satisfies the equation $D_x \eta = \zeta$, and $\g$ as a
$D_x$-extension over $\f$ is a strongly normal extension in the
sense of Kolchin.  The Galois group ${\rm Isom}^{D_x}(\g/\f) =
{\rm Aut}^{D_x}(\g \u^{D_x} / \f \u^{D_x})$ is isomorphic to the
additive group $\u^{D_x}$, is defined over $\c$, and will be
denoted by $G_a$ via this identification.  More explicitly,
consider $\sigma \in {\rm Aut}^{D_x}(\g/\f)$.  Then, in order for
$\sigma$ to commute with $D_x$, $\sigma\eta$ must again be a
solution to this differential equation, and therefore $\sigma
\eta$ must equal $\eta + \rho(\sigma)$, where $\rho(\sigma) \in
\u^{D_x}$.  There being no other algebraic conditions on
$\rho(\sigma)$, the map $\rho:  \sigma \mapsto \rho(\sigma)$
defines a group isomorphism between ${\rm Isom}^{D_x}(\g/\f)$ and
the full algebraic group $G_a$.

Let $\gamma = \cos t / \sin t \in \f$. Then $\eta$ also satisfies
the differential equation $D_t \eta - \gamma \eta =0$.  Indeed, in
the $(D_x, D_t)$-differential polynomial ring $\f\mkset{y}$, it is
easy to verify that the two differential polynomials $A(y) = D_x y
- \zeta$ and $B(y) = D_t y - \gamma y$ form a characteristic set
of a linear differential ideal $\p = [A(y), B(y)]$ (relative to
any ranking) with $\eta$ as a generic zero over $\f$.  Consider
$\sigma$ in the subgroup $H={\rm Isom}^{D_t,D_x}(\g/\f) = {\rm
Aut}^{D_t,D_x}(\g \u^{D_x}/\f \u^{D_x})$ of ${\rm
Isom}^{D_x}(\g/\f)=G_a$.  Then $\sigma$ must map $\eta$ to a
generic solution of $\p$.  Thus $0 =B(\sigma(\eta)) = B(\eta +
\rho(\sigma)) = B(\rho(\sigma))$, which implies that $\rho(\sigma)
= c(\sigma) \sin t$, where $c(\sigma)$ is a constant with respect
to $D_t$.  But $\rho(\sigma)$ is a $D_x$-constant, and so
$c(\sigma)$ must be one, too.  Conversely, it is clear that given
any $D_t$ constant $k \in \u^{D_x}$, the map $\sigma$ where
$\sigma(\eta) = \eta + k \sin t$ is the  unique isomorphism of
$\g$ over $\f$ with $c(\sigma) = k$. Therefore, $\rho(H)$ is a
differential algebraic subgroup of $G_a$ defined over the
$D_x$-constant field $\d = \mathbb{C} \langle
 t,\cos t, \sin t \rangle$ of $\f$ by the prime differential
ideal $[B(y)]$  in the $D_t$-differential polynomial ring
$\d\mkset{y}$, or equivalently, the prime differential ideal $[D_x
y, B(y)]$ in the $(D_x, D_t)$-differential polynomial ring
$\f\mkset{y}$. See a proof just after Proposition \ref{Proposition
on the construction of a G(L) ext}.

\section{Group of Isomorphisms}\label{notation}

\subsection{Notation} To define the category of differential rings,
as developed by Ritt and Kolchin, fix a set $\Delta = \{\delta_1,
\dots, \delta_m \}$. The objects, called $\Delta$-{\it rings} or
{\it differential rings}, are rings on which the set $\Delta$ acts
as commuting derivations. The morphisms, called $\Delta$-{\it
homomorphisms} or {\it differential homomorphisms}, are ring
homomorphisms that commute with the action of $\Delta$. Many terms
of algebra, such as ideal, field and extension, have
straightforward interpretations the category of $\Delta$-rings and
are indicated by the modifier ``$\Delta$'' or ``differential''.
However, ``$\Delta$-embeddings'' are referred to as
``$\Delta$-isomorphisms'', and the now standard term ``radical
ideal'' is used in place of Kolchin's ``perfect ideal''.

Henceforth, all rings are assumed to have characteristic zero.
Throughout this chapter, the set of commuting derivations $\Delta
= \{\delta_1, \dots, \delta_m \}$ is fixed, and $\f$ is a
$\Delta$-field.

Standard notation will now be reviewed from \cite{kol73}. The
$\Delta$-polynomial algebra $\f \{y_1,\ldots,y_n\}_{\Delta}$ over
$\f$ in $\Delta$-indeterminates $y_1,\ldots,y_n$ is the polynomial
ring over $\f$ having one indeterminate for each derivative of
$y_1,\ldots,y_n$ on which $\Delta$ operates in the expected
manner. (For details see Kolchin \cite[pages 69-71]{kol73}.) If
${\rm S}$ is a subset of $\f \{y_1,\ldots,y_n\}_{\Delta}$, the
$\Delta$-ideal generated by ${\rm S}$ is denoted by $[{\rm
S}]_{\Delta}$ (or $[s_1,\ldots,s_n]_{\Delta}$ if ${\rm
S}=\{s_1,\ldots,s_n\}$), and the radical $\Delta$-ideal generated
by ${\rm S}$ will be denoted by $\{{\rm S}\}_{\Delta}$.   Let $\g$
be a $\Delta$-field that is a $\Delta$-extension of $\f$, and let
${\rm T}$ be a subset of $\g$. The $\Delta$-ring generated by
${\rm T}$ over $\f$ is denoted by $\f\{{\rm T}\}_{\Delta}$ (or
$\f\{t_1,\ldots,t_n\}_{\Delta}$ if ${\rm T}=\{t_1,\ldots,t_n\}$),
and the $\Delta$-field generated by ${\rm T}$ over $\f$ is denoted
by $\f\langle {\rm T}\rangle_{\Delta}$ (or $\f\langle
t_1,\ldots,t_n\rangle_{\Delta}$ if ${\rm T}=\{t_1,\ldots,t_n\}$).
If ${\rm T}$ is a finite set, the $\Delta$-ring
$\f\{t_1,\ldots,t_n\}_{\Delta}$ and the $\Delta$-field $\f\langle
t_1,\ldots,t_n\rangle _{\Delta}$ are said to be {\it finitely}
$\Delta$-{\it generated by} ${\rm T}$ {\it over} $\f$, or, for
simplicity, {\it $\E$-$\f$-finitely generated}. If $R$ is any
$\E$-ring, the symbol $R^{\Delta}$ denotes the constants of $R$
with respect to $\Delta$, i.e. the elements $\alpha$ of $R$ such
that $\delta\alpha=0$ for every $\delta \in \Delta$.

A $\Delta$-field $\u$ containing a $\Delta$-subfield $\f$ is
called $\Delta$-{\it universal over} $\f$ if the following
conditions hold:  for each $\Delta$-field $\g$ of $\u$ finitely
$\Delta$-generated over $\f$ and for each $\Delta$-field $\h$ (not
necessarily contained in $\u$) finitely $\Delta$-generated over
$\g$, there exists a $\Delta$-isomorphism of $\h$ into $\u$ over
$\g$. The existence of $\Delta$-universal $\Delta$-extension of
any $\Delta$-field is established by Kolchin in \cite[Theorem 2,
page 134]{kol73}. Such an extension contains all the solutions to
differential equations over $\f$ necessary in Kolchin's work.

\subsection{Specializations}\label{Specializations} In this
section, let $\f$ be a $\Delta$-field, and let $\u$ be a
$\Delta$-extension of $\f$ that is $\Delta$-universal over $\f$.
Let $\g$ be a $\Delta$-extension of $\f$ in $\u$ over which $\u$
is universal.

\begin{definition}\textrm{$($Pre-orders on $\u^r)$} \label{pre-order
of zeros of an ideal}For \,$\eta=(\eta_1,\dots,\eta_r)$ and
\,$\xi=(\xi_1,\dots,\xi_r)$ in \,$\u^r$, define the pre-order by
\,$\eta \xrightarrow[\g]{} \xi$, called {\rm
$\Delta$-specialization over $\g$} or {\rm
$\Delta$-$\g$-specialization}, if there exists a
\,$\Delta$-$\g$-homomorphism of \,$\g
\{\eta_1,\dots,\eta_r\}_\Delta$ to
\,$\g\{\xi_1,\dots,\xi_r\}_\Delta$ over \,$\g$ taking \,$\eta_i$
to \,$\xi_i$ for \,$i=1,\ldots,r$.
\end{definition}

\begin{definition}$($\textrm{$\Delta$-$\g$-Specialization of
$\Delta$-$\f$-Isomorphisms}$)$ \label{pre-order no X^r}  Let
$X=\linebreak \Isom_{\f}^{\E}(\g,\u)$.  On the set $X^r =X \times
\cdots \times X$ define a pre-order \,$\xrightarrow[\g]{} $ $($or,
for simplicity, $\rightarrow )$ called \,$\Delta$-$\g$-{\it
specialization} \,$($of elements of $X^r)$ as follows:   for
\,$\sigma=(\sigma_1,\dots,\sigma_r)$ and
\,$\tau=(\tau_1,\dots,\tau_r) \in X^r$, \,$\sigma
\xrightarrow[\g]{} \tau$ if there exists a
\,$\Delta$-$\g$-homomorphism \,$\phi: \g \{\sigma_1 \g \cup \ldots
\cup \sigma_r \g\}_\Delta \rightarrow \g \{\tau_1 \g \cup \ldots
\cup \tau_r \g\}_\Delta$ such that \,$\phi(\alpha)=\alpha$ and
\,$\phi(\sigma_i\alpha)=\tau_i\alpha$ for all \,$\alpha$ in \,$\g$
and \,$i=1,\ldots,r$.\end{definition}

In the above definition, note that the $\E$-rings $\g \{\sigma_1
\g \cup \ldots \cup \sigma_r \g\}_\Delta  \subset \u$ and $\g
\{\tau_1 \g \cup \ldots \cup \tau_r \g\}_\Delta  \subset \u$ are
the same as the rings $\g [\sigma_1 \g \cup \ldots \cup \sigma_r
\g]$ and $\g [\tau_1 \g \cup \ldots \cup \tau_r \g]$. So that
$\phi$ is in fact a $\E$-$\g$-homomorphism from \linebreak$\g
[\sigma_1 \g \cup \ldots \cup \sigma_r \g]$ to $\g [\tau_1 \g \cup
\ldots \cup \tau_r \g]$.  Also, since $\tau_i \circ \sigma_i^{-1}$
is a $\E$-$\f$-isomorphism for all $i$, $\phi$ is a
$\E$-$\g$-homomorphism if and only if it is a $\g$-homomorphism
(See \cite[Lemma 1, page 385]{kol73}).

\begin{lemma}\label{equivalence of specialization of elements
with specialization of isomorphisms} Let
\,$\eta=(\eta_1,\dots,\eta_r)$ be a set of $\Delta$-generators of
\,$\g$ over \,$\f$. For $\sigma, \tau \in X^r$, \,$\sigma
\xrightarrow[\g]{} \tau$ as in Definition \ref{pre-order no X^r}
if and only if \,$(\ldots,\sigma_i \eta_j,\ldots)
\xrightarrow[\g]{} (\ldots,\tau_i \eta_j,\ldots) $ as in
Definition \ref{pre-order of zeros of an ideal}.
\end{lemma}\proof{(See \cite[Lemma 2,
page 386]{kol73} for a statement of the same lemma without a
proof.) If $\sigma \xrightarrow[\g]{} \tau$, the
$\E$-$\g$-homomorphism $\phi: \g \{\sigma_1 \g \cup \ldots \cup
\sigma_r \g\}_\Delta \rightarrow \g \{\tau_1 \g \cup \ldots \cup
\tau_r \g\}_\Delta$ of Definition \ref{pre-order no X^r} restricts
to a $\E$-$\g$-homomorphism $\rho:\g\{\ldots,\sigma_i
\eta_j,\ldots\}_\E \rightarrow \g\{\ldots,\tau_i
\eta_j,\ldots\}_\E $, and $(\ldots,\sigma_i \eta_j,\ldots)
\xrightarrow[\g]{} (\ldots,\tau_i \eta_j,\ldots) $.

On the other hand, if $(\ldots,\sigma_i \eta_j,\ldots)
\xrightarrow[\g]{} (\ldots,\tau_i \eta_j,\ldots) $, then there is
a $\E$-$\g$-homomorphism $\rho:\g\{\ldots,\sigma_i
\eta_j,\ldots\}_\E \rightarrow \g\{\ldots,\tau_i
\eta_j,\ldots\}_\E $. Let $\II$ be the kernel of $\rho$.  Since
the image of $\rho$ is in $\u$ and, therefore, an integral domain,
$\II$ is a prime $\E$-ideal. Let $\g\{\ldots,\sigma_i
\eta_j,\ldots\}_{\E,\II}$ be the localization of $
\g\{\ldots,\sigma_i \eta_j,\ldots\}_\E $ at $\II$, and let the
induced $\E$-$\g$-homomorphism of $\g\{\ldots,\sigma_i
\eta_j,\ldots\}_{\E,\II}$ into the quotient field of
$\g\{\ldots,\tau_i \eta_j,\ldots\}_\E$ be
\[\overline{\rho}:~~\g\{\ldots,\sigma_i \eta_j,\ldots\}_{\E,\II}
\rightarrow QF(\g\{\ldots,\tau_i \eta_j,\ldots\}_\E).\] The
$\E$-$\g$-homomorphism $\overline{\rho}$ restricted to $\f \{
\sigma_i\eta_1,\dots,\sigma_i\eta_n\}_\E$ is the
$\E$-$\f$-isomor-phism $\tau_i \circ \sigma_i^{-1}:\sigma_i\g
\rightarrow \tau_i\g$ restricted to $\f \{
\sigma_i\eta_1,\dots,\sigma_i\eta_n\}_\E$. Therefore,
$\overline{\rho}$ restricted to $\f \{
\sigma_i\eta_1,\dots,\sigma_i\eta_n\}_\E$ is an
$\E$-$\f$-isomorphism.  Consequently, $\f\{
\sigma_i\eta_1,\dots,\sigma_i\eta_n\}_\E \cap \II= \{0\}_\E$, and
the nonzero elements of $\f\{
\sigma_i\eta_1,\dots,\sigma_i\eta_n\}_\E$ are invertible in
$\g\{\ldots,\sigma_i \eta_j,\ldots\}_{\E,\II}$, i.e. $\sigma_i\g
\subseteq \g\{\ldots,\sigma_i \eta_j,\ldots\}_{\E,\II}$ for all
$i$. Since $\eta$ $\E$-generates $\g$ over $\f$, both
$\overline{\rho}$ and $\tau_i \circ \sigma_i^{-1}$ coincide on
$\sigma_i\g \subseteq \g\{\ldots,\sigma_i
\eta_j,\ldots\}_{\E,\II}$. Therefore $\overline{\rho}$ restricted
to $\g \{\sigma_1 \g \cup \ldots \cup \sigma_r \g\}_\Delta$ is a
$\E$-$\g$-homomorphism $\phi: \g \{\sigma_1 \g \cup \ldots \cup
\sigma_r \g\}_\Delta \rightarrow \g \{\tau_1 \g \cup \ldots \cup
\tau_r \g\}_\Delta$ such that \,$\phi(\alpha)=\alpha$ and
\,$\phi(\sigma_i\alpha)=\tau_i\alpha$ for all \,$\alpha$ in \,$\g$
and \,$i=1,\ldots,r$.  Thus, $\rho$ may be extended to the
$\E$-$\g$-homomorphism $\phi$. By the definition of
$\E$-$\g$-specialization of elements of $X^r$ (Definition
\ref{pre-order no X^r}), $\sigma \xrightarrow[\g]{}
\tau$.}\bigskip

\subsection{$\Ep$-Strong Isomorphisms}

Denote by ``$(\Del,\E)$'' the union of two disjoint sets
$\Ep=\{\epsilon_1,\ldots,\epsilon_r \}$ and $\Delta = \{\delta_1,
\dots, \delta_n \}$. However, when this symbol is used as a
subscript or superscript the parenthesis are removed, e.g.,
$\f\{y\}_{\Ep,\E}$ or $\f^{\Ep,\E}$. In this section, $\f$ will
denote an \de field, $\u$ an \de field that is \de universal over
$\f$, and $\c$ the $\Delta$-constants of $\f$. Then $\k=\u^\Delta$
may be considered as an $\Ep$-field.  As such, it is
$\Ep$-universal over $\c$, considered as an $\Ep$-field. The \de
field $\g \subset \u$ will contain $\f$. If $\f$ and $\g$ are
fields contained in a larger field, then $\f \cdot \g$, or more
simply $\f\g$, will denote their compositum.

\medskip
\begin{definition}\label{Definition of E-strong isomorphism}
 Let \,$\g$ be an \,\de subfield of \,$\u$.
 An \,\de isomorphism \,$\sigma$ of \,$\g$ into \,$\u$ is \,$\Ep$-{\rm
 strong}
if it satisfies the following two conditions. 
\renewcommand{\theenumi}{{\rm St\arabic{enumi}}}
\begin{enumerate}
\item \label{st1} $\sigma$ leaves invariant every element of
\,$\g^{\E}$.

\item \label{st2} $\sigma\g\subset\g\cdot\u^{\E}$ and
\,$\g\subset\sigma\g\cdot\u^{\E}$.
\end{enumerate}
\end{definition}

An \dsi~is the same as an $\Del$-homomorphism which is also a
strong $\E$-isomorphism in the sense defined by Kolchin in
\cite[p. 388]{kol73}. Because of this, some of the proofs in this
chapter can often simply quote the results of Kolchin, and, if
$\Ep$ is empty, many results of this paper are those of Kolchin.

Note that \ref{st2} is equivalent to $\g\cdot\u^{\E}=
\sigma\g\cdot\u^{\E}$.  Also it is clear that any \de automorphism
of $\g$ over $\g^{\E}$ is an \dsi.  For any \de isomorphism
$\sigma$ of $\g$, let $\ges=(\gsg)^{\E}$.  The first inclusion of
\ref{st2} is equivalent to $\g \sigma \g \subset \g\cdot \u^{\E}$
which by \cite[Corollary 2, p. 88]{kol73} is equivalent to
$\g\sigma\g=\g\cdot \ges$.  Similarly the second inclusion  is
equivalent to $\g\sigma\g=\sigma\g\cdot \ges$.

If $\g$ is an arbitrary \de extension of $\f$, it may happen that
not all elements $\sigma$ of $\Isom_{\f}^{E,\E}(\g,\u)$ are \dsi~
\cite[Example 3.147]{landesman}. However, if there is one \dsi,
the next proposition shows that all its
$(E,\Delta)$-$\g$-specializations are \dsi.

\begin{proposition}\label{Proposition on specializations of strong
isomorphisms} Every \,$(E,\Delta)$-$\g$-specialization of an
\,$\Ep$-strong \,$(E,\Delta)$-\linebreak isomorphism of \,$\g$ is
\ds.
\end{proposition}

\proof {Let $\sigma'$ be an $(E,\Delta)$-$\g$-specialization of
the $\Ep$-strong $(E,\Delta)$-isomorphism $\sigma$ of \g. By the
definition of $(E,\Delta)$-$\g$-specialization (Example
\ref{pre-order no X^r}), $\sigma'$ is an \de isomorphism. Now
$\sigma$ is a strong $\E$-isomorphism, and hence $\sigma'$ is also
a strong $\E$-isomorphism by \cite[Proposition 6, p. 390]{kol73}.
Since $\sigma'$ is an $\Del$-homomorphism, $\sigma'$ is an \dsi.
}\bigskip

The following propositions will be used to verify under certain
conditions in Theorem \ref{existance of the differential group of
isomorphisms} that the set of $\Ep$-strong
$(E,\Delta)$-isomorphisms of $\g$ over $\f$ verify the axioms of
an $\Ep$-group.


\begin{proposition}\label{constants of gsigmag finitely generated}
Let \,$\g$ be a finitely  \,$(E,\Delta)$-generated
\,$(E,\Delta)$-extension of \,$\f$. Then for every  \dsi~$\sigma$
of \,\gf, \,$\ges=(\gsg)^{\E}$ is a finitely \,$\Del$-generated
field extension of the \,$\Ep$-field \,$\g^{\Delta}$.
\end{proposition}

\proof
 {Let $\eta=(\eta_{1},\ldots,\eta_{n})$
be a finite family of \de generators of \gf. Let $\sigma$ be an
\dsi~of \gf. The extension $\gsg =
\g\langle\sigma\eta\rangle_{E,\E}$ of $\g$ is a finitely \de
generated extension by $\sigma\eta$. Let $\xi=(\xi_{i})_{i\in I}$
be a family of $\Del$-generators of $\ges=(\gsg)^{\E}$ over
$\g^{\Delta}$. Since \linebreak $\gsg=\g\ges=\g\langle\xi\rangle$,
the family $\xi$ also $\Ep$-generates $\gsg$ over $\g$.\bigskip
\[
\begin{CD}
\g               @>>>                  \g \sigma \g =\g \langle \xi \rangle\\
 @AAA                               @AAA  \\
\g^\E    @>>> \ges=\g^\E \langle \xi \rangle\end{CD}
\]
Because this extension is \de finitely generated over $\g$ by
$\sigma\eta$ and each element of $\sigma\eta$ is in an $\Ep$-field
generated by finitely many of the elements of the family $\xi$,
there is a finite subfamily $(\xi_{1},\ldots,\xi_{m})$ of the
family $\xi$ that $(E,\E)$-generates $\gsg$ over $\g$: that is
$\g\ges=\g\langle\xi_{1},\ldots,\xi_{m}\rangle_{E,\E}$ \linebreak
$=\g\cdot\g^\E\langle\xi_{1},\ldots,\xi_{m}\rangle_{E,\E}$. Since
the elements of $\xi$ are $\Delta$-constants, $\ges
=\g^E\langle\xi_{1},\ldots,\xi_{m}\rangle$ by \cite[Corollary 2,
p. 88]{kol73}.}\bigskip

\begin{proposition}\label{Proposition: fields of inverse isom are
equal} Let \,$\sigma$ and \,$\tau$ be two \,$\Ep$-strong \,\de
isomorphisms of \,$\g$. Then \,$\ges \gest = \ges \get =\gest
\get$, and \,$\gesi =\ges$ as $\Del$-fields.
\end{proposition}

\proof{By considering the fields in the statement of the
proposition as just $\E$-fields, and $\sigma$ and $\tau$ as just
strong $\E$-isomorphisms, Kolchin's result \cite[Proposition 5, p.
390]{kol73} may be applied to obtain these equalities as fields in
$\u$.  Because they are also $\Del$-fields, they are equal as
$\Del$-fields.}

\bigskip


\begin{proposition}\label{Proposition to verify diff group axioms}
Let \,$\sigma,\sigma',\tau,\tau'$ be \,\dsi s of \,$\g$.
\begin{enumerate}
\item If \,$(\sigma',\tau')$ is a specialization of
\,$(\sigma,\tau)$ then \,$(\sigma'^{-1},\sigma'^{-1}\tau')$ is a
specialization of \,$(\sigma^{-1},\sigma^{-1}\tau)$.

\item Suppose that \,$\sigma'$ and \,$\tau'$ are generic
specializations of \,$\sigma$ and \,$\tau$, respectively.  If
\,$(\sigma',\tau')$ is a specialization of \,$(\sigma,\tau)$, then
the induced \,$\Del$-isomorphisms \,$\ges\approx\gesp$ ~and
\,$\get\approx\getp$ ~are compatible, and conversely.

\item Suppose that \,$\sigma'$ and \,$\tau'$ are generic
specializations of \,$\sigma$ and \,$\tau$, respectively, let
\,$h:\d \longrightarrow \d'$ be an \,$\Del$-homomorphism between
subrings of \,\,$\ue$. If $h$ and the induced
\,$\Del$-isomorphisms \,$\ges\approx\gesp$ and
\,$\get\approx\getp$ are compatible, then \,$\sigma'^{-1}$ is a
generic specialization of \,$\sigma^{-1}$ and
\,$\sigma'^{-1}\tau'$ is a specialization of \,$\sigma^{-1}\tau$;
when the latter specialization is generic, then \,$h$ and the
induced \,$\Del$-isomorphisms
\,$\ge\langle\sigma^{-1}\rangle\approx$
$\ge\langle\sigma'^{-1}\rangle$  and
$\ge\langle\sigma^{-1}\tau\rangle\approx$
$\ge\langle\sigma'^{-1}\tau'\rangle$  are compatible.

\end{enumerate}
\end{proposition}
\proof{ Since $\sigma,\sigma',\tau$ and $\tau'$ are \dsi s, it
follows from Kolchin's corresponding result \cite[Proposition
8(a), page 391]{kol73} for strong $\E$-isomorphisms, that
 ($\sigma'^{-1},\sigma'^{-1}\tau'$) is a specialization of
($\sigma^{-1},\sigma^{-1}\tau$) over $\g$. This remains a
specialization  over $\g$ when $\sigma,\sigma',\tau$ and $\tau'$
are considered as \de isomorphisms by \cite[Lemma 1, page
385]{kol73}.

Part 2 is proved a manner similar to that of part 1:
 ($\sigma',\tau'$) is a specialization of ($\sigma,\tau$),
when $\sigma,\sigma',\tau$ and $\tau'$ are considered as strong
$\E$-isomorphisms if and only if the induced isomorphisms
considered as non-differential isomorphisms are compatible. The
result then follows when $\sigma,\sigma',\tau$ and $\tau'$ are
again considered as \de isomorphisms.

Part 3 follows from the same considerations as in the previous
part.}

\begin{corollary}\label{Corollary to verify diff group axioms}
\begin{enumerate}
\item

If \,$\sigma'$ is a specialization of \,$\sigma$, then
\,$\sigma'^{-1}$ is a specialization of \,$\sigma^{-1}$. When the
former specialization is generic, then so is the latter, and the
induced isomorphisms \,$\ges\approx\gesp$ and
\,$\ge\langle\sigma^{-1}\rangle\approx$
$\ge\langle\sigma'^{-1}\rangle$ coincide.

\item Suppose that \,$\sigma'$ and \,$\tau'$ are generic
specializations of \,$\sigma$ and \,$\tau$, respectively,  such
that the induced \,$\Del$-isomorphisms \,$\ges\approx\gesp$ and
\,$\get\approx\getp$ are compatible, then \,$\sigma'\tau'$ is a
specialization of \,$\sigma\tau$. When the last specialization is
generic, and \,$h:\d\to\d'$ is an \,$\Del$-homomorphism between
subrings of \,$\ue$ such that \,$h$ and the induced
\,$\Del$-isomorphisms\, $\ge\langle\sigma\rangle\approx$
$\ge\langle\sigma'\rangle$  and
\,$\ge\langle\tau\rangle\approx\ge\langle\tau'\rangle$ are
compatible, then \,$h$ and the induced \,$\Del$-isomorphism
\,$\ge\langle\sigma\tau\rangle\approx$
$\ge\langle\sigma'\tau'\rangle$  are compatible.

\end{enumerate}
\end{corollary}

\proof{ The first assertion follows from part 1 of the
proposition, in the special case in which $\tau = \sigma, \tau' =
\sigma'$. Since $\gesi =\ges$ (Proposition \ref{Proposition:
fields of inverse isom are equal}), the second assertion follows
from part 3 of the proposition, in the special case in which $\tau
= \sigma,\tau' = \sigma'$, and $h$ is the induced
$\Del$-isomorphism $\ges\approx\gesp$.

 Because of part 1, one may replace $\sigma,\sigma'$ by
$\sigma^{-1},\sigma'^{-1}$.  Part 2 then follows from part 3 of
the proposition.}

\subsection{$\Ep$-Strongly Normal Extensions}\label{Strongly Normal Extensions}

\begin{definition}\label{Definition of E-strongly normal
extension} An \,$\Ep$-strongly normal extension \,$\g$ of the
\,$(E,\E)$-field \,$\f$ is a  finitely \,$(E,\E)$-generated
extension \,$\g$ of \,$\f$ such that every
\,$(E,\E)$-$\f$-isomorphism of \,$\g$ is \,$\Ep$-strong
$($Definition \ref{Definition of E-strong isomorphism} $)$.
\end{definition}

\begin{remark} If  \,$\g$ over \,$\f$ is \,$\Ep$-strongly normal,
it is not necessarily a strongly normal extension for \,$\E$
because  \,$\g$ over \,$\f$ might not be \,finitely
$\E$-generated. A strongly normal extension for \,$(\Del, \E)$ is
an \,$\Del$-strongly normal extension if each \de isomorphism
leaves invariant not only every element of \,$\g^{E,\E}$ but also
those of \,$\g^\E$.
\end{remark}

\begin{proposition}\label{Proposition that there are no new
constants in an E strongly normal extension} If \,$\g$ is an
\,$\Ep$-strongly normal extension of \,$\f$, then \,$\f$ and
\,$\g$ have the same field of \,$\E$-constants.
\end{proposition}

\proof {By Definition \ref{Definition of E-strong isomorphism}
\ref{st1}, the $\E$-constants in $\g$ are invariant under every
isomorphism of \gf.  Since any element of $\g$ fixed by all
$\Ep$-$\f$-isomorphisms of $\g$ is in $\f$ \cite[Corollary, page
388]{kol73}, the $\E$-constants of $\g$ are contained in $\f$.}

\begin{proposition}\label{Proposition on the one sided condition
for E-strongly normal} Let \,$\g$ be a finitely
\,$(E,\E)$-generated extension of \,$\f$ having the same field of
\,$\E$-constants as \,$\f$. Let \,$\sigma_{1} \ldots \sigma_{r}$
be \,$(E,\E)$-$\f$-isomor-phisms of \,$\g$ such that every
\,$(E,\E)$-$\f$-isomorphism of \,$\g$ is an
\,$(E,\E)$-$\g$-specialization of one of these. If \,$\sigma_{k}\g
\subset \g\ue$ for all \,$k$, {\rm (}$1 \leq k \leq r${\rm )},
then \,$\g$ is \,$\Ep$-strongly normal over \,$\f$.
\end{proposition}

\proof { Let $\sigma$ be any $(E,\E)$-$\f$-isomorphism of $\g$.
Since $\g^\E=\f^\E$, $\sigma$ fixes $\g^\E$. By considering
$\sigma$ as a $\Delta$-homomorphism and the remark after
\cite[Proposition 6, page 390]{kol73}, $\sigma\g \subset \g\ue$
since $\sigma_{i}\g \subset \g\ue$.

To prove that $\sigma$ is $\Ep$-strong, it remains to show $\g
\subset \sigma \g \ue$. Following the technique of the proof in
\cite[Proposition 10, page 393]{kol73}, one may show that the
$(E,\E)$-$\f$-isomorphism $\sigma^{-1}:\sigma\g\approx\g$ can be
extended to an $(E,\E)$-$\f$-isomorphism $\varphi$ of $\gsg$
because $\u$ is \de universal over $\f$. The restriction of
$\varphi$ to $\g$ is an $(E,\E)$-$\f$-isomorphism $\tau$ of \gf.
Thus, $\varphi:\gsg\approx\tau\g\cdot\g$ is an
$(E,\E)$-$\f$-isomorphism, $\varphi\g =\tau\g$, $\varphi(\sigma\g)
= \g$, and $\varphi(\ges)=\get$.  By the final result of the last
paragraph, $\tau\g\subseteq \g\get$. Therefore
$\g=\varphi^{-1}(\tau\g)\subseteq\varphi^{-1}(\g\get)=
\varphi^{-1}\g\cdot\varphi^{-1}(\get)=\sigma\g\cdot\ges \subset
\sigma \g \ue$.}

\begin{corollary}\label{Corollary on compositum of E-strong
extensions} Let \,$\g_{1}$ and \,$\g_{2}$ be extensions of \,$\f$
such that \,$\g_{1} \g_{2}$ has the same field of \,$\E$-constants
as \,$\f$. If \,$\g_{1}$ and \,$\g_{2}$ are \,\dsn ~over \,$\f$,
then so is \,$\g_{1}\g_{2}$.
\end{corollary}

\proof { Obviously $\g_{1} \g_{2}$ is a finitely \de generated
extension of $\f$.  If $\sigma$ is any isomorphism of $\g_{1}
\g_{2}$ over $\f$, then the restriction $\sigma_{i}$ of $\sigma$
to $\g_{i}$ is an \dsi ~of $\g_{i}$ so that $\sigma(\g_{1}
\g_{2})= \sigma_{1}\g_{1} \cdot \sigma_{2} \g_{2} \subset
\g_{1}\ue\cdot \g_{2}\ue= (\g_{1} \g_{2})\cdot\ue$. It follows by
Proposition \ref{Proposition on the one sided condition for
E-strongly normal} that $\g_{1} \g_{2}$ is an \dsn ~extension of
$\f$.}\bigskip

\begin{proposition}\label{Identifies strong isomorphisms with automorphisms}
  Let \,$\g$ be any \,$\Delta$-field in \,$\u$.
Each \,$\Ep$-strong \de \linebreak isomorphism of \,$\g$ can be
extended to a unique \,\de-automorphism of \,$\g\ue$ over \,$\ue$.
Conversely, the restriction to \,$\g$ of each \,\de-automorphism
of \,$\g\ue$ over \,$\ue$ is a \,\dsi ~of \,$\g$.
\end{proposition}

\proof{ By \cite[Corollary 1, page 87]{kol73}, $\g$ and $\ue$ are
linearly disjoint over $\ge$.  Also, if $\sigma$ is any \dsi~of
$\g$ (Definition \ref{Definition of E-strong isomorphism}), then
$\sg$ and $\ue$ are also linearly disjoint over $\ge$. Therefore
$\sigma$ can be extended to a unique \de isomorphism $s: \g \ue
\approx \sigma \g \cdot \ue$ over $\ue$. Because $\sigma$ is \dsi,
$\sigma \g \ue = \g \ue$, and $s$ is an \de automorphism of $\g
\ue$ over $\ue$. The converse is clear. }
\bigskip

This proposition canonically identifies the set of all \dsi s~of
$\g$ with the set of all \de automorphisms of $\g\ue$ over $\ue$.
Because the set of all \de automorphisms of $\g\ue$ over $\ue$ has
a natural group structure, this identification induces a group
structure on the set of all \dsi s of $\g$.  If $\f$ is an \de
subfield of $\g$, the set of all \dsi s of \gf~ can be canonically
identified with the group $G$ of all \de automorphisms of $\g \ue$
over $\f\ue$, which is a subgroup of the group of all \de
automorphisms of $\g\ue$ over $\ue$.

Recall the definitions of the $\Ep$-type, $\Del$-dimension and
typical $\Ep$-dimension  of a pre $\Ep$-set in \cite[page
31]{kol85}. If $\h$ over $\f$ (considered as an $\Ep$-field) is
$\Ep$-extension that is finitely $\Ep$-generated by
$\rho=(\rho_1,\ldots,\rho_n)$, $\omega_{\rho/\f}$ will denote the
$\Ep$-transcendence polynomial of $\rho$ over $\f$ \cite[page
117]{kol73}.

\begin{proposition}\label{finitely generated}
Let \,$\g$ be an \,$\Del$-strongly normal extension of \,$\f$, and
let \,$\c$ denote the field of \,$\E$-constants of \,$\f$. For
every isomorphism \,$\sigma$ of \,$\g$ over \,$\f$, define
\,$\cs=(\g\sigma\g)^\E$. Then \,$\cs$, as an $\Ep$-field extension
of $\c$, is finitely $\Ep$-generated over $\c$.  Moreover, \,$\g$
is \,finitely $\Del$-generated over \,$\f$, and, for every
isolated isomorphism \,$\sigma$ of \,$\g$ over \,$\f$, the
\,$\Ep$-type $(${\it resp.} $\Del$-dimension, typical
$\Ep$-dimension$)$ of \,$\cs$ over \,$\c$ is equal to the
\,$\Ep$-type \,$(${\it resp.} \,$\Del$-dimension, typical
$\Ep$-dimension$)$ of  \,$\g$ over \,$\f$.
\end{proposition}
\proof{That $\cs$ is a finitely $\Del$-generated field extension
of $\c$ for every isomorphism $\sigma$ of \gf~is Proposition
\ref{constants of gsigmag finitely generated}.  To show $\g$ is
finitely $\Del$-generated over $\f$, let $\sigma$ be an isolated
\de isomorphism of \gf~that specializes to the identity
isomorphism.  By part b of \cite[Corollary , page 388]{kol85},
$\sigma$ leaves fixed the algebraic closure $\fo$ of $\f$ in $\g$.
Let $\eta=(\eta_{1},\ldots,\eta_{n})$ be a family of \de
generators of \gf, i.e. $\g =\f\langle\eta\rangle_{\Ep,\E}$, and
let $\xi=(\xi_1,\ldots,\xi_r)$ be a family of $\Del$-generators of
$\cs$ over $\c$, i.e. $\cs=\c\langle \xi\rangle_\Ep$. Since
$\g\langle \sigma\eta
\rangle_{\Ep,\E}=\gsg=\g\cs=\g\langle\xi\rangle_{\Ep,\E}$, each
coordinate of $\xi$ is in the $\Del$-field generated over $\g$ by
a finite number of $\E$-derivatives of $\sigma\eta$. Denote the
set $\E$-derivatives of $\sigma\eta$ by
$\vartheta=(\vartheta_{1},\ldots,\vartheta_{s})$. Then $\vartheta$
$\Del$-generates $\gsg$ over $\g$.
\begin{claim}
\,$\fo\langle\vartheta\rangle_\Ep=\sigma\g$\end{claim}\proof{ By
the definition of $\vartheta$,
$\fo\langle\vartheta\rangle_\Ep\subset\sigma\g$. Let
$\alpha\in\sigma\g$. Then
$\alpha\in\gsg=\g\langle\sigma\eta\rangle_{\Ep,\E}
=\g\cdot\fo\langle\vartheta\rangle_\Ep $. If $(\gamma_{i})_{i\in
I}$ is a basis for $\fo\langle\vartheta\rangle_\Ep$ over $\fo$,
$\alpha=(\Sigma~g_{i}\gamma_{i})/(\Sigma~g'_{j}\gamma_{j})$, with
$g_{i}$ and $g'_{j}$ in $\g$ and not all the $g'_{j}$ are $0$.
  Therefore,
$\Sigma~g'_{j}(\gamma_{j}\alpha)-\Sigma~g_{i}\gamma_{i}=0$, and
the family $(\gamma_{j}\alpha, \gamma_{i})$ of elements of
$\sigma\g$ is linearly dependent over $\g$. Since $\sigma$ is
isolated, $\sigma\g$ and $\g$ are algebraically disjoint over $\f$
\cite[Comment on page 387]{kol73}.  {\it A fortiori}, they are
also algebraically disjoint over $\fo$. Since $\g$ is regular over
$\fo$, $\sigma\g$ and $\g$ are linearly disjoint over $\fo$
\cite[Theorem 3, page 57]{langalggeom}. By this disjointness, the
family $(\gamma_{j}\alpha, \gamma_{i})$ is linearly dependent over
$\fo$. So there exists $f_{i}$ and $f'_{i}$ elements of $\fo$, not
all $0$, such that
$\Sigma~f'_{j}(\gamma_{j}\alpha)-\Sigma~f_{i}\gamma_{i}=0$.
Because the $\gamma_{j}$ are linearly independent over $\fo$,
$\Sigma~f'_{j}\gamma_{j}\not=0$.   Therefore
$\alpha=(\Sigma~f_{i}\gamma_{i})/(\Sigma~f'_{j}\gamma_{j})
\in\fo\langle\vartheta\rangle$.}\bigskip

Since the $\Ep$-field $\sigma\g = \fo\langle\vartheta\rangle_\Ep$
is finitely $\Del$-generated over $\fo$,
$\g=\fo\langle\sigma^{-1}\vartheta\rangle_\Ep$ (as $\Ep$-fields)
is also finitely $\Del$-generated over $\fo$. Because any
intermediate extension of a finitely $\Ep$-generated extension is
finitely $\Ep$-generated \cite[Chapter 2, Proposition 14, p.
112]{kol73}, it follows that $\fo$ is finitely \de generated over
$\f$, and, hence, also finitely $\Del$-generated  over $\f$
(because $\fo$ is algebraic). Thus $\g$ is finitely
$\Del$-generated over $\f$.

  Then
\[\omega_{\sigma^{-1}\vartheta/\f}
=\omega_{\vartheta/\f}=\omega_{\vartheta/\g}\] since the first
equality would be true for any $\Ep$-$\f$-isomorphism $\sigma$
\cite[page 387]{kol73} and the second equality holds by
\cite[Comment on page 117]{kol73} because $\sigma \g$ and $\g$ are
algebraically disjoint over $\f$ \cite[Comment on page
387]{kol73}. Also,\[\omega_{\xi/\g}=\omega_{\xi/\c}\] because $\g$
and $\c\langle \sigma \rangle$ are linearly disjoint over $\c$
 \cite[Corollary 1,page 87]{kol73} and \cite[Comment on page 117]{kol73}.
Because $\vartheta$ and $\xi$ both $\Ep$-generate $\gsg$ over
$\g$, the $\Ep$-birational invariants ($\Ep$-type,
$\Del$-dimension, typical $\Ep$-dimension) of
$\omega_{\vartheta/\g}$ and $\omega_{\zeta/\g}$ are equal
(\cite[page 118]{kol73} or \cite[page 7]{kol85}).  By utilizing
the above equalities, the $\Ep$-birational invariants of
$\omega_{\sigma^{-1}\vartheta/\g}$ and $\omega_{\zeta/\c}$ are
also the same.  Thus the $\Ep$-type ({\it resp.} $\Del$-dimension,
typical $\Ep$-dimension) of $\cs$ over $\c$ is equal to the
$\Ep$-type ({\it resp.} $\Del$-dimension, typical $\Ep$-dimension)
of  \,$\g$ over $\f$.}

\subsection{ Pre $\Del$-Sets and $\Ep$-Groups}\label{pre E-sets}

The objects in the category of pre $\Del$-$\c$-sets \cite[Chapter
1]{kol85} are defined as follows.
\begin{definition}\label{definition of presets}\label{new definition of pre sets}Let \,$\c$ be an \,$\Del$-field and let \,$\v$ be a
universal \,$\Ep$-field extension of \,$\c$.  A {\it pre
$\Del$-$\c$-set} {\rm(}relative to $\v${\rm)} is a set \,$A$ for
which there are given
\begin{enumerate}
\item for each element \,$x\in A$, an \,$\Del$-finitely generated
field extension \,$\c \langle x \rangle$ over \,$\c$, \item a pre
order on \,$A$ called \,{\rm $\Ep$-specialization over $\c$ or,
more simply, $\Ep$-$\c$-specialization} {\rm(}which shall be
indicated by the notation $x\rightarrow x'${\rm)}, \item for each
pair \,$(x,x')$ in \,$A^2$ with \,$x\leftrightarrow x'$, an
\,$\Del$-isomorphism $\,S_{x',x}: \c\langle x \rangle \approx \c
\langle x' \rangle$ over \,$\c$,

\bigskip
all subject to the following axioms.
\begin{enumerate}
\item[DAS1] $A$ has a finite subset $\Phi$ such that, for each
$x'\in A$, there exists an $x \in \Phi$ with $x\rightarrow x'$ .

\item[DAS2a]  If $x,x',x'' \in A, x\leftrightarrow x'$, and $x'
\leftrightarrow x''$, then $S_{x'',x'} \circ S_{x',x} =S_{x'',x}$.

\item[DAS2b] If $x \in A$ and $S: \f\langle x \rangle \approx \c'$
is a $\Del$-field isomorphism over $\c$, then there exists a
unique $x' \in A$ with $x\leftrightarrow x'$ such that $ \f\langle
x' \rangle =\c' $ and $S_{x',x} = S$.
\end{enumerate}
\end{enumerate}
\end{definition}

It can be shown \cite{landesman} that the $\v$-valued points of
any $\Ep$-scheme in the sense of Kovacic \cite{kov1} over $\c$ is
a pre $\Del$-$\f$-set.

\begin{definition}\label{Definition of an F-generic element}
A subset \,$B$ of the pre \,$\Del$-$\c$-set \,$A$ is called
$\c$-{\rm irreducible} $(${\rm in} $A)$ if there exists an $x \in
A$ such that $B$ is the set of all elements of \,$A$ that are
\,$\Ep$-$\c$-specializations of \,$x$. Such an $x$ will be called
an {\rm $\c$-generic element} of \,$B$. A maximal $\c$-irreducible
subset of \,$A$ is called an $\c$-{\rm component} of \,$A$.
\end{definition}

Kolchin defines pre $\Del$-$\c$-maps, as below, in a manner such
that the composition of two is not necessarily a third.

\begin{definition}\label{premapping}  Let \,$A$ and \,$B$ be pre \,$\Del$-$\c$-sets.
A pre \,$\Del$-$\c$-mapping of \,$A$ to \,$B$ is a mapping \,$f$
of a subset \,$A_f$ of \,$A$ into \,$B$ with the following four
properties:

\begin{enumerate}
\item  the \,$\c$-generic elements of the components of \,$A$ are
contained in \,$A_f$; \item if \,$x \in A_f$, then \,$\c \langle
f(x) \rangle \subset \f\langle x \rangle $; \item  if \,$x \in A,
x' \in A_f$, and \,$x \rightarrow x'$,  then \,$x \in A_f$ and
\,$f(x) \rightarrow f(x')$; \item  if \,$x, x' \in A_f$ and \,$x
\leftrightarrow x'$, then \,$S_{x',x}$ extends \,$S_{f(x'),f(x)}$.
See Diagram \ref{def of pre E-map} below.
\end{enumerate}

\end{definition}
\begin{diagram}\label{def of pre E-map}
\[
\begin{CD}
\c\langle x \rangle   @+>{S_{x',x}}>>[60]     \c\langle x' \rangle \\
  @AA{inclusion}A                             @AA{inclusion}A         \\
\c\langle f(x) \rangle  @+>{S_{f(x'),f(x)}}>>[60]     \c\langle
f(x') \rangle
\end{CD}
\]
\end{diagram}

\bigskip
 To have morphisms that are composable, pre
$\Del$-$\c$-mappings from $A$ to $B$ that are everywhere defined
(that is $A_f=A$) are taken to be the morphisms in the category of
pre $\Del$-$\c$-sets.

\begin{definition}
The category of pre \,$\Del$-$\c$-sets {\rm(}relative the
universal \,$\Del$ -field $\v${\rm)} is the category with pre
\,$\Del$-$\c$-sets as objects and with everywhere defined
\,$\Del$-$\c$-mappings as morphisms.
\end{definition}

It can be shown \cite{landesman} that the functor of $\v$-valued
points is a functor from the the category of $\Ep$-$\c$-schemes to
category of pre $\Del$-$\c$-sets {\rm(}relative the universal
\,$\Del$ -field $\v${\rm)}.

\begin{definition}\label{Definition of E-group}
{\rm \cite[page 33]{kol85}} An \,$\Del$-$\c$-group {\rm(}relative
to the universal \,$\Del$-field \,$\v${\rm)} is a set \,$G$ which
has both a group structure {\rm(}usually written
multiplicatively{\rm)} and a pre \,$\Del$-$\c$-set structure
relative to the universal \,$\Del$-field \,$\v$, subject to the
following axioms.

\begin{enumerate}

\item[DAG1a] If \,$x_1,x_2 \in G$, then \,$\c \langle x_1 x_2
\rangle \subset \c \langle x_1 \rangle \c \langle x_2 \rangle$.

\item[DAG1b] If \,$x_1,x_2 \in G$, then \,$\c \langle {x_1}^{-1}
x_2 \rangle \subset \c \langle x_1 \rangle \c \langle x_2
\rangle$.

\item[DAG2a]  If \,$x_1 ,x_2 , x_1' ,x_2' \in G$ and $x_1
\leftrightarrow x_1'$ , $x_2 \leftrightarrow x_2'$, and
$S_{x_1',x_1} , S_{x_2',x_2}$ are compatible\footnote{Let $R_1$
and $R_2$ be rings over the ring $R$, all in a common field, and
let $\phi_1$ and $\phi_2$ be ring homomorphisms of $R_1$ and
$R_2$, respectively, into another field over $R$. Then $\phi_1$
and $\phi_2$ are {\it compatible}, if there exists an extension of
$\phi_1$ and $\phi_2$ to the ring $R[R_1,R_2]$.},  then \,$x_1x_2
\rightarrow x_1'x_2'$. If moreover  \,$x_1x_2 \leftrightarrow
x_1'x_2'$, and \,$h$ is an \,$\Ep$-$\c$-homomorphism of finitely
\,$\Ep$-generated \,$\Del$-overrings of \,$\c$ in \,$\u$ such that
\,$h, S_{x_1',x_1} , S_{x_2',x_2}$ are compatible, then \,$h$ and
\,$S_{x_1'x_2', x_1x_2}$ are compatible.

\item[DAG2b]  If \,$x_1 ,x_2 , x_1' ,x_2' \in G$ and \,$x_1
\rightarrow x_1'$ , \,$x_2 \rightarrow x_2'$, then there exist
elements \,$x_1^*,x_2^* \in G$ with \,$x_1 \leftrightarrow x_1^*$,
$x_2 \leftrightarrow x_2^*$ such that \,$x_1^*, x_2^*$ are
algebraically disjoint over \,$\c$ and \,$x_1^*x_2^* \rightarrow
x_1'x_2'$ \,$($i.e., \,$\c\langle x_1^*\rangle$ and \,$\c\langle
x_2^*\rangle$ are algebraically disjoint over \,$\c)$, and such
that, if \,$x_1^*x_2^* \leftrightarrow x_1'x_2'$ and \,$x_2^*
\leftrightarrow x_2'$, then \,$S_{x_1'x_2',x_1^*x_2^*}$,
\,$S_{x_2',x_2^*}$ are compatible.

\item[DAG2c]  If \,$x_1 ,x_2 , x_1' ,x_2' \in G$ and \,$x_1
\leftrightarrow x_1'$ , $x_2 \leftrightarrow x_2'$, and
\,$S_{x_1',x_1}$,  \,$S_{x_2',x_2}$ are compatible, then
\,$x_1^{-1}x_2 \rightarrow x_1'^{-1}x_2'$. If moreover
\,$x_1^{-1}x_2 \leftrightarrow x_1'^{-1}x_2'$, and \,$h$ is an
$\Ep$-$\c$-homomorphism of finitely \,$\Ep$-generated
\,$\Del$-overrings of \,$\c$ in \,$\u$ such that
\,$h$,\,$S_{x_1',x_1}$,\,$S_{x_2',x_2}$ are compatible, then
\,$h$, \,$S_{x_1'^{-1}x_2',x_1^{-1}x_2}$ are compatible.

\item[DAG2d]  If \,$x_1 ,x_2 , x_1' ,x_2' \in G$ and \,$x_1
\rightarrow x_1'$ , \,$x_2 \rightarrow x_2'$, then there exist
elements \,$x_1^*, x_2^* \in G$ with \,$x_1 \leftrightarrow x_2^*
,~ x_2 \leftrightarrow x_2^*$ such that \,$x_1^*$ and \,$x_2^*$
are algebraically disjoint over \,$\c$ and \,$x_1^{* -1}x_2^*
\rightarrow x_1'^{-1}x_2'$.

\item[DAG3] The unity element 1 of \,$G$ is contained in an
\,$\c$-component $($Definition \ref{Definition of an F-generic
element}$)$ of \,$G$ having an \,$\c$-generic element $x$ that is
regular over \,$\c$, i.e. \,$\c$ is algebraically closed in
\,$\c\langle x \rangle$.

\end {enumerate}

\end{definition}

It can be shown \cite{landesman} that the functor of $\v$-valued
points applied to an $\Ep$-$\c$-group scheme of $\Ep$-$\c$-finite
type is an $\Del$-$\c$-group {\rm(}relative the universal \,$\Del$
-field $\v${\rm)}.

\subsection{$\Ep$-Groups of Isomorphisms}\label{group of automorphisms}

\begin{theorem}\label{existance of the differential group of
isomorphisms}
  Let \,$\g$ be an \dsn extension of the $\Delta$-field \,$\f$ with field
of \,$\E$-constants \,$\c$, and let
\,$G=\Isom^{\Ep,\E}_\f(\g,\v)$. With the pre \,$\Ep$-$\c$-set
structure defined above, \,$G$ is an \,$\Ep$-$\c$-group.
Furthermore, the field \,$\g$ is finitely $\Ep$-generated, and, as
such, the \,$\Ep$-type $(${\rm resp.} \,$\Del$-dimension, typical
\,$\Ep$-dimension$)$ of \,$\g$ over \,$\f$ equals the \,$\Ep$-type
$(${\rm resp.} \,$\Del$-dimension, typical \,$\Ep$-dimension$)$ of
the \,$\Del$-$\c$-group \,$G$.
\end{theorem}

Proof: It will be verified that $G$ satisfies the properties of an
$\Ep$-$\c$-group \ref{Definition of E-group}. By Proposition
\ref{Identifies strong isomorphisms with automorphisms} and the
subsequent remark, the set $G$ has the structure of a group.

Endow $G$ with the following pre E-C-set
structure.\begin{enumerate}\item To each \,$\sigma \in G$
associate \,$ \c\langle \sigma \rangle = (\gsg)^{\E}$, considered
as an \,$\Del$-field extension of \,$\c=\g^{\Delta}$. This is
finitely $\Del$-generated by Proposition \ref{constants of gsigmag
finitely generated}.
 \item For each \,$(\sigma,\sigma')\in
G^{2}$, let \,$\sigma\to\sigma'$ mean that \,$\sigma'$ is an
\,$(\Ep,\E)$-$\f$-specialization of \,$\sigma$ \,$($Definition
\ref{pre-order no X^r}$)$. \item For each \,$(\sigma,\sigma')\in
G^{2}$ with \,$\sigma\leftrightarrow\sigma'$ {\rm(}that is, with
\,$\sigma'$ a generic \,$\Ep$-$\f$-specialization of
\,$\sigma${\rm)}, then there exists a unique $\g$-\de isomorphism
$\g \sigma \g \approx \g \sigma' \g$ that, for each $\alpha \in
\g$, maps $\alpha$ to $\alpha$ and $\sigma\alpha$ onto
$\sigma'\alpha$. The restriction of this
$(E,\Delta)$-$\g$-isomorphism to the $\Delta$-constants yields an
$\Del$-$\c$-isomorphism $S_{\sigma',\sigma}: \c\langle \sigma
\rangle\approx\c\langle \sigma' \rangle$ over $\c$ which is called
the {\it $\Del$-$\c$-isomorphism induced by the generic
$(E,\Delta)$-$\g$-specialization}.\end{enumerate}

By \cite[Proposition 1(c), page 387]{kol73}, this pre
$\Ep$-$\c$-set structure satisfies {\it DAS1}. Axioms {\it DAS2a}
and {\it DAS2b} follow from the next proposition.

\begin{proposition}\label{DAS2}
Let \,$\sigma$ be an \dsi~of \,$\g$.
\begin{enumerate}
\item If \,$\sigma'$ is a generic
\,$(E,\Delta)$-$\g$-specialization of \,$\sigma$, and \,$\sigma''$
is a generic \,$(E,\Delta)$-$\g$-specialization of \,$\sigma'$
{\rm (}and therefore of \,$\sigma${\rm )}, then the composite of
the induced \,$\Del$-$\g^\E$-isomorphisms \,$\ges\approx\gesp$ and
\,$\gesp\approx\gespp$ is the induced \,$\Del$-$\g^\E$-isomorphism
\,$\ges\approx\gespp$.

\item If \,$ S: \ges \approx \c'$ is any \,$\Del$-isomorphism over
\,$\ge$, then there exists a unique generic
\,$(E,\Delta)$-$\g$-specialization \,$\sigma'$  of \,$\sigma$ such
that \,$\gesp =\c'$, and \,$S$ is the induced
\,$\Del$-$\g^\E$-isomorphism \,$\ges \approx \gesp$.
\end{enumerate}
\end{proposition}

\proof{
\begin{enumerate}
\item This follows from the corresponding facts about the
$(E,\Delta)$-$\g$-isomor-\linebreak phisms $\g \sigma \g \approx
\g \sigma' \g$, $\g \sigma' \g \approx \g \sigma'' \g$ and $\g
\sigma \g \approx \g \sigma'' \g$.

\item
  $\ges$~ and $\g$ are linearly disjoint over $\ge$,
as are $\c'$ and $\g$:  therefore $S$ can be extended to an
$(E,\Delta)$-$\g$-isomorphism $T:  \g\ges\approx\g\c'$. The
composite mapping $\g\approx^{\sigma}
 \sigma \g \subset \g \sigma \g=\g\ges\approx^{T}
\g\c'$  yields an $(E,\Delta)$-$\ge$-isomorphism  $\sigma': \g
\approx T(\sigma \g)$. Since $T\sigma\g=\sigma'\g$,\linebreak $T:
 \g \sigma \g \approx \g \sigma'
\g$. Therefore $\sigma'$ is a generic
$(E,\Delta)$-$\g$-specialization of $\sigma$, $\c'=\gesp$
\cite[Corollary 2, p. 88]{kol73}, and $S$ is the induced
$\Ep$-$\g^\E$-isomorphism $S_{\sigma',\sigma}$. The uniqueness is
clear.\end{enumerate}}\bigskip

Axiom {\it DAG 1} follows from Proposition \ref{Proposition:
fields of inverse isom are equal}. Axiom {\it DAG 2a} follows from
part 2 of Corollary \ref{Corollary to verify diff group axioms}.
Part 3 of Proposition \ref{Proposition to verify diff group
axioms} implies Axiom {\it DAG 2c}.

To prove parts {\it DAG2b} and {\it DAG2d}, let
$\sigma,\sigma',\tau,\tau'$ be \dsi s of \gf ~ with
$\sigma\to\sigma'$ and  $\tau\to\tau'$. Fix  a family  $\eta
=(\eta_{1}\ldots\eta_{n})$ of \de generators of \gf, and let $p$
({\it resp.} $q$)  denote the defining \de ideal of
$\sigma^{-1}\eta$ ({\it resp.} $\tau\eta$) in the \de-polynomial
algebra $\g\lbrace y_{1},\ldots,y_{n}\rbrace_{\Ep,\E}$ ({\it
resp.} $\g\lbrace z_{1},\ldots,z_{n}\rbrace_{\Ep,\E}$). Let
$\g_{a}$ denote the algebraic closure of $\g$ in $\u$. Then
$\g_{a}p$ and $\g_{a}q$ have components $p_{1},\ldots,p_{r}$ and
$q_{1},\ldots,q_{s} $ such that the quotient fields
$QF(\g_a\lbrace y_{1},\ldots,y_{n}\rbrace_{\Ep,\E}/p_{i})$ for
$i=1,\ldots, r$ and $QF(\g_a\lbrace
z_{1},\ldots,z_{n}\rbrace_{\Ep,\E}/q_{j})$ for $j=1,\ldots, s$ are
regular over $\g_{a}$ \cite[Proposition 3, page 131]{kol73}. By
\cite[Corollary, page 132]{kol85}, each \de-ideal $r_{kl}=
\{p_{k}\cup q_{l}\}_{(\Ep,\Delta)}$ of $\g_{a}\lbrace
y_{1},\ldots,y_{n},z_{1},\ldots,z_{n}\rbrace_{\Ep,\E}$ is prime.
Therefore, $r_{kl}$ has a $\g_a$-generic \de zero
$(\eta^{(k,l)},\xi^{(k,l)})$ where $\eta^{(k,l)}$ is a generic
zero of $r_{kl}\cap\g_{a}\lbrace
y_{1},\ldots,y_{n}\rbrace_{\Ep,\E}=p_{k}$ and therefore of
$p_{k}\cap\g\lbrace y_{1},\ldots,y_{n}\rbrace_{\Ep,\E}=p$, so that
$\eta^{(k,l)}$ is a $\g$-generic \de specialization   of
$\sigma^{-1}\eta$ over $\g$ and hence over $\f$.  Therefore
$\eta^{(k,l)}$ is the image of $\eta$ by an \dsi ~of \gf, which is
denoted by $\sigma_{kl}^{-1}$ and is defined by
$\eta^{(k,l)}=\sigma_{kl}^{-1}\eta$. By \cite[Lemma 2, page
386]{kol73}, $\sigma^{-1}\leftrightarrow\sigma_{kl}^{-1}$.
Similarly $\xi^{(k,l)}=\tau_{kl}\eta$ for some \dsi ~$\tau_{kl}$
of \gf ~with $\tau\leftrightarrow\tau_{kl}$. By hypothesis
$\sigma\to\sigma'$, whence $\sigma^{-1}\to\sigma'^{-1}$ (part 1 of
the Corollary \ref{Corollary to verify diff group axioms}) so that
$\sigma'^{-1}\eta$ is an \de zero of p and hence of some p$_{k}$.
Similarly, $\tau'\eta$ is an \de zero of some $q_{l}$.  Thus,
($\sigma'^{-1}\eta,\tau'\eta)$ is an \de zero of $r_{kl}$, thus
$(\sigma_{kl}^{-1}\eta,\tau_{kl}\eta) \rightarrow_{\g_{a}}
(\sigma'^{-1}\eta,\tau'\eta)$ and hence over $\g$. It follows
\cite[Lemma 2, page 386]{kol73}, that
$(\tau_{kl},\sigma_{kl}^{-1}) \rightarrow_\g
(\tau',\sigma'^{-1})$, and hence by Proposition \ref{Proposition
to verify diff group axioms} 1 and 2, that
$(\tau_{kl}^{-1},\tau_{kl}^{-1}\sigma_{kl}^{-1}) \rightarrow_\g
(\tau'^{-1},\tau'^{-1}\sigma'^{-1})$ and that if
$\tau_{kl}^{-1}\sigma_{kl}^{-1}\leftrightarrow\tau'^{-1}\sigma'^{-1}$
and $\tau_{kl}^{-1}\leftrightarrow\tau'^{-1}$, then the induced
$\Del$-$\c$-isomorphisms
$\c\langle\tau_{kl}^{-1}\sigma_{kl}^{-1}\rangle\approx\c
\langle\tau'^{-1}\sigma'^{-1}\rangle$ and
$\c\langle\tau_{kl}^{-1}\rangle\approx\c \langle\tau'^{-1}\rangle$
are compatible. By part 1 of the Corollary \ref{Corollary to
verify diff group axioms}, then
$\sigma_{kl}\tau_{kl}\to\sigma'\tau'$ and if
$\sigma_{kl}\tau_{kl}\leftrightarrow\sigma'\tau'$ and
$\tau_{kl}\leftrightarrow\tau'$, then the induced
$\Del$-$\c$-isomorphisms
$\c\langle\sigma_{kl}\tau_{kl}\rangle\approx\c
\langle\sigma'\tau'\rangle$ and
$\c\langle\tau_{kl}\rangle\approx\c \langle\tau'\rangle$ are
compatible. This proves {\it DAG2b}, and (because
$\sigma^{-1}\to\sigma'^{-1}$  whenever $\sigma\to\sigma'$) also
part {\it DAG2d}.

To prove axiom {\it DAG3}, one must show that if $\sigma$ is an
isolated isomorphism of \gf ~with $\sigma\to id_{\g}$, then $\cs$
is regular over $\c$. Since $\sigma\to id_{\g}$, $\sigma \fo =\fo$
\cite[Proposition 2(b), page 388]{kol73}.  Since $\g$ is regular
over $\fo$, clearly $\sigma \g$ is regular over $\sigma \fo =\fo$.
By \cite[Remark, page 387]{kol73}, $\sigma \g$ is algebraically
disjoint from $\g$ over $\f$, and, {\it a fortiori}, they are
algebraically disjoint over $\fo$. Because $\g$ is regular over
$\fo$, $\sigma \g$ is linearly disjoint from $\g$ over $\fo$
\cite[Theorem 3, page
57]{langalggeom}.\\
\[
\begin{CD}
\g              @>>>  \g\sigma\g\\
    @AAA                      @AAA \\
 \fo       @>>> \sigma\g,     \\
\end{CD}
\]\bigskip

Recall \cite[Corollary 6, page 58]{langalggeom}, that if $K$ and
$L$ are field extensions of field $k$ in a larger field and if
they linearly disjoint over $k$, then $K$ is regular over $k$ if
and only if $KL$ is regular over $L$.  Therefore, $\gsg$ is
regular over $\g$. Since $\g$ and $\cs$ are linearly disjoint over
$\c$ \cite[Corollary 2, page 88]{kol73}

\[
\begin{CD}
\g              @>>>  \g \c\langle\sigma\rangle =\g\sigma\g\\
    @AAA                      @AAA \\
 \c        @>>> \c\langle\sigma\rangle.     \\
\end{CD}
\] \\
and $\g \c\langle\sigma\rangle =\g\sigma\g$, that $\g\sigma\g$ is
regular over $\g$ implies $\c\langle\sigma\rangle$ is regular over
$\c$, which is {\it DAG3}. This establishes $G$ as a
$\Del$-$\c$-group.

\begin{definition}
    By virtue of Theorem \ref{existance of the differential group of
isomorphisms}, the set of \,$\Ep$-strong \,$(\Ep,\E)$-isomorphisms
of the \,$\Ep$-strongly normal extension \,$\g$ over \,$\f$ has a
natural structure of an \,$\Del$-$\c$-group relative to the
\,$\Ep$-universal field \,$\ue$. This \,$\Del$-$\c$-group is
called the Galois group of \,$\g$ over \,$\f$, and it is denoted
by \,$G_{\Del}(\g /\f)$ or \,$G(\g /\f)$.  The \,$\c$-component of
the identity of \,$G_{\Del}(\g /\f)$ is denoted by
\,$G_{\Del}^{\circ}(\g /\f)$ or \,$G^{\circ}(\g /\f)$.
\end{definition}

\begin{definition}
If \,$G$ is any \,\dcg, a \,$G$-{\rm extension of} \,$\f$ is any
\dsn ~extension \,$\g$ of \,$\f$ such that \,$\ggf$ is \dc
isomorphic to an \dcsg~of \,$G_{\k}$. When \,$\ggf$ is \,\dc
isomorphic to \,$G_{\k}$ itself,
 the \de extension \,$\g$ over \,$\f$ is called {\rm full}.  A {\rm linear extension} of $\f$
is an $\Del$-{\rm GL(n)}-{\rm extension of} $\f$ for some natural
number $n$.
\end{definition}

\subsection{Extending the Constants.}

\begin{definition}
{\rm \cite[page 48]{kol85}} Let $\c$ be an \,$\Ep$-field, let
\,$\v$ be another \,$\Ep$-field that is \,$\Ep$-universal over
\,$\f$, and let \,$\d \subset \v$ be an \,$\Ep$-field containing
\,$\f$ over which \,$\v$ is \,$\Ep$-universal. Let \,$G$ be an
\,$\Del$-$\c$-group relative to \,$\v$, and let \,$H$ be an
\,$\Del$-$\d$-group relative to \,$\v$. An {\rm
$\Del$-$(\d,\c)$-homomorphism of $H$ into $G$} is a group
homomorphism \,$f:  H \rightarrow G$ that satisfies the following
three conditions:
\begin{enumerate}
\item if \,$y \in H$, then \,$\d \langle y \rangle \supset
      \c \langle f(y) \rangle$,

\item  if \,$y, y' \in H$ and \,$y \rightarrow y'$ over \,$\d$,
then \,$f(y) \rightarrow f(y')$ over $\c$, \item  if \,$y, y' \in
H$ and \,$y \leftrightarrow y'$ over \,$\d$, then \,$S_{\d ,y',y}$
extends \,$S_{\c ,f(y'),f(y)}$.
\end{enumerate}
\end{definition}

\begin{definition}\label{Definition of the group induced by
extension of constants}{\rm \cite[page 49]{kol85}} An
\,$\Del$-$\d$-group structure on \,$G$ is said to be {\rm induced}
{\rm (by the given $\Del$-$\c$-group structure on $G$)} if the
following two conditions are satisfied:
\begin{enumerate}
\item the identity map \,${\rm id}_G$ on the set \,$G$ is an
\,$\Del$-$(\d,\c)$-homomorphism from \,$G$ with the structure of
an  \,$\Del$-$\d$-group to \,$G$ with the structure of the
\,$\Del$-$\c$-group \,$G$;

\item   every \,$\Del$-$(\d,\c)$-homomorphism of an
\,$\Del$-$\d$-group into \,$G$ is an \,$\Del$-$\d$-\linebreak
homomorphism.

\end{enumerate}

\end{definition}

      The following generalization of \cite[Theorem 2, page 396]{kol73} interprets, for an $\Del$-extension
$\cp$ of $\c$ in $\k$ ($=\u^{\E}$), the induced $\Del$-$\cp$-group
of the \dcg~$\ggf$.

\begin{theorem}\label{Theorem on extension of constant of a strongly
normal extension} Let \,$\g\subset \u$ be an \,$\Ep$-strongly
normal extension of \,$\f$. Denote the field of \,$\E$-constants
of \,$\f$ by \,$\c$, and let \,$\cp \subset \k$ be an \de
extension of \,$\c$ such that \,$\u$ is \,\de universal over
\,$\f\cp$. Then \,$\u$ is \,\de universal over \,$\g\cp$, and
\,$\g\cp$ is an \,\dsn extension of \,$\f\cp$ with field of
\,$\E$-constants \,$\cp$. Furthermore, the \,$\Del$-$\cp$-group
\,$G(\g\cp/\f\cp)$ is the induced \,$\Del$-$\cp$-group of the
\,\dcg~$\ggf$,  both these groups being identified with each other
by means of their canonical identifications with the group of
\,\de automorphisms of \,$\g\k$ over \,$\f\k$ $($See Proposition
\ref{Identifies strong isomorphisms with automorphisms}\,\,$)$.
\end{theorem}

\[
\begin{CD}
  \g      @>>>   \g\c' @>>> \g\k \\
   @AAA                     @AAA                 @AAA\\
  \f    @>>>     \f\c' @>>> \f\k'
\end{CD}
\]\bigskip

\proof{  Since $\g\cp$ is finitely \de generated over $\f\cp$,
\cite[Proposition 4(b), page 133]{kol73} shows that $\u$ is \de
universal over $\g\cp$. That $\cp$ is the field of $\E$-constants
of $\f\cp$ and $\g\cp$ follows from
\cite[Corollary 2, page 88]{kol73}. If $\sigma$ is any \de
isomorphism of $\g\cp$ over $\f\cp$, then the restriction of
$\sigma$ to $\g$ is an \de isomorphism of \gf ~and as such is \ds.
Hence, $\sigma(\g\cp)=\sigma\g\sigma\cp
\subset\g\cdot\k\cdot\cp=\g\cp\cdot\k$, and similarly $\g\cp
\subset\sigma(\g \cp)\cdot\k$:  that is $\sigma$ is \ds. Therefore
$\g\cp$ is \dsn ~over $\f\cp$, and $G(\g\cp/\f\cp)$ is a
$\Del$-$\cp$-group (Theorem \ref{existance of the differential
group of isomorphisms}). Denote by $\cp \langle \sigma \rangle$
the $\Del$-$\cp$-field associated to any $\sigma \in
G(\g\cp/\f\cp)$.

Define ${\rm id}_G: G(\g\cp/\f\cp)   \rightarrow G(\g / \f)$ by
identifying $\sigma\in G(\g\cp/\f\cp)$ with the \de automorphism
of $\g\cp\cdot\k=\g\k$ over $\f\cp\cdot\k=\f\k$ that extends
$\sigma$ (Proposition \ref{Identifies strong isomorphisms with
automorphisms}), and then with the \ds ~\de isomorphism of \gf ~to
which $\sigma$ restricts.  Then $$\g\cp\langle\sigma\rangle=
\g\cp\cdot\cp\langle\sigma\rangle= \g\cp \sigma(\g\cp)= \g  {\rm
id_G} \sigma \g \cdot\cp=\g\c \langle {\rm id_G} \sigma \rangle
\cp,$$ and, by
\cite[Corollary 2, p. 88]{kol73},
\begin{equation}\label{equation on constant extension}\cp\langle \sigma\rangle=\c \langle {\rm
id_G}\sigma \rangle \cp.
\end{equation}  If $\sigma'$ is an $\Ep$-$\c'$-specialization of $\sigma$ in
$G(\g \cp/\f \cp)$, then $(\sigma'\alpha)_{\alpha\in\g}$ is an \de
$\g\cp$-specialization of $(\sigma\alpha)_{\alpha\in\g}$, and
hence over $\g$, so that ${\rm id}_G \sigma'$ is an
$\Ep$-$\c'$-specialization of ${\rm id}_G \sigma$ in $\ggf$. When
the $\Ep$-$\c'$-specialization in $G(\g \cp/\f \cp)$ is
$\c'$-generic, then there exists an \de isomorphism
\begin{equation}\label{isomorophism1}\g\cp\sigma(\g\cp)\approx\g\cp\sigma'(\g\cp)\end{equation}
 over $\g\cp$ mapping $\sigma\alpha$ onto $\sigma'\alpha$ for every
$\alpha\in\g$, and this restricts to an \de isomorphism
\begin{equation}\label{isomorophism2}\g \cdot {\rm id_G} \sigma \g
\approx\g \cdot {\rm id_G} \sigma' \g\end{equation} over $\g$, so
that the $\Ep$-$\c$-specialization in $\ggf$ is $\c$-generic. This
restricts to the induced $\Del$-$\c$-isomorphism
\begin{equation}\label{isomorophism3}S_{{\rm id_G} \sigma',{\rm
id_G} \sigma}^{\c}: \c\langle {\rm id_G}
\sigma\rangle\approx\c\langle {\rm id_G}
\sigma'\rangle,\end{equation} which is also a restriction of the
\de isomorphism \ref{isomorophism1}.  Moreover, the \de
isomorphism \ref{isomorophism1} also restricts to the induced
$\Del$-$\c$-isomorphism
\begin{equation}\label{isomorophism4}S_{\sigma',\sigma}^{\c'}:
\cp\langle\sigma\rangle\approx\cp\langle\sigma'\rangle.\end{equation}
Therefore, the restriction of $S_{\sigma',\sigma}^{\c'}$ to
$\c\langle {\rm id_G} \sigma\rangle$ is the induced
$\Del$-$\c$-isomorphism $S_{{\rm id_G} \sigma',{\rm id_G}
\sigma}^{\c}: \c\langle {\rm id_G} \sigma\rangle\approx\c\langle
{\rm id_G} \sigma'\rangle$. Therefore, $S_{\sigma',\sigma}^{\c'}$
is an extension of $S_{{\rm id_G} \sigma',{\rm id_G}
\sigma}^{\c}$. This shows that ${\rm id}_G$
is an $\Del$-$(\cp,\c)$-homomorphism.

    Now let $H$ be any $\Del$-$\cp$-group relative to the universal
$\Del$-field $\k$, and let $f: H\to\ggf$ be any
$\Del$-$(\cp,\c)$-homomorphism.  To complete the proof of the
theorem, it must be shown that $f'= {\rm {\rm id_G}}^{-1}f$ from
$H$ to $G(\g \cp/\f \cp)$ is an
$$
\begin{CD}
 H @+>f'>>[90] G(\g
\cp/\f \cp)\\
  @+se f dd(2,1)[100]  @+VV{\rm {\rm id_G}}V[40] \\
 {} @. \ggf
\end{CD}$$ \\ $\Del$-$\cp$-homomorphism \cite[Chapter 1, Section 2, p. 37]{kol85};
that is $f'$ is a homomorphism of groups and an everywhere defined
pre $\Ep$-$\cp$-mapping (Definition \ref{premapping}). Clearly
$f'$ is a homomorphism of groups. By \cite[Corollary 1, p.
90]{kol85}, it suffices to show that the restriction, also denoted
by $f'$, of $f'$ to the $\cp$-generic elements of $H$ is a pre
$\Ep$-$\cp$-mapping.

Property 1 of the definition of pre $\Ep$-$\cp$-mapping is clear,
i.e. the domain of definition of $f$ contains the $\cp$-generic
elements of $H$. For any $y\in H$, $\cp\langle y \rangle\supset\c
\langle f(y) \rangle$ because $f$ is an
$\Del$-$(\cp,\c)$-homomorphism. From this and the equation
\ref{equation on constant extension}, the following containment
may be deduced:
\[\cp\langle y \rangle\supset\c \langle f(y) \rangle \cp = C\langle
{\rm id_G} f'(y) \rangle \cdot \cp = \cp \langle f'(y) \rangle.\]
This is property 2 of the definition of a pre $\Ep$-$\cp$-mapping.

 For properties 3 and 4, if $y \leftrightarrow y'$ in
$H$, then $f(y) \leftrightarrow f(y')$ in \ggf~because $f$ is an
$\Del$-$(\cp,\c)$-homomorphism.
For the same reason, $S_{y',y}^{\c'}$ extends the induced
$\Del$-$\c$-isomorphism $S_{f(y'),f(y)}^{\c}$
\[
\begin{CD}
 \c'\langle y\rangle @>S_{y',y}^{\c'}>> \c'\langle y'\rangle\\
 @A\cup AA   @A\cup AA \\
 \c\langle f(y)\rangle @>S_{f(y'),f(y)}^{\c}>> \c\langle
f(y')\rangle,
\end{CD}\]\\ and
hence $S_{f(y'),f(y)}^{\c}$ and $id_{\c'}$ are bicompatible.
Therefore, the following diagram commutes:
\[
\begin{CD}
 \c'\langle y\rangle @>S_{y',y}^{\c'}>> \c'\langle y'\rangle\\
 @A\cup AA   @A\cup AA \\
 \c\langle f(y)\rangle\cdot \c' @>\varphi>> \c\langle f(y')\rangle\cdot
\c',
\end{CD}\]\\
where the $\Ep$-isomorphism $\varphi$ extends ${\rm id}_{\c'}$ and
$S_{f(y'),f(y)}^{\c}$.

Since $\g$ and $\c[\c'\langle y\rangle]$ are linearly disjoint
over $\c$, as are $\g$ and $\c[\c'\langle y'\rangle]$, it follows
that $id_{\g}$ and $S_{y',y}^{\c'}$ are bicompatible. Therefore,
the top square of the following diagram commutes:\bigskip
\[
\begin{CD}
\g \c'\langle y\rangle @>\alpha>> \g\c'\langle y'\rangle\\
 @A\cup AA   @A\cup AA \\
 \g\c\langle f(y)\rangle\cdot \c' @>\beta>> \g\c\langle f(y')\rangle\cdot \c'\\
@A=AA   @A= AA \\
\g\c\langle {\rm id_G}f'(y)\rangle\c' @>\beta>> \g\c\langle {\rm
id_G}f'(y')\rangle\c'\\
@A=AA   @A= AA \\
\g\c'\langle f'(y)\rangle @>\gamma>> \g\c'\langle f'(y')\rangle
\\
@A=AA   @A= AA \\
\g\c'\cdot\c'\langle f'(y)\rangle @>\lambda>> \g\c'\cdot\c'\langle
f'(y')\rangle
\\
@A=AA   @A= AA \\
\g \cp \cdot f'(y)(\g\cp) @>\mu>> \g \cp \cdot f'(y')(\g\cp)
\end{CD}\]
\\
where the \de isomorphism $\alpha$ extends ${\rm
id}_{\g}$ and $S_{y',y}^{\c'}$ and the \de isomor-phism $\beta$
extends $\varphi$, ${\rm id}_{\g}$, $S_{f(y'),f(y)}^{\c}$ and
${\rm id}_{\c'}$.
 Since $f={\rm id_G} f'$, the third line of this diagram is also
$$\beta:~ \g \c\langle {\rm id_G} f'(y) \rangle \cp \longrightarrow \g \c
\langle {\rm id_G} f'(y')\rangle\cp$$ extending ${\rm id}_\g$,
$S_{{\rm id_G} f'(y'),{\rm id_G} f'(y)}$ and ${\rm id}_{\c'}$.  By
equation \ref{equation on constant extension}, $\c\langle {\rm
id_G}f'(y)\rangle\c'=\c'\langle f'(y)\rangle$, and $\c\langle {\rm
id_G}f'(y')\rangle\c'=\c'\langle f'(y')\rangle$. Because ${\rm
id}_G$ is an $\Del$-$(\cp,\c)$-homomorphism, as was shown in the
first part of this proof, the forth line
$$\gamma:~ \g  \cdot \c'\langle f'(y) \rangle \longrightarrow \g
\cdot \c' \langle f'(y')\rangle$$ is an \de isomorphism extending
${\rm id}_{\g}$ and $S_{f'(y'),f'(y)}^{\c'}$. The fifth line of
the diagram is the \de isomorphism $$\lambda: ~
\g\c'\cdot\c'\langle f'(y)\rangle \longrightarrow
\g\c'\cdot\c'\langle f'(y')\rangle,
$$ which is obtained by writing `$\g\c'$' instead of `$\g$'.  Clearly, the \de
isomorphism $\lambda$ extends ${\rm id}_{\g\c'}$ and
$S_{f'(y'),f'(y)}^{\c'}$. By the $\Del$-strong normality of
$\g\c'$ over $\f\c'$, $\lambda$ is the same as the \de
isomorphism, in the sixth line,
$$\mu: ~\g \cp \cdot f'(y)(\g\cp) \longrightarrow \g \cp \cdot
f'(y')(\g\cp)$$ that extends ${\rm id}_{\g\c'}$ and that maps
$f'(y)\alpha$ onto $f'(y')\alpha$ for every $\alpha\in\g\cp$.
Therefore $f'(y) \leftrightarrow f'(y')$ in $G(\g \cp/\f \cp)$,
which is property 3. Property 4 is obtained by restricting the top
and bottom lines in the above diagram to the $\Delta$-constants,
i.e. $S_{y',y}^{\c'}$ extends $S_{f'(y'),f'(y)}^{\c'}$. }

\begin{proposition}\label{E-strongly normal by isomorphism}
Let \,$\g$ be an \dsn extension of \,$\f$, with \,$\c$ the
subfield of \,$\E$-constants, and let \,$\varphi$ be an \,\de
isomorphism of \,$\g$ over \,$\c$ such that \,$\u$ is universal
over \,$\varphi\g$. Then \,$\varphi\g$ is an \,\dsn ~extension of
\,$\varphi\f$. There is a unique \,\de isomorphism
\,$\g\cdot\k\approx\varphi\g\cdot\k$ over \,$\k$ that extends
\,$\varphi$ $($that also shall be denoted by $\varphi)$.  When
\,$\ggf$, respectively \,$G(\varphi\g /\varphi\f)$, is canonically
identified with the group of \,\de automorphisms of \,$\g\cdot\k$
over \,$\f\cdot\k$, respectively \,$\varphi\g\cdot\k$ over
\,$\varphi\f\cdot\k$, the formula
\,$T_{\varphi}(\sigma)=\varphi\cdot\sigma\cdot\varphi^{-1}$
defines an \dc isomorphism \,$T_{\varphi}: \ggf\approx G(\varphi\g
/\varphi\f)$.
\end{proposition}

\begin{remark} When \,$\varphi$ is an \,\de isomorphism of \,$\g$ over \,$\f$, then
\,$\varphi\in G(\g/\f)$.  After \,$G(\g/\f)$ and \,$G(\varphi\g /
\varphi\f)$ are canonically identified with the group of
automorphisms of the differential field \,$\g \k = \varphi\g
\cdot\k$ over \,$\f \k$, then they coincide as groups $($but not
necessarily as \,$\c$-groups$)$, and \,$T_{\varphi}$ is the inner
\,$\Del$-automorphism determined by \,$\varphi$.
\end{remark}

\proof{Let $\tau$ be any \de isomorphism of $\varphi\g$ over
$\varphi\f$.   The \de isomor-phism $\varphi^{-1} :  \varphi \g
\approx \g$ can be extended to some \de isomorphism $\psi: \varphi
\g \cdot \tau\varphi \g \approx \g \cdot\psi\tau\varphi\g$, and
evidently the formula $\alpha\mapsto \psi\tau\varphi\alpha$
defines an isomorphism of \gf.  Therefore, since $\psi\tau\varphi$
is $\Ep$-strong, the field of constants $\cp$ of
$\g\cdot\psi\tau\varphi\g$ has the property that
\begin{equation}\label{equation22}\g\cp= \g\cdot\psi\tau\varphi\g=
\psi\tau\varphi\g\cdot\cp.\end{equation} By applying $\psi^{-1}$
to equation \ref{equation22},
$$\varphi\g\cdot\ct=\varphi\g\cdot\tau\varphi\g=
\tau\varphi\g\cdot\ct$$ since $\psi^{-1}$ maps $\g$ onto
$\varphi\g$ and $\cp$ onto the field of $\E$-constants $\ct$ of
$\varphi\g\cdot\tau\varphi\g$. Therefore, $\tau$ is \ds, and,
hence, $\varphi\g$ is an \dsn ~extension of $\varphi\f$.

      Since $\g$ and $\k$ are linearly disjoint over $\c$, as are
$\varphi\g$ and $\k$, $\varphi$ can be extended to a unique \de
isomorphism $\g\k\approx\varphi\g\cdot\k$ over $\k$, and denote
it, too, by $\varphi$.   Making the canonical identifications, one
can see that for each $\sigma\in\ggf$,
$\varphi\cdot\sigma\cdot\varphi^{-1} \in G(\varphi\g /
\varphi\f)$. Therefore one can define a mapping $T_{\varphi}: \ggf
\approx G(\varphi\g / \varphi\f)$ by the formula
$T_{\varphi}(\sigma) = \varphi\cdot\sigma\cdot\varphi^{-1}$, and
it is clear that $T_{\varphi}$ is a group isomorphism.  Since
$\varphi\g\cdot\c\langle T_{\varphi}(\sigma) \rangle =
\varphi\g\cdot(\varphi\cdot\sigma\cdot\varphi^{-1})\varphi\g=
\varphi(\gsg)=\varphi(\g \cs)= \varphi\g\cdot\cs$, one may infer
that $\c\langle T_{\varphi}(\sigma\rangle)=\cs$. Furthermore, if
$\sigma\leftrightarrow\sigma'$, then there exists an \de
isomorphism $\gsg\approx\gspg$ over $\g$ mapping $\sigma\alpha$
onto $\sigma\alpha$ ($\alpha\in G$) and inducing the
$\Del$-$\c$-isomorphism $S_{\sigma' ,\sigma}: \cs\approx\csp$.
Since $\varphi$ maps $\gsg$, respectively $\gspg$, onto
$\varphi\g\cdot T_{\varphi}(\sigma)\varphi\g$, respectively
$\varphi\g\cdot T_{\varphi}(\sigma')\varphi\g$, and leaves
$\E$-constants fixed, one obtains an \de isomorphism
$\varphi\g\cdot T_{\varphi}(\sigma)\varphi\g\approx \varphi\g\cdot
T_{\varphi}(\sigma')\varphi\g$ over $\varphi\g$ mapping
$T_{\varphi}(\sigma)\varphi\alpha$ onto
$T_{\varphi}(\sigma')\varphi\alpha$ ($\alpha\in G$), so that
$T_{\varphi}(\sigma)\leftrightarrow T_{\varphi}(\sigma')$ and
$S_{T_{\varphi}(\sigma'),T_{\varphi}(\sigma)}=
S_{\sigma',\sigma}$.  Thus, $T_\varphi$ restricted to the
$\c$-generic elements of $G$ is a pre $\Ep$-$\c$-map, and
$T_{\varphi}$ is an \dc isomorphism by \cite[Corollary 1, p.
90]{kol85}. }

 \section{The Fundamental Theorems}\label{fund}

\subsection{The Topology on $\Ep$-Sets}\label{The Topology on
E-Sets} In this section, let $\f$ be an $\Ep$-field, and let $\v$ be
an $\Ep$-extension of $\f$ that is $\Ep$-universal over $\f$. And
consider $\h$ an $\Ep$-extension of $\f$ over which $\v$ need not be
$\Ep$-universal.  Also, let $A$ be a pre $\Ep$-$\f$-set relative to
$\v$ (Section \ref{pre E-sets} or \cite[page 29]{kol73}).  Then
$x\in A$ is defined to be {\it rational} over $\h$ if $\f\langle x
\rangle \subset \h$ \cite[page 29]{kol85}. In a similar manner,
define \,$x$ to be {\it algebraic} ({\it resp.}, $\Ep$-{\it
algebraic} or {\it regular}) if $\h\f\langle x \rangle$ is an
algebraic ({\it resp.}, \,$\Ep$-algebraic or regular) extension of
$\h$. Denote by $A_{\h}$ the set of elements of $A$ rational over
$\h$. In particular, $A_{\v}$ is the set $A$.

Let $G$ be an $\Ep$-$\f$-group (Section \ref{Definition of
E-group} or \cite[page 33]{kol85}). A {\it homogeneous
\,$\Ep$-$\f$-space for \,$G$} is a set $M$ on which is given a
structure of a homogeneous space for the group G and a structure
of a pre $\Ep$-$\f$-set subject to axioms, which are similar to
those for an $\E$-$\f$-group \cite[page 34]{kol85}.  The
homogeneous $\Ep$-$\f$-space $M$ for $G$ is {\it principal} \,if
it is principal as a homogeneous for $G$ and satisfies additional
axioms \cite[page 35]{kol85}.

A subset $V$ of the pre $\Ep$-$\f$-subset $A$ is {\it
\,$\f$-irreducible $($in \,$A)$} if there exists $x \in V$ such
that $V$ is the set of all $\Ep$-specializations of $x$ over $\f$
\cite[page 30]{kol85}. Such an $x$ is called an $\Ep$-$\f$-generic
element of $V$.  If the set $B$ of $A$ is the union of finitely
many $\f$-irreducible subsets of $A$, then $B$ has the structure
of a pre $\Ep$-$\f$-set that is induced by the restriction of the
pre $\Ep$-$\f$-set structure on $A$.  Such a $B$ is called a {\it
pre \,$\Ep$-$\f$-subset \,$($of \,$A)$}.  A maximal
$\f$-irreducible subset of $A$ is called an {\it $\f$-component
$($of $A)$}

 An
$\Ep$-$\f$-set is a pre $\Ep$-$\f$-subset of a homogeneous
$\Ep$-$\f$-space for an $\Ep$-$\f$-group \cite[page 37]{kol85}.
Then the $\Ep$-$\h$-subsets of $M$ are the closed subsets of a
Noetherian topology on $M$ \cite[Theorem 1 page 72]{kol85}, which
is called the {\it \,$\Ep$-Zariski topology relative to \,$\h$} or
more simply the {\it \,$\Ep$-$\h$-topology}.  If $\h=\v$, the
reference to $\v$ is usually omitted, and it is called the {\it
\,$\Ep$-Zariski topology} or more simply the {\it
\,$\Ep$-topology}. Each $\Ep$-$\f$-set will be considered to have
the topology induced from the $\Ep$-$\h$-topology on its the
ambient homogeneous $\Ep$-$\f$-space for an $\Ep$-$\f$-group. For
an $\Ep$-$\f$-set $A$, the subset $A_\h= \{v \in A \mid \f\langle
v \rangle \subseteq \h\}$ will be called {\it \,$\Ep$-dense in
$A$} if, for each closed $\Ep$-closed subset $C$ of $A$ with
$A\neq C$, $A_\h$ is not contained in $C$. Kolchin shows that, if
$\h$ is constrainedly closed \cite[page 79]{kol85}, then $A_\h$ is
$\Ep$-dense in $A$ \cite[Proposition 3, page 84]{kol85}.

Any $\Ep$-$\f$-group $G$ has a natural structure of a principal
homogeneous $\Ep$-$\f$-space for $G$, which is called the {\it
regular \,$\Ep$-$\f$-space for \,$G$}.  Consequently, any pre
$\Ep$-$\f$-set contained in the $\Ep$-$\f$-group $G$ is an
$\Ep$-$\f$-subset.  An {\it$\Ep$-$\f$-subgroup} is a subgroup of
$G$ that is an $\Ep$-$\f$-subset and satisfies all the
$\Ep$-$\f$-group axioms \cite[page 37]{kol85}.  By
\cite[Proposition 1, page 87]{kol85}, a subgroup that is also an
$\Ep$-$\f$-subset is an $\Ep$-$\f$-subgroup.

\begin{definition}\label{Definition of the connected component of an E-group}
The $\f$-component of the identity of an \,$\Ep$-group \,$G$ is
denoted by \,$G^\circ$.\end{definition}

\subsection{Fundamental Theorems}
In this the rest of this chapter, let $\f$ be an $(E,\E)$-field,
and let $\u$ be an $(E,\E)$-extension of $\f$ which is
$(E,\E)$-universal over $\f$. Then $\k=\u^\E$ considered as an
$\Ep$-field is clearly $\Ep$-universal over $\c=\f^\E$ considered
as an $\Ep$-field and, thus, constrainedly closed. Also, $\g$ will
denote an \dsn extension of $\f$.

 The following
theorem establishes a Galois correspondence between the set of
intermediate differential fields of a \dsn extension and the set
of \dcsg s of its Galois group when the field of
$\Delta$-constants is constrainedly closed. The proofs are very
similar to \cite[Chapter 6, Section 4]{kol73}.

\begin{theorem}[First Fundamental Theorem]\label{First
Fundamental Theorem} Let \,$\g$ be an \,\dsn extension of \,$\f$
with field of \,$\E$-constants \,$\c$.

\begin{enumerate}
\item If \,$\f_{1}$ is an \,\de field with
\,$\f\subset\f_{1}\subset\g$, then \,$\g$ is \,\dsn over
\,$\f_{1}, \,G(\g /\f_{1})$ is an \,\dcsg~ of \,$\ggf$, and the
set of invariants of \,$G(\g /\f_{1})$ in \,$\g$ is \,$\f_{1}$.

\item  If \,$G_{1}$ is an \,\dcsg~of \,$\ggf$ and \,$\f_{1}$
denotes the set of invariants of \,$G_{1}$ in \,$\g$, then
\,$\f_{1}$ is an \,\de field with \,$\f\subset\f_{1}\subset\g$,
and, if the elements of \,$G_1$ rational over \,$\c$ are
\,$\Ep$-dense in \,$G_1$,
then \,$G(\g/\f_{1})=G_{1}$.

 \item If $\c$ is constrainedly closed {\rm \cite[page 79]{kol85}} as an \,$\Ep$-field,
parts 1 and 2 establish a bijective correspondence between \,\de
subfields \,$\f_1$ with \,$\f\subset\f_{1}\subset\g$ and
\,$\Ep$-subgroups \,$G_1 \subseteq \ggf$.

\end{enumerate}
\end{theorem}

\begin{remark} It would be preferable to remove the hypothesis of constrained-ly
closed from part 3. For a certain type of small \,\dc subgroup,
this is accomplished in Corollary \ref{Second Fundamental
Theorem}, and, for subgroups of \,$G_a$ and \,$G_m$, in
Proposition \ref{Proposition exhibiting the Galois correspondence
for Ga} and Proposition \ref{Proposition exhibiting the Galois
correspondence for Gm}. Also, if \,$\c$ is \,$\Ep$-universal over
some \,$\Ep$-field, then \,$\c$ is constrainedly closed.
\end{remark}

\proof{
 To prove part 1, let $\f_{1}$ be an \de field with
$\f\subset\f_{1}\subset\g$. Every \de isomorphism of $\g$ over
$\f_{1}$ is over $\f$, too, and hence is \ds.   Therefore $\g$ is
\dsn over $\f_{1}$, and the Galois group $G(\g /\f_{1})$ is an
$\Ep$-$\c$-group by Theorem \ref{existance of the differential
group of isomorphisms}. It is obviously a subgroup and an \dc
subset \cite[page 30 and 37]{kol85} of $\ggf$.
Thus, $G(\g /\f_{1})$ is an $\Ep$-$\c$-subgroup of $\ggf$. By
definition, every element of $\f_{1}$ is an invariant of $G(\g
/\f_{1})$ in $\g$, and, by Proposition \cite[Corollary, page
388]{kol73}, every such invariant is in $\f_{1}$.

 For part 2, let $\f_1$ be the set of invariants of $G_1$ in $G$. It is obvious that $\f_{1}$ is a \de field with
$\f\subset\f_{1}\subset\g$, and therefore, by part 1, $G(\g
/\f_{1})$ is \dcsg ~of $\ggf$.  Of course $G_{1}\subset G (\g
/\f_{1})$.  It must be shown that $G_{1}=G(\g /\f_{1})$  under the
hypothesis that the elements of $G_1$ rational over $\c$ are
$\Ep$-dense in $G_1$.

Assume that $G_{1}\not=G(\g /\f_{1})$. Fix \dc generic elements
$\sigma_{1}\ldots\sigma_{r}$ of the \dc components of $G_{1}$. By
assumption, there exists an element $\tau\in G( \g /\f_{1})$ that
is not a $\Del$-specialization of any $\sigma_{k}$.  Fixing
elements $\eta_{1},\ldots,\eta_{n}\in\g$ with
$\f\langle\eta_{1},\ldots,\eta_{n}\rangle_{E,\E}=\g$, by Lemma
\ref{equivalence of specialization of elements with specialization
of isomorphisms}, for each index k there exists a differential
polynomial $F_{k}\in \g \lbrace y_
{1},\ldots,y_{n}\rbrace_{(\Del,\E)}$ that vanishes at
$(\sigma_{k}\eta_{1},\ldots,\sigma_{k}\eta_{n})$ but not at
($\tau\eta_{1},\ldots,\tau\eta_{n})$.  Then $F_{k}$ vanishes at
$(\sigma\eta_{1},\ldots,\sigma\eta_{n})$ for all $\sigma$ in the
component of $\sigma_k$. The product $\Pi_i F_{i}$ is a
differential polynomial in $\g\lbrace y_ {1},\ldots,
y_{n}\rbrace_{(\Del,\E)}$ that vanishes at
$(\sigma\eta_{1},\ldots,\sigma\eta_{n})$ for every $\sigma\in
G_{1}$ but not for every $\sigma\in G(\g / \f_{1})$.
Let F be such a differential polynomial with as few non-zero terms
as possible.  Also suppose that one of the coefficients in F is 1.
Consider any $\sigma'\in (G_{1})_{\c}$.  Then $\sigma'$ is an \de
automorphism of \gf.   Since
$F^{\sigma'}(\sigma\eta_{1},\ldots,\sigma\eta_{n})= \sigma'
(F(\sigma'^{-1}\sigma\eta_{1},\ldots,\sigma'^{-1}\sigma\eta_{n})),
F^{\sigma'}$ vanishes at $(\sigma\eta_{1},\ldots,\sigma\eta_{n})$
for every $\sigma\in G_{1}$, because $\sigma'^{-1}\sigma \in G_1$.
And therefore $F-F^{\sigma'}$ does too.  Since $F-F^{\sigma'}$ has
fewer terms than $F$, $F-F^{\sigma'}$ must vanish at
$(\sigma\eta_{1},\ldots,\sigma\eta_{n})$ for every $\sigma\in G(\g
/ \f_{1})$.  Hence for any $\alpha\in \g$,
$F-\alpha(F-F^{\sigma'})$ vanishes at
$(\sigma\eta_{1},\ldots,\sigma\eta_{n})$ for every $\sigma\in
\g_{1}$ but not for every $\sigma\in G(\g / \f_{1})$. If
$F-F^{\sigma'}$ were not zero, one could choose $\alpha$ so that
$F-\alpha(F-F^{\sigma'})$ has fewer terms than $F$ and is nonzero.
Therefore $F-F^{\sigma'}=0$ for $\sigma'\in(G_{1})_{\c}$.

By part 1, the set $\{\sigma \in G(\g / \f) \mid F=F^{\sigma}\}$
is the $\Ep$-$\c$-group leaving invariant the
$(\Ep,\E)$-$\f$-field generated by the coefficients of $F$. In
particular, it is an $\Ep$-$\c$-subset of $G(\g / \f)$ and a
closed subset of the $\Ep$-$\c$-topology on $G(\g / \f)$.  If the
closed set $\{\sigma \in G(\g / \f) \mid F=F^{\sigma}\} \cap G_1$
were not all of $G_1$, there would be an element of $(G_{1})_{\c}$
not in $\{\sigma \in G(\g / \f) \mid F=F^{\sigma}\} \cap G_1$
since (by hypothesis) $(G_{1})_{\c}$ is $\Ep$-dense in G$_{1}$.
Therefore, $G_1 \subseteq \{\sigma \in G(\g / \f) \mid
F=F^{\sigma}\} $, or $F=F^{\sigma}$ for all $\sigma \in G_{1}$.
Since $\f_1$ is the $(\Ep,\E)$-field invariant under the action of
$G_1$, $F\in\f_{1}\lbrace y_ {1},\ldots,y_{n}\rbrace$. However,
then $F^{\sigma}=F$ for every $\sigma\in G(\g / \f_{1})$, so that
$F(\sigma\eta_{1}\ldots\sigma\eta_{n})= \sigma F(e\eta_{1}\ldots
e\eta_{n})=0$, since the identity $e$ of $G_1$ is contained in
$G_1$.  This contradiction shows that $G_{1}=G(\g / \f_{1})$ under
the hypothesis that the elements of $G_1$ rational over $\c$ are
$\Ep$-dense in $G_1$.

For part 3, the hypothesis that $\c$ is constrainedly closed
implies that the elements of $G_1$ rational over $\c$ are
$\Ep$-dense in $G_1$ (\cite[Proposition 3, page 84]{kol85}).
}\bigskip

  After two preliminary
lemmas, the next corollary characterizes the $\Ep$-$\c$-subgroups
of $\ggf$ with fixed field $\h$ having the property that $\g$ over
$\h$ is strongly normal (in the sense of Kolchin).

\begin{lemma}\label{Lemma on compositum of 2 linear disjoint
extensions being equal} Let \,$G$ and \,$K$ be field extensions of
\,$F$. Let \,$H'$ be a subfield of \,$GK$ containing \,$K$. Put
\,$H= G\cap H'$. If \,$H'$ and \,$G$ are linearly disjoint over
\,$H$ and if \,$K$ and \,$H$ are linearly disjoint over \,$F$,
then \,$H'=HK$.\end{lemma}
\[
\begin{CD}
 G @>>> GK\\
 @AAA   @AAA \\
H = G\cap H' @>>> H'\\
@AAA   @AAA \\
F @>>> K
\end{CD}\]
\bigskip

\proof{  Evidently  $HK \subset H'$. Consider any element
$\varphi\in H'$. Fix a basis $(c_{i})$ of $K$ over $F$. By
considering $\varphi$ as an element of $GK$, one may write
$\varphi=(\Sigma\beta_{i} c_{i})/(\Sigma\gamma_{j} c_{j})$, where
the $\beta_{i}$ and $\gamma_{j}$ are elements of $G$, and
therefore $\Sigma\gamma_{j}(c_{j}\varphi)-\Sigma\beta_{i}c_{i}=0$.
Thus the elements $c_{j}\varphi$ and $c_{i}$ of $H'$ are linearly
dependent over $G$.  By the first hypothesis, they must be
linearly dependent over $H$, that is there exist elements
$\beta_{i}'$ and $\gamma'_{j}$ of $H$, not all $0$, such that
$\Sigma\gamma'_{j}(c_{j}\varphi)-\Sigma\beta_{i}'c_{i}=0$. By the
second hypothesis, the elements $c_{j}$ of $K$ are linearly
independent over $H$, and therefore
$\Sigma\gamma'_{j}c_{j}\not=0$, so that
$\varphi=(\Sigma\beta_{i}'c_{i})/(\Sigma\gamma'_{j}c_{j})\in HK$.
This shows that $HK=H'$.}

\begin{lemma}\label{Lemma on condition that implies strong
normality} Let \,$\g$ over \,$\f$ be an $\,\Ep$-strongly normal
extension of \,$(\Ep,\E)$-fields, and let \,$G=G(\g/\f)$, the
associated \,$\Ep$-$\c$-group of \,$(\Ep,\E)$-isomor-phisms. Let
\,$H$ be an \,$\Ep$-$\c$-subgroup of \,$G$ and \,$\h$ be the \,\de
field of invariants of \,$H$ in \,$\g$. If \,$\c\langle \sigma
\rangle \subset \c\u^{\Ep,\E}$ for all \,$\sigma \in H$, then
\,$\g$ over \,$\h$ as an \de extension is strongly normal in the
sense of Kolchin.
\end{lemma}

\proof{ For all $\sigma\in H$, $\c\langle \sigma
\rangle\subset\c\u^{\Ep,\E}$ implies $\sigma\g \subset \gsg = \g
\c\langle \sigma \rangle\subset\g(\c\u^{\Ep,\E}) =\g\u^{\Ep,\E}$.
Since $\sigma$ leaves invariant $\Delta$-constants, it also leaves
invariant \de constants. By \cite[Propostion 10, page 393]{kol73},
$\g$ over $\h$ as an \de extension is strongly normal as an \de
extension in the sense of Kolchin. }

\begin{lemma}\label{Lemma on denseness over constants}Let \,$F$ be an \,$\Ep$-field, and let \,$\v\supset F$
 be an \,$\Ep$-field that is \,$\Ep$-universal over \,$F$. Let \,$B$ be
an \,$\Ep$-$\f$-set. Let \,$C$=$F^\Ep$, and let \,$C_a$ be the
algebraic closure of \,$C$ in \,$\v^\Ep$. If \,$B_{\v} \subset
B_{F \v^\Ep}$, then \,$B_{F C_a}$ is \,$\Ep$-dense in \,$B$.
\end{lemma}\proof{Since $\v$ is a constrainedly closed extension of
$F$ (\cite[Proposition 3, page 84]{kol85}), $B_{\v}$ is dense in
$B$ \cite[Proposition 3, page 84]{kol85}. However, each point of
$B$ rational over $\v$ is rational over $F\v^\Ep$ by assumption.
But an element constrained over $F$ rational over $F\v^\Ep$ is, in
fact, rational over $F C_a$ because an $\Ep$-extension constrained
over $\c$ has $\Ep$-constants algebraic over $\c$
\cite[Proposition 7(d), page 142]{kol73}. Therefore, the set $B_{F
C_a}$ is $\Ep$-dense in $B$.}\bigskip

The formulation of the following corollary was influenced by
Chapter 3 of Sit's thesis \cite{sit75}, in which he considers
$\E$-subfields of $\f\langle t\rangle_\E$ over which $\f\langle
t\rangle_\E$ is strongly normal in the sense of Kolchin, where $t$
is a $\E$-indeterminant over the $\E$-field $\f$.  For instance,
the previous lemma is a generalization of \cite [Lemma 2.1, page
652]{sit75} from an affine $\Ep$-Zariski closed subset of $\v^n$
to an $\Ep$-$\f$-subset that is not necessarily affine.  In this
corollary, these ideas have been combined with those of Kolchin in
the second part of his proof of the fundamental theorem for
strongly normal extensions (\cite[Theorem 3, page 398]{kol73}).

\begin{corollary}\label{Second
Fundamental Theorem} Let \,$\l= \u^{\Del,\E}$ and let \,$G=G(\g
/\f)$. Let \,$\i$ be the set of \,$(\Del,\E)$-subfields \,$\h$ of
\,$\g$ containing \,$\f$ such that \,$\g$ over \,$\h$ is strongly
normal as an\, \de field extension \,{\rm(}in the sense of
Kolchin{\rm)}, and let \,$\s$ be the set of \,\dc subgroups \,$H$
of \,$G$ such that \,$H_{\u^{\E} }\subset H_{\c\l}$. Then there is
a Galois correspondence between \,$\i$ and \,$\s$.
\end{corollary}

\proof{  Let $\h\in \i$.  Then by part 1 of the First Fundamental
Theorem \ref{First Fundamental Theorem}, there exists an \dc
subgroup $H=G(\g/\h)$ of $G$ such that the $(\Del,\E)$-field of
invariants of $H$ is $\h$. Let $\sigma\in H_{\u^{\E}}$.
Because $\g$ over $\h$ is strongly normal as an \de extension (in
the sense of Kolchin), $\sigma$ is a strong (in the sense of
Kolchin) \de isomorphism of $\g$ over $\h$, and $\gsg \subset \g
\cdot\l$.  \smallskip \[
\begin{CD}
 \g @>>> \g\l\\
 @AAA   @AAA \\
(\g\l)^{\E}\cap \g =\g^{\E} @>>> (\g\l)^{\E}\\
@AAA   @AAA \\
\g^{\Ep,\E} @>>> \l
\end{CD}\] \smallskip Apply Lemma \ref{Lemma on compositum of 2 linear
disjoint extensions being equal} to the case $G=\g$, $K=\l$,
$F=\g^{\Ep,\E}$, $H'=(\g\l)^{\E}$ and $H=\g^{\E}$. By the linear
disjointness of $\Ep$-constants \cite[Corollary 1, page
87]{kol73}, $\g^{\E}$ and $\l$ are linearly disjoint over
$\g^{\Ep,\E}$, and $\g$ and $(\g\l)^{\E}$ are linearly disjoint
over $\g^{\E}$ by the linear disjointness of $\E$-constants. This
lemma then implies $(\g\l)^{\E}=\g^{\E}\l$. Then, $\c\langle
\sigma \rangle =(\gsg)^{\E}\subset (\g
\cdot\l)^{\E}=\ge\cdot\l=\c\l$. Therefore, $\sigma\in H_{\c \l}$,
and $H \in \s$.

Let $H \in \s$, and let $\h$ be the corresponding subfield of
invariants of $H$ in $\g$. If the elements of $H$ rational over
$\c$ are $\Ep$-dense in $H$, by part 1 of the First Fundamental
Theorem \ref{First Fundamental Theorem}, $\h$ is a differential
field with $\f\subset \h\subset \g$ and $H=G(\g/\h)$. By Lemma
\ref{Lemma on condition that implies strong normality}, $\g$ over
$\h$ is strongly normal as an \de extension (in the sense of
Kolchin), and $\h \in \i$.

Let $\d=\c^{E}=\g^{(E,\E)}$, and let $\da$ be the algebraic
closure of $\d$ in $\l$.  For all $H \in \s$, the set of elements
of $H$ rational over $\c\da$ is $\Ep$-dense in $H$ by Lemma
\ref{Lemma on denseness over constants}. (For the affine case, see
a lemma of Sit \cite [Chapter 2, Section 2]{sit75}.)  By results
in the two paragraphs above, if $\da\subseteq\c$ or equivalently
if $\da\c = \c$, then there is a Galois correspondence between
$\i$ and $\s$.

To prove the corollary without assuming $\da \subseteq \c$, let $H
\in \s$, and let $\h$ be the set of invariants of $H$ in $\g$. It
will be shown that $H=G(\g/\h)$. Let $\hp$ denote the set of
invariants of $H$ in $\g\da$. Then $\hp$ is an \de field with
$\f\da\subset\hp\subset\g\da$ and $\g\cap\hp=\h$.

\smallskip \[
\begin{CD}
 \g @>>> \g\da\\
 @AAA   @AAA \\
\g\cap\hp=\h @>>> \hp\\
@AAA   @AAA \\
\f @>>> \f\da\\
@AAA   @AAA \\
\d @>>> \da
\end{CD}\] \smallskip

\begin{claim}
The fields \,$\g$ and \,$\hp$ are linearly disjoint over \,$\h$.
\end{claim}
To prove this, consider elements
$\varphi_{1},\ldots\varphi_{s}\in$ $\hp$ that are linearly
dependent over $\g$.  It must be shown that they are linearly
dependent over $\h$.  It may be assumed that $s>1$ and no $s-1$ of
them are linearly dependent over $\g$. Then there exist nonzero
elements $\alpha_{1},\ldots,\alpha_{s}\in\g$ with $\underset{1\leq
j \leq s}\Sigma\alpha_{j}\varphi_{j}=0$. Dividing by $\alpha_{s}$
one may suppose that $\alpha_{s}=1$. For any $\sigma\in H$, since
$\h'$ is invariant under $H$, $\underset{1\leq j \leq
s}\Sigma(\sigma\alpha_{j})\varphi_{j}=0$, and therefore
$\underset{1\leq j \leq
s-1}\Sigma(\sigma\alpha_{j}-\alpha_{j})\varphi_{j}=0$.

Take $\sigma\in H_{\c\da}$ so that, by definition, $\c\langle
\sigma \rangle$ is algebraic over $\c$. By part 1 of the
Definition \label{definition of presets} of a pre $\Ep$-set,
$\c\langle \sigma \rangle$ is finitely $\Ep$-generated over $\c$.
Therefore, the degree of $\g\c\langle \sigma \rangle$ over $\g$ is
finite. Let $f_i : \g\c\langle \sigma \rangle \to \g\da$ for
$i=1,\ldots,t$ denote the finite set of isomorphisms (not
necessarily differential) over $\g$. Suppose $f_1 = id$. It is
simple to show that each $f_i$ is, in fact, an \de isomorphism
because $\g\c\langle \sigma \rangle$ over $\g$ is algebraic. By
Lemma \ref{Lemma on condition that implies strong normality},
$\sigma$ is a strong isomorphism of $\g$ over $\f$ (in the sense
of Kolchin) such that $(\g\sigma\g)^{\E}=\c\langle \sigma \rangle
\subset \c\da$ and $\gsg = \g \c \langle \sigma \rangle \subset
\g\da$. Consider the \de isomorphisms of $\g$ defined as $\sigma_i
= f_i \sigma$ for $i=1,\ldots,t$. So $\sigma=\sigma_1$. For each
$i$, by the definitions of $f_i$ and $\sigma_i$, there exist
$(\Ep,\E)$-$\g$-isomorphisms $\psi_i:\g\sigma \g \rightarrow \g
\sigma_i \g$ such that $\psi_i(\alpha)=\alpha$ and $\psi_i(\sigma
\alpha)=\sigma_i \alpha$ for all $\alpha \in \g$. That is  $\sigma
\leftrightarrow \sigma_i$ in $G$. Since $\sigma \rightarrow
\sigma_i$ and since $H$ is $\Ep$-closed in $G(\g/\f)$, $\sigma_i
\in H$.

So that $\underset{1\leq j \leq
s-1}\Sigma(\sigma_{k}\alpha_{j}-\alpha_{j})\varphi_{j}=0$ (for
$1\leq k \leq t$). If $\sigma\alpha_{1}-\alpha_{1} \neq 0$, then,
because $f_k$ is an isomorphism over $\g$, $0 \neq
f_k(\sigma\alpha_{1}-\alpha_{1})= f_k\sigma\alpha_1 -f_k\alpha_1
=\sigma_k\alpha_1 -\alpha_1$ . So, one may divide by
$\sigma_k\alpha_1 -\alpha_1$ for each $k$ to obtain
\begin{equation}\label{equations to be summed} \underset{1\leq j
\leq s-1}\Sigma(\sigma_{k}\alpha_{1}-\alpha_{1})^{-1}
(\sigma_{k}\alpha_{j}-\alpha_{j})\varphi_{j}=0 ~~~({\rm for}~ 1
\leq k \leq t).\end{equation} Set $\alpha'_{j} = \underset{1\leq k
\leq t}\Sigma (\sigma_{k}\alpha_{1}-\alpha_{1})^{-1}
(\sigma_{k}\alpha_{j}-\alpha_{j}) = {\rm Tr}~
(\sigma\alpha_{1}-\alpha_{1})^{-1} (\sigma\alpha_{j}-\alpha_{j})$
\linebreak(${\rm Tr}$ is the trace of $\g\da$ over $\g$). By
summing the equations \ref{equations to be summed}, one would have
$\underset{1\leq j \leq s-1}\Sigma\alpha'_{j}\varphi_{j}=0$,
$\alpha'_{j}\in \g (1\leq j \leq s-1)$, $\alpha'_{1} ={\rm Tr}~ 1
\not= 0$. This contradicts the linear independence of
$\varphi_{1},\ldots,\varphi_{s-1}$ over $\g$. Therefore,
$\sigma\alpha_{1} =\alpha_{1}$ for every $\sigma\in H_{\c\da}$.
Since
$H_{\c \da}$ is $\Ep$-dense in $H$, $\sigma\alpha_{1} =\alpha_{1}$
for every $\sigma\in H_{\ue}$. Therefore, $\alpha_{1}\in\h$.
Similarly, $\alpha_{k}\in\h$ for every index $k$, so that
$\varphi_{1},\ldots,\varphi_{s}$ are linearly dependent over $\h$.
This establishes the claim.

By the  claim and Lemma \ref{Lemma on compositum of 2 linear
disjoint extensions being equal}, $\h'=\h\da$. It follows from
Theorem \ref{Theorem on extension of constant of a strongly normal
extension} that $G(\g/\h)=G(\g \da/\h \da)=G(\g \da/\hp)$. Because
it has been shown that $H_{\c\da}$ is $\Ep$-dense in $H$ and $\h'$
is the \de subfield of invariants of $H$ in $\g\da$, the Galois
correspondence (part 2 of the First Fundamental Theorem \ref{First
Fundamental Theorem}) implies $G(\g \da/\hp) =H$ and, thus,
$G(\g/\h)=H$.  Since $\h \in \i$, this establishes the Galois
correspondence of the theorem.}

\begin{corollary}
Assume that \,$\c$ is constrainedly closed over \,$\f$ as an
\,$\Ep$-field. Let \,$\fon$ and \,$\ft$ be \,\de differential
fields contained in \,$\g$ and containing \,$\f$. Then
\,$G(\g/\fon\ft)=G(\g/\fon)\cap G(\g/\ft)$, and
\,$G(\g/\fon\cap\ft)$ is the smallest \,\dcsg ~of \,$G(\g/\f)$
containing \,$G(\g/\fon)G(\g/\ft)$.
\end{corollary}

\proof{  Observe that an \de isomorphism of $\g$ leaves invariant
every element of $\fon\ft$ if and only if it leaves invariant
every element of $\fon$  and every element of $\ft$.  Thus the
first assertion is true, because, under the assumptions of this
corollary, the Galois correspondences of the First Fundamental
Theorem and Corollary \ref{Second Fundamental Theorem} imply the
subgroups Galois groups are uniquely determined by the invariant
subfields.

For the second assertion, the smallest \dcsg~ of $G(\g/\f)$
containing $G(\g/\fon)G(\g/\ft)$ is of the form $G(\g/\fp)$, where
$\fp\subset\fon\cap\ft$, so that $G(\g/\fp)\supset
G(\g/\fon\cap\ft)$.  On the other hand, $G(\g/\fon\cap\ft)$ is a
\dcsg ~of $G(\g/\f)$ containing $G(\g/\fon)$ and $G(\g/\ft)$, so
that $G(\g/\fp)\subset G(\g/\fon\cap\ft)$. }

\begin{theorem}\label{Galois Theorem on Normal subgroups}
Assume that \,$\c$ is constrainedly closed over \,$\f$ as an
\,$\Ep$-field. Let \,$\fon \subseteq \g$ be an \,\de field
containing \,$\f$. Then the following four conditions are
equivalent.
\begin{enumerate}

\item  \,$\fon$ is an \,$\Ep$-strongly normal extension of \,$\f$.

\item   For each element \,$\alpha\in\fon$ with
\,$\alpha\notin\f$, there exists an \,$\Ep$-strong isomorphism
\,$\sigma_{1}$ of \,$\fon$ over \,$\f$ such that
\,$\sigma_{1}\alpha\not=\alpha$.

\item \,$G(\g/\fon)$ is a normal \,$\Del$-subgroup of
\,$G(\g/\f)$.

\item  \,$\sigma\fon\subset\fon\ue$ for every \,$\sigma\in
G(\g/\f)$.

\end{enumerate}

        When these conditions are satisfied, then, for each
\,$\sigma\in G(\g/\f)$, the restriction \,$\sigma_{1}$ of
\,$\sigma$ to \,$\fon$ is an element of\, $G(\fon/\f)$, and the
formula \,$\sigma\mapsto\sigma_{1}$ defines a surjective
\,$\Del$-$\c$-homomorphism \,$G(\g/\f)\rightarrow G(\fon/\f)$ with
kernel \,$G(\g/\fon)$.
\end{theorem}
\begin{remark}
In the proof below, only the implication condition 4 implies
condition 3 uses part 3 of the First Fundamental Theorem.
\end{remark}
\proof{ If condition 1 is satisfied, then, by part 1 of the First
Fundamental Theorem \ref{First Fundamental Theorem}, the set of
invariants of $G(\fon/\f)$ in $\fon$ is $\f$, so that part 2 is
satisfied.  Let condition 2 be satisfied. The normalizer $N$ of
$G(\fon/\f)$ in $G(\g/\f)$ is a \dcsg ~of $G(\g/\f)$ containing
$G(\fon/\f)$ \cite[Corollary 2, page 103]{kol85}. By the First and
Corollary \ref{Second Fundamental Theorem}, there exists a
differential field $\ft$ with $\f\subset\ft\subset\fon$ such that
$G(\g/\ft)=N$. By the universality of $\u$, if $\sigma_{1}$ is any
\ds ~isomorphism of $\fon$ over $\f$, $\sigma_{1}$ can be extended
to an $\Ep$-$\f$-isomorphism of $\g$, that is, to an element
$\sigma\in G(\g/\f)$. Then for any $\tau\in G(\g/\fon)$ and any
$\beta\in\fon$, $\sigma\beta=\sigma_{1}\beta\in\fon\ue$, hence
$\tau\sigma\beta=\sigma\beta$ and
$\sigma^{-1}\tau\sigma\beta=\beta$, so that
$\sigma^{-1}\tau\sigma\in G(\g/\fon)$.  Thus, $\sigma\in N$, so
that $\sigma_1$ leaves invariant every element of $\ft$. Since
$\sigma$ is an extension of an arbitrary element of $G(\fon/\f)$,
it follows by condition 2 that $\ft=\f$, that is, $N=G(\g/\f)$.
Therefore, condition 3 is proved from condition 2. Next, let
condition 3 be satisfied. Consider any $\sigma\in G(\g/\f)$ and
any $\beta\in\fon$.  For every $\tau\in G(\g/\fon)$,
$\sigma^{-1}\tau\sigma\in G(\g/\fon)$, so that
$\sigma^{-1}\tau\sigma\beta=\beta$ and
$\tau\sigma\beta=\sigma\beta$.  Since by Theorem 4.30,
$G(\g/\fon)=G(\g\c\langle\sigma\rangle/\fon\c\langle\sigma\rangle)$,
and since $\sigma\beta\in\gsg=\g\c\langle\sigma\rangle$,
$\sigma\beta$ is an invariant of
$G(\g\c\langle\sigma\rangle/\fon\c\langle\sigma\rangle)$ in
$\g\c\langle\sigma\rangle$, and hence, by the first part of the
First Fundamental Theorem and Corollary \ref{Second Fundamental
Theorem}, $\sigma\beta\in\fon\c\langle\sigma\rangle$. Therefore
condition 4 is satisfied.  Let condition 4 be satisfied.  If
$\sigma'$ is any isomorphism of $\fon$ over $\f$, then $\sigma'$
can be extended to an element $\sigma\in G(\g/\f)$. Then because
of condition 4 $\sigma'\fon=\sigma\fon\subset\fon\k$. It follows
by \cite[Proposition 10, page 393]{kol73}, that condition 1 is
satisfied. Therefore all four conditions are equivalent.

      Let the conditions be satisfied.  It is obvious that the
restriction mapping defined by the formula
$\sigma\mapsto\sigma_{1}$ is a group homomorphism
$G(\g/\f)\rightarrow G(\g/\fon)$ with kernel $G(\g/\fon)$. It has
already been observed that every isomorphism of $\fon$ over $\f$
can be extended to an isomorphism of $\g$, and this shows that the
homomorphism is surjective. It remains to prove that it is a
$\Del$-$\c$-homomorphism. First of all, $\c\langle\sigma\rangle=
(\gsg)\cap\ue\supset(\fon\sigma_{1}\fon)\cap\ue=
\c\langle\sigma_{1}\rangle$.  Next, if $\sigma'$ is an
$\Ep$-specialization of $\sigma$ in $G(\g/\f)$, then, by
definition, $\sigma \xrightarrow[\g]{} \sigma'$ (Definition
\ref{pre-order no X^r}).  By Lemma \ref{equivalence of
specialization of elements with specialization of isomorphisms},
this is equivalent to $(\sigma'\alpha)_{\alpha\in\g}$ is an
$(\Ep,\E)$-$\g$-specialization of $(\sigma\alpha)_{\alpha\in\g}$
over $\g$, so that a fortiori $(\sigma'_{1}\alpha)_{\alpha\in\g}$
is an $(\Ep,\E)$-$\g$-specialization of
$(\sigma_{1}\alpha)_{\alpha\in\g}$ over $\fon$, that is,
$\sigma'_{1}$ is a differential specialization of $\sigma_{1}$ by
Lemma \ref{equivalence of specialization of elements with
specialization of isomorphisms}. Finally, if $\sigma'$ is a
generic specialization of $\sigma$, then by the above,
$\sigma'_{1}$ is a generic specialization of $\sigma_{1}$. Since
the induced $\Ep$-$\c$-isomorphism $S_{\sigma',\sigma}$:
$\cs\approx\csp$ is a restriction of the
$(\Ep,\E)$-$\g$-isomorphism $\gsg\approx\gspg$ mapping
$\sigma\alpha$ onto $\sigma'\alpha$ $(\alpha\in\g)$, and the
induced $\Ep$-$\c$-isomorphism
$S_{\sigma'_{1},\sigma_{1}}:\c\langle\sigma_{1}\rangle\approx
\c\langle\sigma'_{1}\rangle$ is a restriction of the
$(\Ep,\E)$-$\g$-isomorphism
$\fon\sigma_{1}\fon\approx\fon\sigma'_{1}\fon$ over $\fon$ mapping
$\sigma\alpha$ onto $\sigma'\alpha$ $(\alpha\in\fon)$,  it is
evident that $S_{\sigma',\sigma}$ is an extension of
$S_{\sigma'_{1},\sigma_{1}}$. This shows that the restriction
mapping is a $\Del$-$\c$-homomorphism and completes the proof of
the theorem.\nolinebreak }

\begin{corollary}
Assume that \,$\c$ is constrainedly closed over \,$\f$ as an
\,$\Ep$-field. Let \,$\fo$ denote the algebraic closure of \,$\f$
in \,$\g$.  Then \,$G(\g/\fo)=G^{o}(\g/\f)$ $($Definition
\ref{Definition of the connected component of an E-group}$)$,
\,$\fo$ is an \,$\Ep$-strongly normal extension of \,$\f$, and
\,$G(\fo/\f)\approx G(\g/\f)/G^{o}(\g/\f)$. In particular, the
degree of \,$\fo$ over \,$\f$ equals the index of \,$G^{o}(\g/\f)$
in \,$G(\g/\f)$, so that \,$\f$ is algebraically closed in \,$\g$
if and only if \,$G(\g/\f)$ is connected, and \,$\g$ is algebraic
over \,$\f$ if and only if \,$G(\g/\f)$ is finite.
\end{corollary}

\proof{ By the (Definition \ref{definition of presets}), there
exists an isolated $(\Ep,\E)$-$\f$-isomorphism $\sigma_0 \in
G^{o}(\g/\f)$ such that $\sigma_0 \xrightarrow[\g]{} {\rm id}$. By
part b of \cite[Corollary to Proposition 2, page 388]{kol85}, the
set of invariants of $G^{o} (\g/\f)$ is $\fo$. Therefore, by the
First Fundamental Theorem, $G^{o}(\g/\f)=G(\g/\fo)$.  As
$G^{o}(\g/\f)$ is a normal \dcsg ~of $G(\g/\f)$ \cite[Theorem 1,
page 39]{kol85}, the previous theorem shows that $\fo$ is \dsn
~over $\f$ and $G(\fo/\f)\approx G(\g/\f)/G^{o}\g/\f)$. }

\begin{corollary}
Assume that \,$\c$ is constrainedly closed over \,$\f$ as an
\,$\Ep$-field. Assume that \,$\g\h$ and \,$\f$ have the same field
of \,$\E$-constants. Then \,$\g\cap\h$ is an \,$\Ep$-strongly
normal extension of $\f$.
\end{corollary}

\proof{ By Corollary \ref{Corollary on compositum of E-strong
extensions}, $\g\h$ is \dsn over $\f$.  By Theorem 5.43,
$G(\g\h/\g)$ and $G(\g\h/\h)$ are normal \dcsg s of $G(\g\h/\f)$,
so that their product is also \cite[Corollary 2, page 109]{kol85}.
By Corollary 5.42, the product is $G(\g\h/\g \cap\h)$.  Since it
is normal in $G(\g\h/\f)$, it follows by Theorem 5.43 that
$\g\cap\h$ is \dsn ~over $\f$. }

\begin{theorem}
Assume that \,$\c$ is constrainedly closed over \,$\f$ as an
\,$\Ep$-field. Let \,$\e$ be an extension of \,$\f$ such that
\,$\u$ is universal over \,$\e$ as an \,$\Ep$-strongly normal
extension and the field of \,$\E$-constants of \,$\g\e$ is \,$\c$.
Then \,$\g\e$  is an \,$\Ep$-strongly normal extension of \,$\e$,
for each element \,$\tau\in G(\g\e/\e)$ the restriction
\,$\tau_{1}$ of $\tau$ to \,$\g$ is an element of
\,$G(\g/\g\cap\e)$, and the formula \,$\tau\mapsto\tau_{1}$
defines an \,$\Del$-$\c$-isomorphism $G(\g\e/\e) \approx
G(\g/\g\cap\e)$.
\end{theorem}

\proof{   For any $(\Ep,\E)$-$\e$-isomorphism $\tau$ of $\g\e$,
$\tau_{1}$ is obviously an $(\Ep,\E)$-isomorphism of $\g$ over
$\g\cap\e$ and hence is \ds. Therefore, $$\tau(\g\e)\subseteq
\g\e\cdot\tau(\g\e)=\g\e\tau\g\cdot\e=
\g\tau_{1}\g\cdot\e=\g\c\langle\tau_{1}\rangle\cdot\e=
\g\e\c\langle\tau_{1}\rangle \subset \g\e \u^\Delta.$$ It follows
from Proposition \ref{Proposition on the one sided condition for
E-strongly normal}, $\g\e$ is \dsn ~over $\e$.

Clearly the formula $\tau\mapsto\tau_{1}$ defines an injective
group homomorphism \linebreak $G(\g\e/\e)\rightarrow G(\g/\g
\cap\e)$. It also follows from the above sequence of equalities
that $\g\e\c\langle\tau\rangle= \g\e\c\langle\tau_{1}\rangle$ and
by \cite[Corollary 2, page 88]{kol73}
$\c\langle\tau\rangle=\c\langle\tau_{1}\rangle$. If $\tau$ and
$\tau'$ are elements of $G(\g\e/\e)$ and $\tau\rightarrow\tau'$,
then $(\tau'\beta)_{\beta\in\g\e}$ is an $(\Ep,\E)$-specialization
of $(\tau\beta)_{\beta\in\g\e}$ over $\g\e$, so that
$(\tau'_{1}\beta)_{\beta\in\g}$ is an $(\Ep,\E)$-specialization of
$(\tau_{1}\beta)_{\beta\in\g}$ over $\g$, whence
$\tau_{1}\rightarrow\tau'_{1}$. If moreover
$\tau\leftrightarrow\tau'$, then
$\tau_{1}\leftrightarrow\tau'_{1}$, and the $(\Ep,\E)$-isomorphism
$\g\e\cdot\tau(\g\e)\approx\g\e\cdot\tau'(\g\e)$ over $\g\e$
mapping $\tau\beta$ onto $\tau\beta'( \beta\in\g\e)$ is an
extension of the $(\Ep,\E)$-isomorphism
$\g\tau_{1}\g\approx\g\tau'_{1}\g$ over $\g$ mapping
$\tau_{1}\beta$ onto $\tau'_{1}\beta$ $(\beta\in\g)$.  Since these
two $(\Ep,\E)$-isomorphisms are extensions of the induced
$\Ep$-isomorphisms $S_{\tau',\tau}:
\c\langle\tau\rangle\approx\c\langle\tau'\rangle$ and
$S_{\tau'_{1},\tau_{1}}:
\c\langle\tau_{1}\rangle\approx\c\langle\tau'_{1}\rangle$, and
since $\c\langle\tau\rangle=\c\langle\tau_{1}\rangle$ and
$\c\langle\tau'\rangle=\c\langle\tau'_{1}\rangle$,
$S_{\tau',\tau}=S_{\tau'_{1},\tau_{1}}$. It follows that the
injective group homomorphism is an $\Del$-$\c$-homomorphism.  Its
image is an \dcsg ~$G_{1}$ of $G(\g/\g\cap\e)$.  If an element
$\alpha\in\g$ is an invariant of $G_{1}$, then it is an invariant
of $G(\g\e/\e)$, whence $\alpha\in\e$. Thus, the set of invariants
of $G_{1}$ in $\g$ is $\g\cap\e$, so that $G_{1}=G(\g/\g\cap\e)$
by the First Fundamental Theorem and Corollary \ref{Second
Fundamental Theorem}. }

\section{Disjointness from Derivatives}\label{techniques for constructing
examples}

In this chapter, Kolchin's concept of disjointness is defined and
 used in two ways to construct $\Ep$-strongly
normal extensions. First, an $\Ep$-strongly normal extension will
be constructed with Galois group $\Ep$-isomorphic to any given
connected $\Ep$-group. The method of proof of this result is new
even for algebraic groups in Kolchin's setting \cite[Theorem 2,
page 880]{kol55} and does not require the field of constants to be
algebraically closed as does the result of Kolchin. A second use
of these extensions will be to define a functor from pre
$\Delta'$-sets to pre $\Delta$-sets. This takes a $\Delta'$-group
to a $\Delta$-group and is compatible with the Galois theory
(Section \ref{Section of extension of Galois groups by Delta and
Galois theory}).  By combining this result with the First
Fundamental Theorem, for any $\Delta'$-subgroup of an algebraic
group, a $\E'$-strongly normal extension is obtained with that
$\Delta'$-subgroup as its Galois group.

In this section, $\Delta$ is the union of two disjoint subsets
$\Delta'$ and $\Delta''$ and $\f$ is a $\Delta$-field.

\subsection{Definition of $\Delta''$-Free}

\begin{definition}\label{definition of free}
Let \,$\a$ be a \,$\Delta$-\,$\f$-algebra. Let \,$\Delta'$ be a
finite commuting subset of the vector space of derivations of
\,$\u$ spanned by \,$\Delta$ over \,$\f$. Let \,$\a'$ be a
\,$\Delta'$-$\f$-subalgebra of \,$\a$ such that \,$\a'$ generates
\,$\a$ as a \,$\Delta$-$\f$-algebra. Define \,$\a$ to be
{\rm\,$\Delta /\Delta'$-$\f$-free over \,$\a'$} if any
\,$\Delta'$-$\f$-homomorphism of \,$\a'$ into a \,$\Delta$-field
extension of \,$\f$ can be extended to a
\,$\Delta$-$\f$-homomorphism of \,$\a$.  If \,$\Delta$ is the
disjoint union of two subsets \,$\Delta'$ and \,$\Delta''$, define
\,$\a$ to be {\rm\,$\Delta''$-$\f$-free over \,$\a'$} if \,$\a$ is
\,$\Delta /\Delta'$-$\f$-free over \,$\a'$.
\end{definition}

Kolchin \cite[Section 7, page 19]{kol85} uses the terminology
``$\a'$ and $\Delta$ are $\Delta'$-disjoint over $\f$" instead of
$\a$ is $\Delta /\Delta'$-$\f$-free over $\a'$.  Although
Kolchin's terminology does not refer to the ring $\a$ that $\a'$
$\Delta$-generates. But $\a$ is implicit in Kolchin's definition
because the $\Delta'$-algebra $\a'$ is assumed to be contained in
some larger unspecified $\Delta$-algebra, so that $\a$ is uniquely
determined by $\a'$ and the $\Delta$-algebra containing it.

The following proposition shows that if $\a$ is $\Delta
/\Delta'$-$\f$-free over $\a'$ the $\E'$-$\f$-isomorphism class of
$\a'$ determines the $\E$-$\f$-isomorphism class of
$\a'_{\Delta}$.

\begin{proposition}\label{isomorphism of free extensions}
Let \,$\a$ and \,$\b$ be \,$\E$-$\f$-algebras that are integral
domains.  Let \,$\a'$ and \,$\b'$ be \,$\Delta'$-$\f$-subalgebras
of \,$\a$ and \,$\b$ such that \,$\a$ is \,$\Delta
/\Delta'$-$\f$-free over \,$\a'$ and \,$\b$ is \,$\Delta
/\Delta'$-$\f$-free over \,$\b'$. If \,$\a'$ and \,$\b'$ are
\,$\Delta'$-$\f$-isomorphic, then \,$\a=\a'_{\Delta}$ and
\,$\b=\b'_{\Delta}$ are \,$\Delta$-$\f$-isomorphic.
\end{proposition}
\proof{  In the definition of $\Delta/\Delta''$-free, the
extension $\Delta$-homomorphism is clearly unique because it is
determined by the action of the $\Delta'$-homomorphism on
$\Delta'$-ring generators. Let $\varphi':\a' \rightarrow \b'$ be a
given $\Delta'$-$\f$-isomorphism, and let $\chi': \b' \rightarrow
\a'$ be inverse $\Delta'$-$\f$-isomorphism. Then $\varphi'$ and
$\chi'$ extend to unique $\Delta$-$\f$-homomorphisms $\varphi: \a
\rightarrow \b$ and $\chi: \b \rightarrow \a$. The composite
$\Delta$-$\f$-homomorphism $\chi \varphi: \a \rightarrow \a$ is
the unique $\Delta$-$\f$-homomorphism extending the identity
$\E'$-$\f$-isomorphism of $\a'$ and, therefore, is the identity
$\E$-$\f$-isomorphism of $\a$. Similarly, $\varphi \chi$ is the
identity, and, therefore, $\varphi$ is a $\E$-$\f$-isomorphism. }

\begin{corollary}\label{Corollary
on the extension of automorphisms of free rings} Let \,$\a$ be an
integral domain and \,$\Delta /\Delta'$-$\f$-free over \,$\a'$.
Then each \,$\Delta'$-automorphism of \,$\a'$ extends uniquely to
a \,$\Delta$-automorphism of \,$\a=\a'_{\Delta}$.
\end{corollary}

The following is the first basic proposition of Kolchin
 about this concept of $\Delta
/\Delta'$-$\f$-free extensions \cite[Proposition 9, page
20]{kol85}.
\begin{proposition}\label{Proposition from Kolchin on Delta
freeness} Let \,$\eta=(\eta_j)_{j \in J}$ be a family of elements
of a \,$\Delta$-field extension \,$\u$ that is
\,$\Delta$-universal over \,$\f$, let \,$\Delta'$ be a commutative
linearly independent subset of \,$\f\Delta$, and let \,$\p'$ and
\,$\p$ denote, respectively, the defining \,$\Delta'$-ideal of
\,$\eta$ in \,$\f\{(y_j)_{j \in J}\}_{\Delta'}$ and the defining
\,$\Delta$-ideal of \,$\eta$ in \,$\f\{(y_j)_{j \in
J}\}_{\Delta}$. Then the following three conditions are
equivalent.
\begin{enumerate}
\item \,$\f\{\eta\}_{\Delta}$ is \,$\Delta''$-free over
\,$\f\{\eta\}_{\Delta'}$. \item \,$\f\{\eta\}_{\Delta}$ is
\,$\Delta''$-free over \,$\f\langle\eta \rangle_{\Delta'}$.\item
\,$\p=\{\p'\}_\Delta$.
\end{enumerate}

\end{proposition}

The equivalence of condition 1 and condition 3 in this proposition
shows that $\f\{\eta\}_{\Delta}$ is $\Delta''$-free over
$\f\{\eta\}_{\Delta'}$ if and only if $\{\p'\}_\Delta$ is the
defining $\Delta$-ideal of $\eta$ in $\f\{(y_j)_{j \in
J}\}_{\Delta}$. This observation enables one to construct a
$\Delta'$-$\f$-algebra $\b' \subset \u$ which is
$\Delta'$-isomorphic over $\f$ to $\a'$ and such that
$\b'_{\Delta}$ is $\Delta/\Delta'$-$\f$-free over $\b'$.
\cite[Proposition 11, page 22]{kol85}. Just take $\xi=(\xi_j)_{j
\in J }$ to be an $\f$-generic $\Delta$-zero of $\{ \p'
\}_{\Delta}$. Then, by the equivalence stated, $\f \{ \xi
\}_{\Delta}$ is $\Delta/\Delta'$-$\f$-free over $\b'$, and $\b'$
is $\Delta'$-isomorphic over $\f$ to $\a'$ because $\xi$, as is
$\eta$, is an $\f$-generic $\Delta'$-zero of $\p'$.

The proof that condition 3 implies condition 1 is straight forward
application of the definition of $\Delta''$-freeness. For
simplicity, assume that the indexing set $J$ is finite, i.e.
$\eta=( \eta_1, \ldots, \eta_n)$. Let $\varphi': \a' \rightarrow
\u$ be a $\Delta'$-$\f$-homomorphism. Then, $\varphi'(\eta)$ is a
$\Delta'$-zero of $\p'$ and, thus, a $\Delta$-zero of
$\p=\{\p'\}_\Delta$.  Since $\eta$ is an $\f$-generic
$\Delta$-zero of its defining ideal $\p$, $\eta \rightarrow
\varphi'(\eta)$, and, thus, $\varphi'$ extends to a
$\Delta$-$\f$-homomorphism $\varphi: \a_\Delta=\f\{\eta\}_\E
\rightarrow \f\{\varphi'\eta\}_\E$.

The following proof that condition 1 implies condition 3 is
slightly different than that of Kolchin and will serve to motivate
the next proposition.  Clearly, $\p\supset\{\p'\}_\Delta$. Assume
that there exists $F \in \p \subset \f \{ y_1,\ldots, y_n
\}_{\Delta}$ with $F \notin \{\p'\}_\Delta$.  Since $\p \cap \f \{
y_1,\ldots, y_n \}_{\Delta'}= \p'$ \cite[Proposition 8, page
16]{kol85}, the $\Delta$-polynomial $F \notin \f \{ y_1,\ldots,
y_n \}_{\Delta'}$ and, thus, must involve some
$\Delta''$-derivatives of some $y_i$. Since $\u$ is a
$\E$-universal over $\f$, one may choose $\Delta$-zero $\xi= (
\xi_1, \ldots, \xi_n)\in  \u^n$ of $\{\p'\}_\Delta \subset \f \{
y_1,\ldots, y_n \}_{\Delta}$ such that $F(\xi) \neq 0$. Because
$\xi$ is a zero of $\p'$, there is a $\Delta'$-homomorphism of
$\f\{ \eta \}_{\Delta'}$ onto $\f\{ \xi \}_{\Delta'}$ sending
$\eta$ to $\xi$. This $\Delta'$-homomorphism cannot extend to a
$\Delta$-homomorphism from $\f\{ \eta \}_{\Delta}$ to $\f\{ \xi
\}_{\Delta}$ because $F(\eta)=0$ and $F(\xi)\neq 0$. Therefore,
$\a$ is not $\Delta /\Delta'$-$\f$-free over $\a'$.

The existence of the $\Delta$-polynomial $F$ `prevents' $\a$ from
being $\Delta /\Delta'$-$\f$-free over $\a'$.  Since $F \notin \f
\{ y_1,\ldots, y_n \}_{\Delta'}$, proper $\Delta''$-derivatives of
$\Delta'$-derivatives of $( y_1,\ldots, y_n )$ are present in $F$.
Since $\eta$ is a $\Delta$-zero of $F$, some
$\Delta''$-derivatives of $\Delta'$ derivatives of $\eta$ are
algebraically dependent over $\a'$. Thus, the algebraic
independence of certain of the ring generators of $\a$ over $\a'$
is a necessary condition for freeness. This is made precise in the
following proposition, which is a generalization of the results of
Sit (with $\E'$ empty) \cite[Corollaries 1 and 2, page 25]{sit78}.

\begin{proposition}\label{equivalent properties of freeness}
Let \,$\xi=(\xi_{1},\ldots,\xi_{n})$ be \,$\Delta$-generators of a
\,$\Delta$-field over \,$\f$ . Assume that \,$\Delta$ is the union
of two disjoint subsets \,$\Delta'$ and \,$\Delta''$. Then the
following statements are equivalent.
\begin{enumerate}
\item The \,$\Delta$-$\f$-algebra \,$\f \{\f\langle \xi
\rangle_{\Delta'} \}_{\Delta}$ is \,$\Delta''$-$\f$-free over
\,$\f\langle \xi \rangle_{\Delta'}$. \item Every transcendence
basis
for the field \,$\f \langle \xi \rangle_{\E'}$ over \,$\f$ is
\,$\Delta''$-algebraically independent over \,$\f$. \item There
exists one transcendence basis
for the field \,$\f \langle \xi \rangle_{\E'}$ over \,$\f$ that is
\,$\Delta''$-algebraically independent over \,$\f$.

\end{enumerate}
\end{proposition}
\proof{ Because there always exists a transcendence basis
for the field $\f \langle \xi \rangle_{\E'}$ over $\f$, condition
2 implies condition 3.  Assuming condition 3, let the
transcendence basis $(t_i)_{i \in I}$ for the field $\f \langle
\xi \rangle_{\E'}$ over $\f$ be $\E''$-algebraically independent
over $\f$.
\begin{claim}
$\f \{\f\langle \xi \rangle_{\Delta'} \}_{\Delta}= \f\langle \xi
\rangle_{\Delta'}[ (\theta'' t_i)_{i \in I, \theta'' \in ~
\Theta_{\Delta''}}]$\end{claim} \proof{The right hand side is
clearly contained in the left. To prove the claim, it must be
shown that all the $\Delta''$-derivatives of $\alpha \in \f\langle
\xi \rangle_{\Delta'}$ are in the right hand side.  If $\alpha \in
\f((t_i)_{i \in I})$, this is clear by the formula for the
derivative of a quotient. If $\alpha$ is algebraic over
$\f((t_i)_{i \in I})$ and not in $\f((t_i)_{i \in I})$, let $f(x)
\in\f ((t_i)_{i \in I})[x]$ be the minimal polynomial for
$\alpha$.  For $\delta''\in\Delta''$, let  $S_f(x)=df/ dx$, and
let $f^{\delta''}(x)$ be the polynomial obtained from $f(x)$ by
differentiating the coefficients of $f(x)$ with respect to
$\delta''$. Then, $S_f (\alpha) \delta''\alpha +
f^{\delta''}(\alpha)=0$.  Since the degree of $S_f(x)$ in $x$ is
less than than the degree of the minimal polynomial, $S_f
(\alpha)\neq 0$. Then,
 $\delta''\alpha =
-f^{\delta''}(\alpha) /S_f (\alpha)$ is an element of the right
hand side because the coefficients of $f^{\delta''}(x)$,
$f^{\delta''}(\alpha)$ and $1/S_f (\alpha)$ are in the right hand
side. \nolinebreak}\bigskip

To show condition 1, let $\rho: \f\langle \xi \rangle_{\Delta'}
\rightarrow \h$ be a $\E'$-$\f$-homomorphism to an
$\Delta$-$\f$-field $\h$. The $\theta'' t_i$ for all $i \in I$ and
all $\theta'' \in \Theta_{\E''}$ of positive order, in addition to
being algebraically independent over $\f$, are algebraically
independent over $\f \langle \xi \rangle_{\E'}$ because an
algebraic relation over $\f \langle \xi \rangle_{\E'}$ would
contradict the algebraic independence of the family $(\theta''
t_i)_{i \in I, \theta'' \in \Theta''}$ over $\f$. Therefore, one
may extend $\rho$ to an $\f$-homomorphism of $\f\langle \xi
\rangle_{\Delta'} [ (\theta'' t_i)_{i \in I, \theta'' \in ~
\Theta_{\Delta''}}] $ by defining $\rho (\theta'' t_i )=\theta''
\rho (t_i)$ for all $i \in I$ and all $\theta'' \in
\Theta_{\E''}$.  To complete the proof of condition 1, it will be
shown that $\rho$ is a $\Delta$-$\f$-isomorphism.

To show $\rho$ is an $\E''$-$\f$-homomorphism, since $\rho$
restricted to \linebreak$\f [ (\theta'' t_i)_{i \in I, \theta''
\in ~ \Theta_{\Delta''}}]$ clearly is, it must be shown that $\rho
\delta''\alpha =\delta'' \rho \alpha$ for all $\delta''$ in
$\Theta_{\Delta''}$ and for $\alpha \in \f \langle \xi
\rangle_{\E'}$ algebraic over $\f ((t_i)_{i \in I})$. If $\alpha$
is not in $\f ((t_i)_{i \in I})$, as before, let $f(x) \in\f
((t_i)_{i \in I})[x]$ be the minimal polynomial for $\alpha$.
Then, for $\delta \in \Delta''$, $S_f (\alpha) \delta''\alpha +
f^{\delta''}(\alpha)=0$,  $S_f (\alpha)\neq 0$, and
$\delta''\alpha = -f^{\delta''}(\alpha) /S_f (\alpha)$ is an
element of $\f \langle \xi \rangle_{\E'} [ (\theta'' t_i)_{i \in
I, \theta'' \in ~ \Theta_{\Delta''}}]$, the domain of $\rho$.
Since $\rho$ restricted to $\f \langle \xi \rangle_{\Delta'} $ is
an isomorphism, $\rho\alpha $ satisfies $(\rho f)(x)$ and $
S_{\rho f} (\rho\alpha)=\rho(S_f (\alpha)) \neq 0$. Apply
$\delta''$ to $(\rho f)(\alpha)=0$ to obtain $S_{\rho f}
(\rho\alpha) \delta''\rho\alpha + (\rho
f)^{\delta''}(\rho\alpha)=0$ and $\delta''\rho\alpha = -(\rho
f)^{\delta''}(\rho\alpha) /S_{\rho f} (\rho\alpha)$. Since the
coefficients of $f$ are in $\f ((t_i)_{i \in I})$ where $\rho$ and
$\delta''$ commute, $(\rho f)^{\delta''}(x) =
\rho(f^{\delta''})(x)$. Therefore,
$$\delta''\rho\alpha = -(\rho f)^{\delta''}(\rho\alpha) /S_{\rho f}
(\rho\alpha) = -\rho (f^{\delta''})(\rho\alpha) /S_{\rho f}
(\rho\alpha)$$ $$= -\rho ((f^{\delta''})(\alpha)) /\rho(S_{f}
(\alpha))=-\rho ((f^{\delta''})(\alpha) /S_{f} (\alpha))= \rho
\delta''\alpha.$$

This $\Delta''$-$\f$-homomorphism $\rho$ is also a
$\Delta'$-$\f$-homomorphism because $\rho$ restricted to $\f
\langle \xi \rangle_{\E'}$ was assumed to be a
$\Delta'$-$\f$-isomorphism and because, for all $\theta''$ in
$\Theta_{\Delta''}$ and all $\delta'$ in $\Delta'$, $$\rho(\delta'
\theta'' t_i)=\rho(\theta''\delta' t_i)=\theta''\rho(\delta'
t_i)=\theta''\delta' \rho (t_i)=\delta'\theta'' \rho
(t_i)=\delta'\rho (\theta'' t_i).$$ Therefore, $\rho$ is a
$\Delta$-homomorphism of $\f \{\f\langle \xi \rangle_{\Delta'}
\}_{\Delta}$ .  This shows $\f \{\f\langle \xi \rangle_{\Delta'}
\}_{\Delta}$ is $\Delta''$-$\f$-free over $\f\langle \xi
\rangle_{\Delta'}$.

Assume condition 1. Let  $(t_i)_{i \in I}$ be a transcendence
basis of $\f\langle \xi \rangle_{\Delta'}$ over $\f$. Let
$(y_i)_{i \in I}$ be a family of $\Delta''$-indeterminates over
$\f\langle \xi \rangle_{\Delta'}$. Define an isomorphism over $\f$
of fields $\varphi : \f ((t_i)_{i \in I} ) \mapsto \f ((y_i)_{i
\in I} )$ such that $\varphi(t_i)=y_i$ for each $i\in I$. Then
because each element of $\f \langle \xi \rangle_{\E'}$ is
algebraic over $\f ((t_i)_{i \in I} )$, $\varphi$ extends to an
isomorphism of $\f \langle \xi \rangle_{\E'}$ into an
algebraically closed field containing $\f \langle(y_i)_{i \in
I}\rangle_{\Delta''}$. Endow the image $\h$ of $\varphi$ with the
unique $\Delta'$-structure such that $\varphi$ is a $\Delta'$-
$\f$-isomorphism mapping each $t_i$ to $y_i$ for $i$ in $I$. Then
$\h \{ (y_{i})_{i \in I} \}_{\Delta''}$ has a structure of a
$\Delta''$-$\f$-algebra because the elements in $\h$ not in $\f (
(y_i)_{i \in I} )$ are algebraic over $\f ( (y_i)_{i \in I} )$
and, as shown in the proof of the claim, have uniquely determined
$\Delta''$-derivatives in $\h \{ (y_{i})_{i \in I} \}_{\Delta''}$.
The $\E'$-structure on $\h$ may be extended to all of $\h \{
(y_{i})_{i \in I} \}_{\Delta''}$ by defining $\delta' (\theta''
y_i) = \theta'' \delta' y_i$ for each $\theta''$ in
$\Theta_{\Delta''}$, each $ \delta' \in \E' $ and $i \in I $.
Because $\delta' \delta'' y_i = \delta'' \delta' y_i$, the
derivation $\delta' \delta'' - \delta'' \delta' $ on $\f (
(y_i)_{i \in I} )$ is the zero derivation. Since it extends
uniquely to the zero derivation on $\h$, $\delta' \delta'' \beta =
\delta'' \delta' \beta$ for $\beta$ in $\h$ not in $\f(y_i)_{i \in
I}$. This shows that there is a well-defined $\Delta$-structure on
$\h _{\E''}$.

Because condition 3 implies condition 1, $\h_{\Delta''}$ is
$\Delta''$-$\f$-free over $\h$. By Lemma 8.59, since $\varphi$
from $\f\langle \xi \rangle_{\Delta'}$ to $\h$ is an
$\E'$-isomorphism, $(\f\langle \xi \rangle_{\Delta'})_{\Delta}$
and $\h \{ (y_{i})_{i \in I} \}_{\Delta''}$ are $\E$-isomorphic
over $\f$ by an isomorphism that sends $t_i$ to $y_i$. Because the
$(y_i)_{i \in I}$ are $\E''$-algebraically independent over $\f$,
the $(t_i)_{i \in I}$ are $\E''$-algebraically independent over
$\f$ also. }\bigskip

The goal of the rest of this section is to analyze the constants
of free extensions.

\begin{proposition}\label{Exercise 1 of Kolchin}{\rm \cite[Exercise 8, page
159]{kol73}}
  Let \,$\u$ a \,$\E$-field
universal over \,$\f$.  Let \,$t_1 ,\ldots, t_n \in \u$ be
\,$\E$-algebraically independent over \,$\f$. Then each element
\,$u$ of \,$\f\langle t_1 ,\ldots, t_n \rangle_\E$ not in \,$\f$
is \,$\E$-transcendental over \,$\f$.
\end{proposition}
\proof{Let $u=P(t_1,\ldots,t_n)/ Q(t_1,\ldots,t_n)$ where $P,Q\in
\f \{y_1,\ldots,y_n \}_\E$ such that \,$PQ \notin \f$ and \,$ {\rm
gcd}(P,Q)=1$. Choose orderly rankings for $\f \{y_1,\ldots,y_n
\}_\E$ and $\f \{z \}_\E$. Assume $g \in \f \{z \}_\E$ is of
lowest rank among the non-zero $\E$-polynomials satisfied by $u$.
Let $g=I_dv_g^d +I_{d-1}v_g^{d-1}+\ldots+I_0$ where $d$ is a
positive integer, $v_g$ is the leader of $g$, and the $I_k$ are
$\E$-polynomials in $\f \{z \}_\E$ of lower rank than $v_g$.
Because $I_0$ and $I_d$ are of lower rank than $g$, $I_0 (u)\neq
0$ and $I_d (u)\neq 0$. If ord $v_g = 0$, substitute $P/Q$ for
$z$, clear denominators and observe $P$ divides $Q$. But ${\rm
gcd}(P,Q)=1$, so it may be assumed that ord $v > 0$. Let $v_g$,
$v_P$ and $v_Q$ be the leaders of $g$, $P$ and $Q$, respectively,
and $S_g$, $S_P$ and $S_Q$ the separants. Write $v_g =\theta z$,
where $\theta$ is the non-empty product of $r$ derivations from
$\E$.
\begin{claim}\label{Claim in the proof of Kolchin's exercise}
$v_g(P/Q)=\theta (P/Q)= [Q^{r-1}(S_P \theta v_P Q - P S_Q \theta
v_Q) + W] / Q^{r+1}$ such that \,$W$ is the sum of terms of rank
lower than the maximum rank of \,$\theta v_P$ and \,$\theta v_Q$.
\end{claim} \proof{ The claim is clearly true for $r=1$. Assume the claim is true
for $r$. By differentiating $v_g(P/Q)=\theta (P/Q)= (S_P \theta
v_P Q - P S_Q \theta v_Q)/ Q^2 + W / Q^{r+1}$ with respect to one
of the $\delta \in \Delta$, $\delta v_g(P/Q)=$ $$\delta\theta
(P/Q)= (S_P \delta\theta v_P Q - P S_Q \delta\theta v_Q)/ Q^2
+V/Q^3 +(\delta W Q-(r+1)W \delta Q) / Q^{r+2}$$ such that the
rank of $V$ is lower than the maximum rank of $\delta \theta v_P$
and $\delta \theta v_Q$. Since $\delta W Q$ and $(r+1)W\delta Q$
also have lower than the maximum rank of $\delta \theta v_P$ and
$\delta \theta v_Q$, after adding the three fractions, the claim
is true for $r+1$. }\bigskip

Let $t$ be a positive integer such that $Q^t\cdot
I_{j}(P/Q)v_g^{j}(P/Q)$ is a $\E$-poly-nomial, in $\f
\{y_1,\ldots,y_n \}_\E$, for each $j=0,\ldots,d$. By substituting
$u$ into $Q^t g(z)$, one obtains the zero $\Delta$-polynomial
$$Q^t g(u)=Q^t (I_d(P/Q) v_g^d(P/Q) +I_{d-1}(P/Q) v_g^{d-1}(P/Q) +
\ldots +I_0(P/Q)).$$ If ${\rm rank} P > {\rm rank} Q$, then, by
the claim, the sum of the highest ranking terms of $Q^{t} g(P/Q)$
is the $\E$-polynomial $Q^t I_n(P/Q) (Q^{r-1} S_{P} \theta v_P
Q)^d$ which is equal to zero because $Q^tg(u)=0$. So that, since
$I_n(P/Q)\neq 0$ and $Q \neq 0$, it follows that $S_P=0$ and $P\in
\f$. Thus, $Q\in \f$ because ${\rm rank} P > {\rm rank} Q$. This
is contrary to the assumption $PQ \notin \f$. If ${\rm rank} Q >
{\rm rank} P$, the same type of contradiction results.

If ${\rm ord} ~P = {\rm ord} ~Q$, by the claim, $$Q^t
I_n(P/Q)Q^{(r-1)d}(S_P \theta v_P Q - P S_Q \theta v_Q)^d$$ $$ =
Q^t I_n(P/Q)Q^{(r-1)d}(S_P Q - P S_Q)^d (\theta v_P )$$ is the sum
of the highest ranking terms of $Q^tg$ and is equal to $0$.
Therefore, $(S_P Q - P S_Q)=0$.  Then, $P$ divides $S_P$ because
${\rm gcd}(P,Q)=1$. But, this is impossible because $S_P$ has
lower rank than $P$.}\bigskip

\begin{corollary}\label{Delta constants of a transext}
 Let \,$\u$ a \,$\E$-field
\,$\E$-universal over \,$\f$.  Let \,$t_1 ,\ldots, t_n \in \u$ be
\,$\E$-algebraically independent over \,$\f$.  Then \,$\f \langle
t_1 ,\ldots, t_n \rangle^{\Delta}=\f^{\Delta}$.
\end{corollary}
\proof{ The condition that an element be a $\Delta$-constant is a
$\Delta$-relation on that element.  This is impossible by the
previous proposition.}\bigskip

The next lemma is well-known.

\begin{lemma}\label{The Algebraic Constant Lemma}
{{\rm(}The Algebraic Constant Lemma{\rm)}} Let \,$\g$ over \,$\f$
be an extension of \,$\Delta$-fields. A \,$\Delta$-constant of
\,$\g$ algebraic over \,$\f$ is algebraic over the
\,$\Delta$-constants of \,$\f$.

\end{lemma}
\proof{ Let $\alpha$ be a $\Delta$-constant of $\g$ algebraic over
$\f$.  Let $f(x) \in \f[x]$ be the minimal polynomial of $\alpha$
over $\f$.  Write $f(x)= \displaystyle\Sigma_{i=1,\ldots,d}~ a_i
x^i$ for $a_i \in \f$. Then, for each $\delta \in \Delta$,
$S_f(\alpha) \delta \alpha + f^{\delta}(\alpha)=0$, where $S_f(x)$
is the derivative of $f$ with respect to $x$ and $f^{\delta}(x)$
is the polynomial obtained by applying $\delta$ to the
coefficients of $f(x)$. Since $\delta \alpha=0$,
$f^{\delta}(\alpha)=0$. Because the leading coefficient of $f(x)$
is $1$, the degree of $f^{\delta}(x)$ is less than that of $f(x)$.
Since $f(x)$ is the minimal polynomial of $\alpha$,
$f^{\delta}(x)=0$. Consequently, $\delta a_i=0$ for $i=1,\ldots,d$
and all $\delta \in \Delta$. Therefore, the coefficients of $f(x)$
are $\delta$-constants in $\f$, and $\alpha$ is algebraic over
$\f^\Delta$.}

\begin{lemma}{{\rm(}No New $\E''$-Constant Lemma{\rm)}}\label{No New Delta''-Constant
Lemma} Assume that \,$\Delta$ is the union of two disjoint subsets
\,$\Delta'$ and \,$\Delta''$. Let \,$\xi=(\xi_{1},\ldots,\xi_{n})$
be a finite family of elements of \,$\u$. If the \,$\Delta$-ring
\,$\f \{\f\langle \xi \rangle_{\Delta'} \}_{\Delta}$ is
\,$\Delta''$-$\f$-free over \,$\f\langle \xi \rangle_{\Delta'}$,
then the \,$\Delta''$-constants of \,$\f \langle \xi \rangle_{\E}$
are contained in the algebraic closure of \,$\f^{\E''}$ in $\f
\langle \xi \rangle_{\E'}$.  If \,$\f \langle \xi \rangle_{\E'}$
is a regular extension of \,$\f$, \,$\f \langle \xi \rangle_{\E}$
and \,$\f$ have the same \,$\Delta''$-constants.
\end{lemma}

\proof{By Proposition \ref{equivalent properties of freeness},
there is a transcendence basis $(t_i)_{i \in I}$ for the field $\f
\langle \xi \rangle_{\E'}$ over $\f$ that is
$\Delta''$-algebraically independent over $\f$. By Corollary
\ref{Delta constants of a transext}, the $\E''$-constants of $\f
\langle (t_i)_{i \in I} \rangle_{\Delta''}$ are in $\f$.

Let $\gamma \in \f \langle \xi \rangle_{\E}$ be a
$\Delta''$-constant and assume $\gamma \notin \f \langle (t_i)_{i
\in I} \rangle_{\Delta''}$. Then $\gamma$ is algebraic over $\f
\langle (t_i)_{i \in I} \rangle_{\E''}$ because $\xi$ and all its
$\Delta''$-derivatives are algebraic over $\f \langle (t_i)_{i \in
I} \rangle_{\Delta''}$. The Algebraic Constant Lemma \ref{The
Algebraic Constant Lemma} can then be applied to show $\gamma$ is
algebraic over the $\Delta''$-constants of $\f \langle (t_i)_{i
\in I} \rangle_{\E''}$, which is equal to $\f^{\Delta''}$ by
Corollary \ref{Delta constants of a transext}. If $\f \langle \xi
\rangle_{\E'}$ is regular over $\f$, then $\f \langle \xi
\rangle_{\E}$ is regular over $\f$ (\cite[Proposition 10(c), page
21]{kol85}) and, therefore, $\gamma \in \f$. }\bigskip

If the $\Delta$-ring $\f \{\f\langle \xi \rangle_{\Delta'}
\}_{\Delta}$ is $\Delta''$-$\f$-free over $\f\langle \xi
\rangle_{\Delta'}$ and if $\f \langle \xi \rangle_{\E}$
is a not regular extension of $\f$, there may be some
$\Delta''$-constants in $\f\langle \xi \rangle_{\Delta'}$
algebraic over $\f$. For example, take $\f=\Qset$, $\E'=\emptyset$
and $\p_{\E'} =(y^2 +1) \subset \Qset[y]$ a prime ideal. Then $ \{
\p \}  _{\E',\E''} =  \{y^2 +1\}_{\E''} \subset \Qset\{ y
\}_{\E''}$ is a $\Delta$-prime ideal ({\rm\cite[Proposition 8,
page 16 ]{kol85}}). Let $\xi$ be a $\Qset$-generic zero of $\{y^2
+1\}_{\E''}$ in $\u$. Then, by Proposition \ref{Proposition from
Kolchin on Delta freeness}, $\Qset(\xi)_{\Delta}$ is
$\Delta$-$\Qset$-free over $\Qset(\xi)$. And, since $\delta'' y
\in \{y^2 +1\}_{\E}$ for $\delta'' \in \Delta''$, $\xi$ is a
$\Delta''$-constant of $\Qset(\xi)$. In fact, the same technique
shows that, if $\p_{\E'} =(f)$ where $f \in \Qset[y]$ is an
irreducible polynomial, $\delta''\xi=0$ for a $\Qset$-generic zero
$\xi$ of $\{f\}_{\E''}$ in $\u$.

The next two propositions analyze $\E'$-constants of $\f \langle
\xi \rangle_{\E}$ instead of the $\E''$-constants.

\begin{proposition}
Let \,$\xi=(\xi_{1},\ldots,\xi_{n})$ be a finite family of
elements of \,$\u$. If the $\Delta$-ring \,$\f \{\f\langle \xi
\rangle_{\Delta'} \}_{\Delta}$ is \,$\Delta''$-$\f$-free over
$\f\langle \xi \rangle_{\Delta'}$ and if \,$\xi$ are
\,$\E'$-independent over \,$\f$, then
\,$(\f\langle\xi\rangle_{\Delta'})^{\E'}=((\f\langle\xi\rangle_{\Delta}))^{\E'}=\f^{\Delta'}$.
\end{proposition}
\proof{The set of all the $\Delta'$-derivatives of $\xi$ is a
transcendence basis for $\f\langle\xi\rangle_{\Delta'}$ over $\f$.
By Proposition \ref{equivalent properties of freeness}, they are
$\Delta''$-algebraically independent over $\f$, and all the
$\Delta''$-derivatives of $\xi$ are $\Delta'$-independent. By
Corollary \ref{Delta constants of a transext}, there are no new
$\Delta'$-constants, and the conclusion follows.}

\begin{proposition}\label{new constants found}
Let \,${\rm card} ~\E'={\rm card} ~\E''=1$ and \,$\xi=(\xi_{1})$,
and let \,$\xi \in \u$. Let \,$f(y)\in \f^{\Delta}\{y\}_{\Delta'}$
such that \,$f(y)=\Sigma a_{ij}y^i(\delta'y)^j$ with \,$a_{ij} \in
\f^{\Delta}$. Assume \,$f(\xi)=0$ and \,$S(\xi)\neq 0$ where
\,$S(y)$ is the separant of \,$f$ relative to an orderly ranking
of \,$\f^{\Delta}\{y\}_{\Delta'}$. Also assume the \,$\Delta$-ring
$\f \{\f\langle \xi \rangle_{\Delta'} \}_{\Delta}$ is
$\Delta''$-$\f$-free over \,$\f\langle \xi \rangle_{\Delta'}$. If
\,$f(y)$ is of order zero, i.e. \,$a_{ij}=0$ for \,$j>0$, then
$\delta'\xi=0$ and $\delta''\xi=0$. If not, then \,$\delta'\xi_{1}
/\delta''\xi_{1}$ is a \,$\E'$-constant of
\,$\f\langle\xi\rangle_{\E}$ not in \,$\f\langle\xi\rangle_{\E'}$.

\end{proposition}
\proof{Let $\p'$ and $\p$ denote, respectively, the prime defining
$\Delta'$-ideals of $\xi$ in $\f\{y\}_{\Delta'}$ and the defining
$\Delta$-ideal of $\xi$ in $\f\{y\}_{\Delta}$. By Proposition
\ref{Proposition from Kolchin on Delta freeness},
$\p=\{\p'\}_\Delta$.

If $\delta'y$ is not present in $f$, then $\xi$ is algebraic over
$\f^\Delta$. Let $g \in \f^\E[y]$ be the minimal polynomial for
$\xi$.  Clearly, $g \in \p'$, and $\delta' g \in \p'$.  Let $S(y)$
be $d g/ d y$. Because $g$ is the minimal polynomial, $S(y) \notin
\p'$. Since $\delta' g = S(y) \delta' y$ and since $\p'$ is prime,
$\delta' y \in \p' \subset \p$, and $\delta' \xi =0$. Similarly,
$\delta'' \xi =0$.

If $\delta'y$ is present in $f$, $\delta' f=S(y)\delta'^2y +
(\partial f/\partial y) \delta' y$ and $\delta''f =S(y)
\delta''\delta'y + (\partial f/\partial y) \delta'' y$ are
elements of $\p=\{\p'\}_\Delta$, where $S(y)=\partial f/\partial
\delta' y$ and $S(y) \notin \p$.  Then,$$\delta''y \cdot\delta' f
- \delta'y \cdot \delta'' f = S(y)( \delta''y   \delta'^2y -
\delta'y\delta''\delta'y)$$ is also an element of $\p$. Since
$S(y) \notin \p$, $\delta''y   \delta'^2y -
\delta'y\delta''\delta'y $ is.  Because $\xi$ is a $\Delta$-zero
of $\p$, $\delta''\xi \delta'^2\xi - \delta'\xi\delta''\delta'\xi
=0$, and $\delta'(\delta'\xi /\delta''\xi)=0$. Since $\delta'\xi
\in \f\langle\xi\rangle_{\E'}$ and $\delta''\xi \notin
\f\langle\xi\rangle_{\E'}$, clearly $\delta'\xi /\delta''\xi
\notin \f\langle\xi\rangle_{\E'}$.}\bigskip

The last proposition applies to the familiar Weiestrass
$\wp$-function (a $\Delta$-zero of $f(y)= (\delta'y)^2 -y^3-ay-b$)
and the exponential function (a $\Delta$-zero of $f(y)=y -
\delta'y$), in which case a new $\Delta'$-constant is $\delta'\xi
/\delta''\xi=\xi /\delta''\xi$.

\subsection{The $\Ep$-Group Induced from an Algebraic Group.}
\label{The E-Group Induced from an Algebraic Group.} In this
section, let $\f$ be a $\Delta$-field and let $\Delta'$ be a
commutative linearly independent subset of the vector space
spanned by $\Delta$ over $\f$. Let $\u$ be a $\Delta$-universal
extension of $\f$. In \cite[Chapter 2, Section 3, page 56]{kol85},
Kolchin develops a procedure for associating to each
$\Delta'$-$\f$-group $G$ (relative to the $\Delta'$-field $\u$) a
$\Delta$-$\f$-group $G_{\Delta}$ (relative to the $\Delta$-field
$\u$) which is called the induced $\Delta$-$\f$-group. The
elements of $G_{\Delta}$ are defined to be the same as those of
$G$.  If the $\Delta'$-subfield of $\u$ associated to $x$ in $G$
is $\f \langle x \rangle_{\Delta'}$, the $\Delta$-subfield of $\u$
associated to $x$ in $G_{\Delta}$ is $\f\langle \f \langle x
\rangle_{\Delta'}\rangle_{\Delta}$.

Heuristically, to each open affine $B$ of $G$ defined by a
$\Delta'$-ideal $\p'$ of $\f \{y_1,\ldots,y_n \}_{\Delta'}$, one
may associate the open affine $B_\E$ of $G_{\Delta}$ defined by
the $\Delta$-ideal $\{\p'\}_{\Delta}$ of $\f \{y_1,\ldots,y_n
\}_{\Delta}$. To the element $x$ of $G$, thought of as a
$\Delta'$-zero in $\u^n$ of $\p'$, corresponds the element $x$ of
$G_{\Delta}$, thought of as a $\E$-zero of $ \{\p'\}_{\Delta}$.
The $\Delta'$-rational functions giving the group law on $G$ are
also $\Delta$-rational functions on $G_{\Delta}$ and give the
group law on $G_{\Delta}$.  An $\f$-generic element $v$ of $G$,
which is a generic zero of some $\p'$ as above, will be an
$\f$-generic element of $G_{\Delta}$ if and only if it is a
generic zero of $\{\p'\}_{\Delta}$ \cite[Theorem 3(2c), page
58]{kol85}.  The discussion in the last section implies $v$ will
be an $\f$-generic element of $G_{\Delta}$ if $v$ is a
$\f$-generic element of $G$ and $(\f \langle x
\rangle_{\Delta'})_{\Delta}$ is $\Delta / \Delta'$-$\f$-free over
$\f \langle x \rangle_{\Delta'}$.

\begin{definition}\label{definition of Delta Delta' homomorphism}
{\rm \cite[page 56]{kol85}} Let $\Delta'$ be a commutative
linearly independent subset of $\f \Delta$. Let $G$ be a
$\E'$-$\f$-group $($relative to the $\Delta'$-field $\u$ $)$, and
let $H$ be an $\E$-$\f$-group $($relative to the $\Delta$-field
$\u$$)$. A {\it $(\Delta, \Delta')$-$\f$-homomorphism of $H$ into
$G$} is a group homomorphism $f:  H \rightarrow G$ that satisfies
the following three conditions:
\begin{enumerate}
\item if $y \in H$, then $\f \langle f(y) \rangle_{\Delta'} \subset
      \f \langle y \rangle_{\Delta}$,

\item  if $y, y' \in H$ and $y \rightarrow^{\Delta} y'$,
then $f(y) \rightarrow^{\Delta'} f(y')$,
\item  if $y, y' \in H$ and $y \leftrightarrow^{\Delta} y'$,
then $S_{\Delta,y',y}$ extends $S_{\Delta',f(y'),f(y)}$.
\end{enumerate}

\end{definition}

\begin{definition}\label{definition of group induced by the
extension of derivations} {\rm \cite[page 57]{kol85}} Let $G$ be a
$\E'$-$\f$-group relative to the universe $\u$. A
$\Delta$-$\f$-group structure on $G$, denoted by $G_{\Delta}$, is
said to be {\it induced} {\rm (}by the given $\Delta'$-$\f$-group
structure on $G${\rm)} if the following two conditions are
satisfied:
\begin{enumerate}
\item ${\rm id}_G$ is a $(\Delta, \Delta')$-$\f$-homomorphism;

\item   every $(\Delta, \Delta')$-$\f$-homomorphism of a
$\Delta$-$\f$-group into $G$ is a \linebreak
$\Delta$-$\f$-homomorphism.

\end{enumerate}

\end{definition}

\subsection{Varying the Universal field}\label{Varying the
Universal field} For $\f$ a $\Delta$-field. the functor "extending
the universal field of $\f$", has been developed by Kolchin. (See
\cite[Chapter 2, Section 1, Varying the universal differential
field, page 45]{kol85} and \cite[Chapter 8, Section 10, The
Lie-Cassidy-Kovacic method, page 247]{kol85}).  Let $\v$ and $\u$
be $\Delta$-extensions of $\f$ that are $\E$-universal over $\f$
and such that $\u \subseteq \v$. The functor "extending the
universal field of $\f$" takes the category of
$\Delta$-$\f$-groups (relative to $\u$) and $\Delta$-$\f$-group
homomorphisms to the category of $\Delta$-$\f$-groups (relative to
$\v$) and $\Delta$-$\f$-groups homomorphisms. Heuristically, a set
defined as the $\E$-zeros in $\u$ of a system of $\E$-equations is
associated to the set of $\E$-zeros in $\v$ of the same system of
$\E$-equations.

\subsection{The Existence Theorem} The purpose of this section
is to prove every connected $\Ep$-group is isomorphic to the
Galois group of an $\Ep$-strongly normal extension.

Let $\f$ be an $\Ep$-field, and let $\v$ be an $\Ep$-extension of
$\f$ that is $\Ep$-universal over $\f$. Let $G$ be a connected
$\Ep$-$\f$-group (relative to the $\Ep$-field $\v$). Let $\h
\subset \v$ an $\Ep$-extension of $\f$, with $\v$ not necessarily
universal over $\h$. Let $\chi$ be an $\Ep$-derivation ($\chi $
commutes with the action of $\Ep$) of $\h$ into $\v$ over $\f$.
For each element $g$ of $G$ rational over $\h$, evaluation at $g$
of $\Ep$-$\f$-functions on $G$ defined at $g$ composed with $\chi$
is local $\Ep$-derivation at $g$.  If $g$ is
$\Ep$-$\h^{\chi}$-affine, this local derivation can be extended to
a unique tangent vector to $G$ at $g$ \cite[Section 8, Chapter
8]{kol85}. By right translating this tangent vector to all of $G$,
one obtains an element $l \chi(g) $ of the Lie algebra $\l
_{\E}(G)$ of invariant $\Ep$-derivations of $G$ which is called
the logarithmic derivative of $g$ relative to $\chi$ \cite[page
236]{kol85}. Thus, for any local derivation $\chi$ at $g\in G$,
there exists a unique element $l\chi (g)$ of the Lie algebra
$\l_\E(G)$ with the property that $$l\chi (g)(f)(g)=\chi(f(g))$$
for every $\Ep$-$\f$-function $f$ defined at $g$.

In the remainder of this section, let $\f$ be an \de field, let
$\c=\f^\E$, and let $\u$ be an \de extension of $\f$ that is \de
universal over $\f$. Let $G$ be a connected $\Ep$-$\c$-group
(relative to the $\Ep$-universal field $\u^{\Delta}$). By
extending the universal $\Ep$-field from $\u^{\Delta}$ to $\u$,
considered as an $\Ep$-field (Section \ref{Varying the Universal
field} or \cite[Chapter 2, Section 1, page 44]{kol85}), $G$ may be
considered as $\Ep$-$\c$-group (relative to the $\Ep$-field $\u$).
For each $\delta$ in $\Delta$ and any $g$ in $G_{\u}$, the
logarithmic derivative is $l\delta(g)\in \l _{\Ep}(G)$.

The following lemma is one of the well known properties of the
logarithmic derivative \cite[Proposition 8, page 236]{kol85} and
will be used a few times.
\begin{lemma}\label
{lemma on two equal logarithmic derivatives}Let \,$x,y \in G_\u$.
If \,$l\delta x=l\delta y$ for all \,$\delta \in \Delta$, there
exist an element \,$c \in G_{\u^{\E}}$ such that \,$c= x^{-1} y $.
\end{lemma}
\proof{Assume $l\delta x=l\delta y$ for all $\delta \in \Delta$.
By \cite[Remark after Theorem 3, page 237]{kol85}, for $w,z \in
G$, $l \delta(wz)= l\delta(w) + \tau_{w}^{\#}(l \delta(z))$ where
$\tau_{w}^{\#}$ is the isomorphism of the Lie algebra induced by
conjugation with $w$. By letting $w=x$ and $ z= x^{-1} y$,
$l\delta (y)= l\delta (x)+ \tau_{x}^{\#}(l \delta (x^{-1} y))$. So
$0=\tau_{x}^{\#}(l \delta (x^{-1} y))$, and $0=l \delta (x^{-1} y
)$. Then $c=x^{-1} y \in G_{\u^{\E}}$ \cite[Proposition 8(c), page
236]{kol85}. }

\begin{definition}
The element \,$\alpha \in G_{\u}$ is a \,$G$-{\rm primitive over}
\,$\f$ if the logarithmic derivative \,$l\delta(\alpha) \in
\l_{\Ep,\f}(G)$ for each \,$\delta \in \E$. A \,$G$-{\rm primitive
extension} is an extension of \,$\f$ of the form \,$\f \langle
\alpha \rangle$ where \,$\alpha $ is a \,$G$-primitive over
\,$\f$.\end{definition}

\begin{proposition}\label{proposition given a primitive of G on the
injective homomorphism from the Galois group the G}  Let
\,$\alpha$ be a \,$G$-primitive over \,$\f$ such that the field of
\,$\E$-constants of \,$\f \langle \alpha \rangle_{\Ep,\E}$ is
\,$\c$. Then \,$\f \langle \alpha \rangle_{\Ep,\E}$ is an \,\dsn
extension of \,$\f$ \,$($relative to \de field $\u$$)$, and the
map \,$c: G(\f \langle \alpha \rangle_{\Ep,\E} / \f) \mapsto G$
defined by \,$c(\sigma)= \alpha^{-1}\sigma \alpha$ defines an
injective \,$\Ep$-$\c$-homomorphism of \,$\Ep$-$\c$-groups
$($relative to the \,$\Ep$-field $\u^{\E}$$)$.\end{proposition}
\proof{ Since $\alpha$ is a $G$-primitive over $\f$,
$l\delta(\alpha) \in \l_{\Ep,\f}(G)$ for each $\delta \in \E$. So
that, for any \de isomorphism $\sigma$ of $\f\langle \alpha\rangle
_{\Ep,\E}$ over $\f$, $\sigma(l\delta(\alpha))=l \delta(\alpha)$
for $\delta \in \Delta$. Also,
$l\delta(\sigma\alpha)=\sigma(l\delta(\alpha))$ for all
$\delta\in\E$ by \cite[Proposition 8(b), page 236]{kol85}.
Therefore, $l\delta(\sigma\alpha) =l\delta(\alpha)$, and, by Lemma
\ref{lemma on two equal logarithmic derivatives}, $c(\sigma)=
\alpha^{-1}\sigma \alpha$ is an element of $G_{\u^{\E}}$.  Since
$$\sigma(\f\langle\alpha\rangle_{\Ep,\E})
\subset\f\langle\alpha\rangle_{\Ep,\E}
\sigma(\f\langle\alpha\rangle_{\Ep,\E})=
\f\langle\alpha,\sigma\alpha\rangle_{\Ep,\E}$$
$$=\f\langle\alpha,c(\sigma))\rangle_{\Ep,\E}
=\f\langle\alpha\rangle_{\Ep,\E}\c\langle c(\sigma)\rangle_{\Ep},$$
$\f\langle \alpha\rangle$ is \dsn over $\f$ by Proposition
\ref{Proposition on the one sided condition for E-strongly normal}.
By definition,  $\c\langle \sigma
\rangle=(\f\langle\alpha\rangle_{\Ep,\E}
\sigma(\f\langle\alpha\rangle_{\Ep,\E}))^\E$. Therefore, $\c\langle
\sigma \rangle= \c\langle c(\sigma)\rangle_\Ep$ by \cite[Corollary 2
to Theorem 1, page 88]{kol73}. For any $\sigma,\tau \in
G(\f\langle\alpha\rangle_{\Ep,\E} / \f)$, $c$ is a group
homomorphism since $\alpha c(\sigma\tau) = \sigma\tau\alpha=
\sigma(\alpha c(\tau)) = \sigma\alpha\circ c(\tau)= \alpha
c(\sigma)c(\tau)$. If $\sigma$ is in the kernel of $c$, $\sigma
\alpha=\alpha c(\sigma)=\alpha$ and, hence, $\sigma = id_{\f\langle
\alpha \rangle}$ because $\alpha$ \de generates $\f\langle \alpha
\rangle_{\Ep,\E}$. Therefore $c$ is injective.

To prove that $c$ is an $\Ep$-$\c$-homomorphism, it will be shown to
be pre $\Ep$-$\c$-mapping (Definition \ref{premapping}). Then, since
$c$ is a homomorphism, \cite[Corollary 1, page 90]{kol85} implies
that $c$ is an $\Ep$-$\c$-homomorphism.  Parts 1,2 and 3 of the
Definition \ref{premapping} follow by taking the domain to consist
only of $\c$-generic elements and from the fact that $\c\langle
\sigma \rangle= \c\langle c(\sigma)\rangle_\Ep$. To show part 4 of
the definition, take $\sigma \leftrightarrow \sigma'$ two
$\c$-generic elements. By the definition of $\c$-generic
$\Ep$-specialization in $G(\f \langle \alpha \rangle_{\Ep,\E} /
\f)$, there exists an \de $\f\langle\alpha\rangle_{\Ep,\E}
$-isomorphism $\varphi :\f\langle\alpha\rangle_{\Ep,\E}
\sigma(\f\langle\alpha\rangle_{\Ep,\E}) \approx
\f\langle\alpha\rangle_{\Ep,\E}
\sigma'(\f\langle\alpha\rangle_{\Ep,\E})$ that maps $\sigma \beta$
onto $\sigma' \beta$ for each $\beta\in \f\langle\alpha\rangle$.
Therefore, $\varphi(c(\sigma)) =\varphi(\alpha^{-1} \sigma \alpha)
=\alpha^{-1} \sigma' \alpha =c(\sigma') $. Thus, the induced
$\Ep$-$\c$-isomorphism $S_{ c(\sigma') ,c(\sigma)}$ obtained by
restricting $\varphi$ to $\c\langle \sigma \rangle= \c\langle
c(\sigma)\rangle_\Ep$, is exactly the induced $\Ep$-$\c$-isomorphism
$S_{\sigma',\sigma}$, and $c(\sigma) \leftrightarrow c(\sigma')$.
}\bigskip

The following Lemma has a pivotal role in the next theorem.

\begin{lemma}\label{lemma on the generic preimage of the
logarithmic derivative} Let \,$G$ be a connected
\,$\Ep$-$\c$-group \,$($relative to the \,$\u$$)$.  Let \,$\eta$
and \,$\xi$ be elements of \,$G_\u$, i.e. elements of \,$G$
rational over \,$\u$. Assume \,$\eta$ is \,$\c$-generic and
\,$\c\langle \eta \rangle ^{\Delta} = \c$. If \,$l\delta(\eta) =
l\delta(\xi)$ for all \,$\delta \in \Delta$, then \,$\xi$ is
\,$\c$-generic, and \,$\eta \leftrightarrow \xi$ in \,$G$.

\end{lemma}
\proof{ By Lemma \ref{lemma on two equal logarithmic derivatives},
there exists $\gamma \in G_{\u^{\Delta}}$ such that $\eta \gamma =
\xi$.  By the theorem on the linear disjointness of $\E$-constants
\cite[Corollary 1, page 87]{kol73}, $\c\langle \eta \rangle$ and
$\c\langle \gamma \rangle$ are linearly disjoint over $\c$. By
\cite[Theorem 1(d), page 39 ]{kol85}, $ \eta \gamma$ is a
$\c$-generic element of $G_{E,\Delta}$.  Since $\eta\gamma=\xi$,
$\xi$ is $\c$-generic. Because $G$ is connected, $\eta
\leftrightarrow \xi$ in $G$.}\bigskip

For the proof the next Theorem, one uses the fact that the
elements of $G$ (relative to the $\Ep$-field $\u^{\Delta}$) are
contained in $(G_{\Ep,\Delta})_{\u^{\Delta}}$, as the following
discussion indicates. An $\Ep$-$\c$-group $G$ (relative to the
$\Ep$-field $\u^{\Delta}$) is given.  Let $G_\u$ (relative to
$\u$) be the $\Ep$-$\c$-group obtained from $G$ (relative to
$\u^{\Delta}$) by extending the universal differential field from
$\u^{\Delta}$ to $\u$.  The elements of $G$ (relative to
$\u^{\Delta}$) are the elements $(G_\u)_{\u^\E}$ of the
$\Ep$-group $G_\u$ (relative to $\u$) rational over $\u^{\Delta}$.
Let $G_{\Ep,\E}$ (relative to the \de field $\u$) be the \de
$\c$-group obtained from the $\Ep$-$\c$-group $G_\u$ (relative to
$\u$) by extending the derivations from $\Ep$ to $(\Ep,\E)$.  From
the discussion in the preceding section on the \de -$\c$-group
$G_{\Ep,\E}$, the elements of the $\Ep$-$\c$-group $G_\u$ are
included in the elements of the \de $\c$-group $G_{\Ep,\E}$.
Therefore the elements of the $\Ep$-$\c$-group $G$ (relative to
$\u^{\E}$) are elements $(G_{\Ep,\E})_{\u^\E}$ of \de -$\c$-group
$G_{\Ep,\E}$ (relative to the \de field $\u$).

\begin{theorem}\label{Theorem on the construction of an
$E$-strongly normal extension from any $E$-group} Let \,$G$ be a
connected \,$\Ep$-$\c$-group \,$($\,relative to the \,$\Ep$-field
\,$\u^{\E})$. Let \,$\eta$ be a \,$\c$-generic element of
\,$G_{\Ep,\E}$. Then, \,$\g=\c\langle \eta \rangle _{\Ep,\E}$ is
\,$\Ep$-strongly normal over \,$\f=\c\langle
l\delta_1\eta\rangle_{\Ep,\E} \cdots \c\langle
l\delta_m\eta\rangle_{\Ep,\E}$ \,$($relative to the \,\de field
\,$\u)$ such that the Galois group \,$G(\g/\f)$ \,$($relative to
the \,$\Ep$-field \,$\u^{\Delta}$$) \linebreak is $
\,$\Ep$-$\c$-isomorphic to \,$G$.
\end{theorem}

\proof{  Since the $\Ep$-$\c$-group $G$ (relative to the
$\Ep$-field $\u^{\E}$) is connected, the $\Ep$-$\c$-group $G$
(relative to the $\Ep$-field $\u$) is connected \cite[Section 1,
page 44]{kol85}. This implies that the \de $\c$-group $G_{E,\E}$
(relative to the \de field $\u$) is connected \cite[Theorem 3,
page 58]{kol85}.

By Proposition \ref{Proposition from Kolchin on Delta freeness},
$\c \{\c \langle \eta \rangle_{\Ep} \} _{\Delta}$ is $\E$-free
over $\c\langle \eta \rangle_{\Ep}$. Because $G_{\Ep,\E}$ is
connected, $\g=\c\langle \eta \rangle_{\Ep,\E}$ is a regular
extension of $\c$ by the third axiom for $\Ep$-groups. The No New
$\Delta''$-Constant Lemma \ref{No New Delta''-Constant Lemma} then
implies that the $\Delta$-constants of $\g=\c\langle \eta
\rangle_{\Ep,\E}$ are in $\c$.

Set $\g=\c\langle \eta \rangle _{\Ep,\E}$ and $\f=\c\langle
l\delta_1\eta\rangle_{\Ep,\E} \cdots \c\langle
l\delta_m\eta\rangle_{\Ep,\E}$. Since for each $\delta \in
\Delta$, $l\delta:G_{\Ep,\E} \rightarrow
(\l_{\Ep,\f}(G))_{\Ep,\E}$ is a pre $(\Ep,\E)$-mapping
\cite[Corollary, page 243]{kol85}, $\c\langle l \delta\eta
\rangle_{\Ep,\E} \subseteq \c\langle \eta \rangle _{\Ep,\E}$ for
each $\delta \Delta$. Therefore, $\f \subset \g$, and
 $\g^{\E}=\f^{\E}=\c$. By
construction, $\eta$ is a $G$-primitive over $\f$. By Proposition
\ref{proposition given a primitive of G on the injective
homomorphism from the Galois group the G}, $\g$ is strongly
$\Ep$-normal over $\f$, and the map $c: G(\g / \f) \mapsto G$
defined by $c(\sigma)= \eta^{-1}\sigma \eta$ is an injective
$\Ep$-$\c$-homomorphism.

To show that $c$ is surjective, let $\beta$ be any element of the
connected $\Ep$-$\c$-group $G$ $($relative to the universal
$\Ep$-field $\u^{\E})$.  Using the identification of the elements
of the $\Ep$-$\c$-group $G$ (relative to the $\Ep$-field
$\u^{\E}$) with the subset $(G_{\Ep,\E})_{\u^{\E}}$ of the
elements of the \de-$\c$-group $G_{\Ep,\E}$ (relative to the \de
field $\u$), consider $\beta$ as an element of $G_{\Ep,\E}$.
Because $l\delta(\eta\beta)=l\delta(\eta)+ \tau^{*}_{\eta}
l\delta(\beta) =l\delta(\eta)$, Lemma \ref{lemma on the generic
preimage of the logarithmic derivative} implies $\eta
\leftrightarrow \eta \beta$. Then, by part 3 in the definition of
a pre set, there is an \de-isomorphism
$S_{(\Ep,\E),\eta\beta,\eta}: \c\langle\eta\rangle_{\Ep,\E}
\approx \c\langle\eta\beta\rangle_{\Ep,\E}$ over $\c$. Let $\sigma
= S_{(\Ep,\E),\eta\beta,\eta }$. By {\it DAS 2b} in the definition
of a pre set, there exist a unique element $x$ of $G_{\Ep,\E}$
such that $\eta \leftrightarrow x$, $S_{(\Ep,\E),x,\eta}= \sigma$
and $\sigma(\c\langle \eta \rangle_{\Ep,\E}) = \c \langle x
\rangle_{\Ep,\E}$. This element $x$ is the definition of $\sigma
\eta$ \cite[page 30]{kol85}. Therefore, $\sigma \eta = \eta
\beta$. For all $\delta \in \Delta$, the computation $\sigma
l\delta (\eta)= l\delta(\sigma \eta)=l\delta( \eta \beta)=
l\delta(\eta)+ \tau^{*}_{\eta} l\delta(\beta)= l\delta(\eta)$
shows that $\f$ is invariant under $\sigma$, and, hence, $\sigma
\in G(\g /\f)$. Then, $c$ is surjective since $c(\sigma)=\eta^{-1}
\sigma\eta =\beta$. Because a bijective $\Ep$-$\c$-homomorphism of
$\Ep$-$\c$-groups is an $\Ep$-$\c$-isomorphism \cite[Corollary 4,
page 97]{kol85}, $c$ is an $\Ep$-$\c$-isomorphism.}\bigskip

 For given $\Ep$-group, the procedure in the next
corollary constructs an $\Ep$-strongly normal extension in two
stages.

\begin{corollary}\label{Corollary on the construction of
an $E$-strongly normal extension from any $E$-group} Assume
\,$\Delta=\{\delta\}$. Let \,$G$ be a connected \,$\Ep$-$\c$-group
\,$($relative to the \,$\Ep$-field \,$\u^{\E})$. Let
\,$G_{\Ep,\Delta}$ be the \,\de \,$\c$-group \,$($relative to the
\de field \,$\u$$)$ obtained by first extending the universal
\,$\Ep$-field from \,$\u^{\Delta}$ to \,$\u$ and then by extending
the the derivations from \,$\Ep$ to \,$(\Ep,\Delta)$. First choose
a \,$\c$-generic element \,$a$ of \,$\l_{\Ep,\c}(G)_{\Ep,\E}$, and
then choose an element \,$b$ of \,$G_{\Ep,\Delta}$ such that
\,$l\delta(b)=a$. Then \,$b$ is a \,$C$-generic element of
\,$G_{\Ep,\E}$, and \,$\c \langle b \rangle_{(\Ep,\E)}$ over \,$\c
\langle a \rangle_{(\Ep,\E)}$ is \,$\Ep$-strongly normal
\,$($relative to the \de field $\u$$)$ with Galois group
\,$\Ep$-$\c$-isomorphic to \,$G$.

\end{corollary}

\proof{ There exist a $\c$-generic element $a$ of
$\l_{\Ep,\c}(G)_{\Ep,\E}$ because of the definition of pre \de
sets. That $b$ exists follows from the surjectivity of the
logarithmic derivative \cite[Proposition 11, page 240]{kol85}.

Let $\eta$ be a $\c$-generic element of $G_{\Ep,\Delta}$. Set
$\g=\c\langle \eta \rangle _{\Ep,\E}$ and $\f= \c\langle
l\delta\eta\rangle_{\Ep,\E}$.  By the previous theorem, $\g$ over
$\f$ is an \dsn extension with Galois group $G(\g/\f)$ which is
$\Ep$-$\c$-isomorphic to $G$ $($relative to the universal
$\Ep$-field $\u^{\E})$. The proof of this corollary will be
accomplished by showing that $\c \langle b \rangle_{\Ep,\E}$ is
\de isomorphic to $\c\langle \eta \rangle _{\Ep,\E}$ over $\c$.

Because $\eta$ is a $\c$-generic element of $G_{\Ep,\Delta}$ and
the logarithmic derivative $l\delta$ is a surjective \de
$\c$-mapping, $l\delta\eta$ is a $\c$-generic element of
$\l_{\Ep,\c}(G)_{\Ep,\E}$ because, if $t$ is any element of
$\l_{\Ep,\c}(G)_{\Ep,\E}$ and $\xi$ is an element of
$G_{\Ep,\Delta}$ such that $l\delta \xi=t$, then $\eta \rightarrow
\xi$ implies $l\delta \eta \rightarrow l\delta \xi=t$ since
$l\delta$ is pre \de mapping \cite[Corollary, page 242]{kol85}.
Because $a$ and $l\delta\eta$ are both $\c$-generic elements of
$\l_{\Ep,\c}(G)_{\Ep,\E}$, there exists an \de-isomorphism
$\varphi$ over $\c$ from $\c\langle a \rangle_{\Ep,\E}$ to $\c
\langle l\delta\eta \rangle_{\Ep,\E}$. Because $\u$ is \de
universal over $\c\langle a \rangle_{\Ep,\E}$, $\varphi$ extends
to an \de $\c$-isomorphism, also called $\varphi$, from $\c\langle
b\rangle_{\Ep,\E}$ to $\u$.

Since $b$ is an element of $G_{\Ep,\Delta}$, by {\it DAS 2b} in
the definition of pre sets, there exist a unique $x$ in
$G_{\Ep,\Delta}$ with $b \leftrightarrow x$ such that $\c\langle x
\rangle_{\Ep,\E}= \varphi(\c\langle b\rangle_{\Ep,\E} )$ and
$S_{(\Ep,\E),\c,b,x}=\varphi$. Since isomorphisms over $\c$
commute with the logarithmic derivative \cite[Proposition 8, page
236]{kol85}, $l\delta(x) =l\delta(\varphi b)=\varphi(l\delta
(b))=\varphi a =l\delta( \eta).$ By Lemma \ref{lemma on the
generic preimage of the logarithmic derivative}, $x$ is a
$\c$-generic element of $G_{\Ep,\Delta}$, and $x \leftrightarrow
\eta$. Therefore, $b \leftrightarrow \eta$, and $b$ is a
$\c$-generic element of $G_{\Ep,\Delta}$.
Because $S_{(\Ep,\E),\eta,b}:\c\langle b \rangle_{\Ep,\E} \approx
\c\langle \eta \rangle_{\Ep,\E}$ is an \de $\c$-isomorphism and
$S_{(\Ep,\E),\eta,b} (\c \langle a \rangle)_{\Ep,\E}= \c\langle
l\delta(\eta) \rangle_{\Ep,\E}$, by Proposition \ref{E-strongly
normal by isomorphism}, $\c \langle b \rangle_{(\Ep,\E)}$ over $\c
\langle a \rangle_{(\Ep,\E)}$ is $\Ep$-strongly normal with Galois
group $\Ep$-$\c$-isomorphic to $G$.}

\subsection{The $\Del$-Strongly Normal Extension Corresponding to
the $\Ep$-Group Induced from an Algebraic Group.}\label{Section of
extension of Galois groups by Delta and Galois theory}

This section is a precise explanation of the heuristics described
in the fourth paragraph of the introduction. In particular, given
a linear differential operator $L$ in the variable $x$ such that
the coefficients are in the $(D_t,D_x)$-field $\f$. Let $\g'$ be
the extension $D_x$-field of the coefficient field $\f$ generated
by a fundamental system of $D_x$-zeros of $L$. Furthermore, assume
the $D_x$-constants of $\g'$ equals those of $\f$ so that the
extension $\g'$ over $\f$ is strongly normal with Galois group
$G$. Let $\g$ be the $(D_t,D_x)$-field generated by $\g'$ such
that the $D_x$-constants of $\g$ equals those of $\f$, which is
true if the function field are analytic functions of two
variables. Then $\g$ is a $D_t$-strongly normal extension of $\f$,
and the Galois groups $H$ is an $D_t$-group. Corollary
\ref{Corollary on embeddability} shows that $H$ is embedded via a
$D_t$-homomorphism to the $D_t$-group $G_{D_t}$ induced from $G$
by the extension of derivations (Section \ref{The E-Group Induced
from an Algebraic Group.}). An open problem is to compute the
$D_t$-Galois groups of classical differential equations depending
on parameters, such as the hypergeometric differential equation.

If $A$ is an $\E$-ring which is a subset of an $(\Ep,\E)$-ring,
$A_\Ep$ will denote the \de ring generated by $A$. If $A$ is an
$\E$-ring which is a subset of an $(\Ep,\E)$-field, $A_{(\Ep)}$
will denote the \de field generated by $A$. Always
$(A^{\E})_{(\Ep)}\subset (A_{(\Ep)})^{\E}$. Also, please note
that, if $A$ and $B$ are two $\E$-rings which are subsets of an
$(\Ep,\E)$-field, $(A [B ] )_{(\Ep)}= A_{(\Ep)}\cdot B_{(\Ep)}$.

In this section, the following notations will be used. Let $\u$ an
\de field that is \de universal over some \de field. Consider $\Ep$
as the union of two disjoint subsets $\Ep'$ and $\Ep''$. Let $\f'$
be an $(\Ep',\E)$-subfield of $\u$ such that $\u$ is universal over
$\f_{(\Ep'')}$ as \de fields. This implies that $\u$ considered as
an $(\Ep',\E)$-field is also $(\Ep',\E)$-universal over $\f'$. Let
$\g'$ be an $(\Ep',\E)$-subfield of $\u$ which is an $\Ep'$-strongly
normal extension of $\f'$ relative to the universal
$(\Ep',\E)$-field $\u$. Also, let $\g=(\g')_{(\Ep'')}$,
$\f=(\f')_{(\Ep'')}$, $\c'=\g'^\E=\f'^\E$ and $\c=\g^\E$. This
definition of $\c$ is a change in notation from the usual
$\c=\f^\E$. (See Remark \ref{Remark on the possibility of new
constants}.)

 \[
\begin{CD}
\g'    @>>>\g   @>>>   \g = \g\c \\
   @AAA        @AAA           @AAA\\
  \f'                     @>>> \f  @>>> \f\c   \\
    @AAA                      @AAA                     @AAA\\
   \c'          @>>>     \f^\E             @>>>      \c\\
\end{CD}
\]\bigskip

All the results in this section relate the Galois groups of the
$(\Ep',\E)$-fields $\g'$ over $\f'$ to the Galois group of the \de
fields $\g\c=\g$ over $\f\c$ and constitute a straight forward
application of basic definitions. In one's first reading of this
material, the reader may assume that $\Ep'$ is empty.
The theorems are presented in the increased generality, with $\Ep'$
not empty, because no extra work is involved and they might be
useful.

\begin{lemma}\label{lemma on extending an E''sn extension}
Let \,$\g'$ be an \,$(\Ep',\E)$-subfield of \,$\u$ which is an
\,$\Ep'$-strongly normal extension of the \,$\Ep'$-field \,$\f'$
relative to the \,$(\Ep',\E)$-universal \,$(\Ep',\E)$-field
\,$\u$. Assume \,$\u$ is \,\de universal over
\,$\g=\g'_{(\Ep'')}$. Then any \,\de isomorphism \,$\sigma$ of
\,$\g=\g\c$ into \,$\u$ over \,$\f\c$ is \,$\Ep$-strong.
Furthermore, $(\g\sigma\g)^{\E} = ( (\g' \sigma
\g')_{(\Ep'')})^{\E} = \c ((\g' \sigma \g')^{\E})_{(\Ep'')} $, and
$\c\langle \sigma \rangle =\c \cdot \c'\langle \sigma
\rangle_{(\Ep'')}$.
\end{lemma}

\begin{remark}\label{Remark on the possibility of new constants}
The field generated by the \,$\Ep''$-derivatives of \,$\g'$ may
contain new \,$\Delta$-constants not in the field generated by the
\,$\Ep''$-derivatives of \,$\f'$.  An example of a strongly normal
extension of \,$\Delta$-fields \,$\g'$ over \,$\f'$ with this
property is any \,$\g'$ generated by a Weierstrassian over a field
of \,$\Delta$-constants $\f'$.  $($See {\rm \cite[Examples, page
405]{kol73}} and Corollary \ref{new constants found}.$)$ This
means that, in the lemma, for \,$\sigma$ to be \,$\Ep$-strong it
must leave fixed a field \,$\c$ of \,$\Delta$-constants that might
include \,$\Delta$-constants not in $\c'$.
\end{remark}
\proof{
Because $\sigma$ is an \de isomorphism of $\g$ over $\f\c$ and
$\c=\g^\E$, $\sigma$ leaves the $\E$-constants $\c$ of $\g$
invariant.
Since $\sigma$ restricted to $\g'$ is $\Ep'$-strong, $\sigma\g'
\subset \g'\u^{\E}$ and $\g' \subset \sigma \g' \u^{\E}$. Then,
$$\sigma \g=\sigma(\g'_{(\Ep'')})=(\sigma\g')_{(\Ep'')}\subset (\g'\u^{\E})_{(\Ep'')}
=\g'_{(\Ep'')}(\u^{\E})_{(\Ep'')}=\g\u^{\E},$$ and
$$\g=\g'_{(\Ep'')} \subset (\sigma \g' \u^{\E})_{(\Ep'')} =(\sigma
\g')_{(\Ep'')}(\u^{\E})_{(\Ep'')}=\sigma(\g'_{(\Ep'')})\u^{\E} =
\sigma(\g)\u^{\E}. $$ Therefore, $\sigma$ is $\Ep$-strong.

For the first equality,
$$(\g\sigma\g)^{\E}=(\g'_{(\Ep'')}\sigma(\g'_{(\Ep'')}))^{\E}
=(\g'_{(\Ep'')}(\sigma\g')_{(\Ep'')})^{\E}
=((\g'\sigma\g')_{(\Ep'')})^{\E}.$$ Since the $\Ep'$-strong
normality of $\sigma$ implies
 $\g'\sigma\g'=\g'(\g'\sigma\g')^{\E}$, above sequence of equalities is equal to$$
((\g'(\g'\sigma\g')^{\E})_{(\Ep'')})^{\E}
=(\g'_{(\Ep'')}((\g'\sigma\g')^{\E})_{(\Ep'')})^{\E}=(\g\cdot
 ((\g' \sigma \g')^{\E}) _{(\Ep'')} )^{\E}$$
 $$=(\g\cdot\c
 ((\g' \sigma \g')^{\E}) _{(\Ep'')} )^{\E}
=\c((\g' \sigma \g')^{\E})_{(\Ep'')}, $$  where the last equality
follows from \cite[Corollary 2, page 88]{kol73} because $\g$ and
the $\E$-constants $\c ((\g \sigma \g)^{\E}) _{(E'')}$ are
linearly disjoint over $\c$. The last equality of the proposition
follows from the first two equalities and the definitions of
$\c\langle \sigma \rangle$ and $ \c'\langle \sigma \rangle$ as
$(\g\sigma\g)^{\E}$ and $(\g'\sigma\g')^{\E}$.}

\begin{proposition}\label{Proposition on the gal of an E-gen
gal}Let \,$\g'$ be an \,$(\Ep',\E)$-subfield of \,$\u$ which is an
\,$\Ep'$-strongly normal extension of \,$\f'$ relative to the
universal \,$(\Ep',\Delta)$-field \,$\u$.
Then \,$\g$ is an \,$\Ep$-strongly normal extension of \,$\f\c$
relative to the universal \,$(\Ep,\E)$-field \,$\u$. Define the
map \,$\rho$ from the \,$\Ep$-$\c$-group $G( \g/ \f\c)$ to the
\,$\Ep'$-$\c$-group $G( \g'\c /\f'\c)$ that associates to an \,\de
$\f\c$-isomorphism of \,$\g$ its restriction to \,$\g'\c$. Then
\,$\rho$ is an injective \,$(\Ep,\Ep')$-$\c$-homomorphism
\,$($Definition \ref{definition of Delta Delta'
homomorphism}\,$)$. Furthermore, $\c\langle \sigma \rangle =\c
\cdot \c'\langle \rho(\sigma) \rangle_{(\Ep'')}$.
\end{proposition}

\proof{ Because $\g'$ over $\f'$ is finitely
$(\Ep',\E)$-generated, $\g$ over $\f$ and, therefore, $\g$ over
$\f\c$ are finitely $(\Ep,\E)$-generated. By Lemma \ref{lemma on
extending an E''sn extension}, any \de isomorphism of $\g$ over
$\f\c$ is $\Ep$-strong. And, since $\g^\E=(\f\c)^\E$, $\g$ over
$\f\c$ is $\Ep$-strongly normal.

By Theorem \ref{Theorem on extension of constant of a strongly
normal extension}, $G( \g'\c/ \f'\c)$ is the induced
$\Ep'$-$\c$-group of the $\Ep'$-$\c$-group $G(\g'/\f')$, both
being identified with each other by means of their canonical
identifications with the group of $(\Ep',\E)$-automorphisms of
$\g'\u^{\E}$ over $\f'\u^{\E}$. That $\rho$ is a group
homomorphism is clear by identifying the $\Ep$-group $G( \g /
\f\c)$ with \de automorphisms of $\g\u^{\E}$ over
$\f\c\u^{\E}=\f\u^{\E}$ and the $\Ep'$-group $G(\g'\c/ \f'\c)$
with $(E',\Delta)$-automorphisms of $\g'\u^{\E}$ over $\f'\u^{\E}$
and observing that the restriction $\rho$ preserves composition in
these groups. Because any set of $(\Ep',\E)$-generators of the
$(\Ep',\E)$-field $\g'\c$ over $\f'\c$ are $(\Ep,\E)$-generators
of the $(\Ep,\E)$-field $\g\c$ over $\f\c$, $\rho$ is injective.

To show $\rho$ is an $(\Ep,\Ep')$-$\c$-homomorphism each part of
Definition \ref{definition of Delta Delta' homomorphism} will be
verified. For $\sigma\in G( \g / \f\c )$, $\c\langle \sigma
\rangle =\c \cdot \c'\langle \rho(\sigma) \rangle_{(\Ep'')}$ by
Lemma \ref{lemma on extending an E''sn extension}.  Since $\c
\cdot \c'\langle \rho(\sigma) \rangle=\c\langle \rho( \sigma
)\rangle$ (Theorem \ref{Theorem on extension of constant of a
strongly normal extension}), it follows that $\c\langle \sigma
\rangle \supset \c\langle \rho (\sigma) \rangle$.  If $\sigma
\rightarrow \tau$ for $\sigma, \tau \in G(\g/\f\c)$, then, by the
definition of specialization, there is an \de homomorphism
$\varphi: \g [\sigma \g]\rightarrow \g[\tau\g]$ over $\g$ such
that $\varphi(\sigma \alpha)=\tau\alpha$ for all $\alpha \in \g$.
Since $\g'\c  \subset \g$, the restriction of $\varphi$ to $\g'\c
[\rho(\sigma)(\g'\c)]$ is an $(\Ep',\E )$-homomorphism $\g'\c
[\rho(\sigma) (\g'\c)] \rightarrow \g'\c[\rho(\sigma)(\g'\c)]$
over $\g'\c$ which takes $\rho(\sigma) \alpha$ to
$\rho(\tau)\alpha$ for all $\alpha \in \g\h$. Therefore, by
definition, $\rho(\sigma) \rightarrow\rho(\tau)$. If $\sigma
\leftrightarrow\tau$, then the \de homomorphism $\varphi$, defined
above, is an \de isomorphism and, therefore, extends to an \de
isomorphism, also denoted by $\varphi$, of the $\Ep$-field $\g
\sigma \g$ to the field $\g\tau\g$. The restriction of this \de
isomorphism to $(\g \sigma \g)^\E= \c \langle \sigma \rangle$ is
the induced $\Ep$-$\c$-isomorphism $S_{\Ep;\tau, \sigma}:\c
\langle \sigma \rangle_{\Ep} \rightarrow \c \langle \tau
\rangle_{\Ep}$. The \de isomorphism $\varphi$ also restricts to an
$(\Ep',\E)$-$\c$-isomorphism from the $(\Ep',\E)$-field $\g'\rho(
\sigma) \g'$ to the $(\Ep',\E)$-field $\g'\rho(\tau)\g'$, which in
turn restricts to the induced $\Ep'$-$\c$-isomorphism
$S_{\Ep';\rho(\tau),\rho(\sigma)}:\c \langle \rho(\sigma)
\rangle_{\Ep'} \rightarrow \c \langle \rho(\tau )\rangle_{\Ep'}$.
Since $\c\langle \rho(\sigma) \rangle_{\Ep'} \subset \c\langle
\sigma \rangle_{E}$, $S_{\Ep;\tau, \sigma}$ extends
$S_{\Ep';\rho(\tau),\rho(\sigma)}$.\nolinebreak}\bigskip

This Proposition, in the case $E'$ is empty, can be used to
produce examples of $\Ep$-strongly normal extensions.  Start with
a $\Delta$-extension $\g'$ over $\f'$ which is strongly normal (in
the sense of Kolchin) such that the coefficients of the
differential equations defining $\g'$ over $\f'$ depend on
parameter $t$. Assume that the $\Delta$-field $\f$ is closed with
respect to differentiation by $t$. Differentiate the elements of
$\g$ with respect to $t$ to generate a $\{ d/dt,\Delta \}$-field
extension $\g$. Then if $(\g)^{\Delta} \subset \f$, $\g$ over $\f$
is $\{ d/dt \}$-strongly normal over $\f$.

\begin{corollary}\label{Corollary on embeddability} In the above proposition, assume $\c=\g^\E
\subset \f^\E=\c'$. Then the injective
\,$(\Ep,\Ep')$-$\c$-homomorphism \,$\rho:G(\g/\f)\rightarrow
G(\g'/\f')$ identifies the \,$\Ep$-$\c$-group \,$G(\g/\f)$ with an
\,$\Ep$-$\c$-subgroup of the \,$\Ep$-$\c$-group \,$G(\g'/\f')_\Ep$
induced from the \,$\Ep'$-$\c$-group \,$G(\g'/\f')$ by extending
the derivations to $\Ep$ $($Definition \ref{definition of group
induced by the extension of derivations}\,$)$.
\end{corollary}
\proof{ Kolchin proved that the induced $\Ep$-$\c$-group
$G(\g'/\f')_\Ep$ always exists \cite[Theorem 3, page 58]{kol85}.
By Definition \ref{definition of group induced by the extension of
derivations} of the induced $\Ep$-group, the
$(\Ep,\Ep')$-$\c$-homomorphism \,$\rho$ of the last proposition
extends to a unique $\Ep$-$\c$-homomorphism
$\overline{\rho}:G(\g/\f)\rightarrow G(\g'/\f')_\Ep$. It is also
injective because $\rho$ and $\overline{\rho}$ are equal on the
elements of $G(\g/\f)$. The image of an $\Ep$-$\c$-group under an
$\Ep$-$\c$ homomorphism is a $\Ep$-$\c$-subgroup \cite[Proposition
4, page 92]{kol85}. Because $\rho$ is a bijective
$\Ep$-$\c$-homomorphism of $G(\g/\f)$ to its image, the
$\Ep$-$\c$-group $G(\g/\f)$ and its image in $G(\g'/\f')_\Ep$ are
$\Ep$-$\c$-isomorphic \cite[Corollary 4, page
97]{kol85}.\nolinebreak}

\section{Examples}\label{examples}
In this chapter, $\f$ will denote an \de field, and $\u$ will
denote an \de field universal over $\f$. The field $\k$ of
$\E$-constants of $\u$ is, as an $\Ep$-field, $\Ep$-universal over
the $\E$-constants $\c$ of $\f$.

\subsection{$G^{\Ep}_a$-extensions}
Denote the additive \de -$\mathbb{Q}$-group \cite[page 28]{kol85}
(relative to $\u$) by the symbol $G^{\Ep,\E}_{a}$. The elements of
$G^{\Ep,\E}_{a}$ are those of $\u$, and its group structure is
that of the field $\u$ under addition. Similarly, $G^{\Ep}_{a}$
will denote the additive $\Ep$-$\mathbb{Q}$-group (relative to
$\k$) with elements those of $\k$. Let $\kappa\in
\FF(G^{\Ep,\E}_{a})$ be the canonical coordinate function on
$G^{\Ep,\E}_{a}$. Then, $\delta_i \kappa \in \FF(G^{\Ep,\E}_{a})$,
and the $\Ep$-$\f$-mapping
$l\E=(\delta_1\kappa,\dots,\delta_m\kappa):G^{\Ep,\E}_{a}\rightarrow(G^{\Ep,\E}_{a})^n$
\cite[Proposition 6, page 129]{kol85} is the logarithmic
derivation on $G^{\Ep,\E}_{a}$ relative to $\E$ \cite[Example 1,
page 352]{kol73}. By \cite[Proposition 3, page 89]{kol85}, it is
an $(\Ep,\E)$-$\f$-homomorphism. The kernel of $l\E$ is the \de
$\f$-subgroup consisting of \de zeros of the \de ideal $[\delta_1
y,\ldots, \delta_m y] \subset \f\{y\}_{\Ep,\E}$ and can be
identified with $G^{\Ep}_{a}$ relative to the $\Ep$-universal
field $\k$.\bigskip
\begin{definition}\label{Definition of Delta-primitive}
 An element  \,$\alpha\in\u$ is \,$\Delta$-primitive over \,$\f$ if
\,$l\E \alpha \in\f^{m}${\rm ;} that is, for suitable elements
\,$a_{1},\ldots ,a_{m}\in\f$, \,$\alpha$ satisfies the system of
differential equations
$$ \delta_{i}\alpha=a_{i}
 ~(1\leq i\leq m).$$
\end{definition}

Let $\alpha$ be $\Delta$-primitive over $\f$, and suppose that the
field of $\E$-constants of $\f\langle\alpha\rangle_{\Del,\E}$ is
$\c=\f
^{\E}$. For any \de isomorphism $\sigma$ of $\f%
\langle\alpha\rangle_{\Del,\E}$ over $\f$,
$(\delta_{1}(\sigma\alpha),\ldots,\delta_{m}(\sigma\alpha)) =
(\sigma(\delta_{1}\alpha),\ldots,\sigma(\delta_{m}\alpha)) =
(\delta_{1}\alpha,\ldots,\delta_{m}\alpha)$; hence the difference
c($\sigma $)=$\sigma\alpha-\alpha$ is in the kernel of the above
homomorphism $l\delta$ and a $\Delta$-constant.
As \[\f%
\langle\alpha\rangle_{\Del,\E}\sigma(\f\langle\alpha\rangle_{\Del,\E})
=\f\langle\alpha\rangle_{\Del,\E}\f\langle\sigma\alpha\rangle_{\Del,\E}
~~~~~~\] \[~~~~~~~=
\f\langle\alpha\rangle_{\Del,\E}\f\langle\alpha+c(\sigma)\rangle_{\Del,\E}
=\f\langle\alpha\rangle_{\Del,\E}\c\langle
c(\sigma)\rangle_{\Del,\E},\] it follows that
$\f\langle\alpha\rangle_{\Del,\E}$ is $\Del$-strongly normal over
$\f$, and  \linebreak$\c\langle\sigma\rangle_\Del=(\f%
\langle\alpha\rangle_{\Del,\E}\sigma(\f\langle\alpha\rangle_{\Del,\E}))^\E=\c\langle
c(\sigma)\rangle_\Del$. For any two elements \linebreak
$\sigma,\sigma^{\prime}\in G( \f\langle\alpha\rangle_{\Del,\E} /
\f)$ (regarded as elements of ${\rm
Aut}_{E,\Delta}(\f\langle\alpha\rangle_{\Del,\E} \k / \f \k)$ by
means of Proposition \ref{Identifies strong isomorphisms with
automorphisms}),
\[\alpha +c( \sigma\sigma^{\prime}) = \sigma\sigma^{\prime}\alpha =
\sigma(\alpha +c(\sigma^{\prime}))=\sigma\alpha +c(\sigma') =
\alpha +c(\sigma)+c(\sigma^{\prime})\]since $c(\sigma^{\prime})
\in \k$ and, thus,
$\sigma(c(\sigma^{\prime}))=c(\sigma^{\prime})$. Therefore,
$c(\sigma\sigma^{\prime}) =c(\sigma)+c(\sigma^{\prime})$, and,
evidently, $c(\sigma)=0$ only when
$\sigma=id_{\f\langle\alpha\rangle_{\Del,\E}}$. This proves the
first part of the following proposition, and the remainder is the
same as that of Proposition \ref{proposition given a primitive of
G on the injective homomorphism from the Galois group the G}.

\begin{proposition}\label{Proposition on Ga extensions} Let \,$\alpha$ be a \,$\E$-primitive
over \,$\f$, and suppose that the field of \,$\E$-constants of
\,$\f\langle\alpha\rangle_{\Del,\E}$ is \,$\c=\f ^{\E}$. Then,
each \,\de $\f$-isomor-phism $\sigma$ of
\,$\f\langle\alpha\rangle_{\Del,\E}$ into $\u$ is of the form
\,$\sigma \alpha=\alpha+c(\sigma)$ for \,$c(\sigma) \in \k$. In
addition, \,$\f\langle\alpha\rangle_{\Del,\E}$ is \,$\Ep$-strongly
normal over \,$\f$, and the mapping \,$c:
G(\f\langle\alpha\rangle_{\Del,\E} / \f)
 \rightarrow  G^{\Ep}_{a}$ defined by \,$c(\sigma)=\sigma \alpha -\alpha$ for
 \,$\sigma \in  G(\f\langle\alpha\rangle_{\Del,\E} / \f)$ is an injective \,$\Del$-$%
\c$-homomorphism of \,$\Ep$-groups relative to the
\,$\Ep$-universal field
\,$\k$. Consequently, \,$\f%
\langle\alpha\rangle_{\Del,\E}$ is a \,${\rm G^{\Ep}_a}$-extension
of \,$\f$.\end{proposition}

\begin{proposition}\label{Galois subgroups of Ga}
Let \,$G$ be an \,$\Del$-$\c$-subgroup of ${G}_a^\Ep$. Let \,$\l
\subseteq \c \left\{ y \right\} _{\Del }$ be the linear
\,$\Del$-ideal defining \,$G$ \,{\rm \cite[page 151]{kol85}}. Let
\,$b \in \u$ be a $\c$-generic \,\de zero of \,$\l_{\E,E} \subset
\c \left\{ y \right\} _{\Del,\E }$. Let \,$a=(a_1,\ldots,a_m) =
l\E b$. Put \,$\f=\c \langle a \rangle_{E,\E}$, and \,$\g=\f
\langle b \rangle_{E,\E}$. Then \,$\g$ over \,$\f$ is an
\,$\Ep$-strongly normal extension with Galois group
$\Ep$-$\c$-isomorphic to \,$G$.

\end{proposition}

\proof{ This is a special case of Theorem \ref{Theorem on the
construction of an $E$-strongly normal extension from any
$E$-group}.}\bigskip

Let $\g$ be $\Ep$-strongly normal over $\f$ with Galois group $G
\subset G^{\Ep}_{a}$. Theorem \ref{existance of the differential
group of isomorphisms} shows that $G$ is an $\Ep$-$\c$-group where
$\c=\f^\E$. By  \cite[page 151]{kol85}, $G$ is set of $\Ep$-zeros
of a linear $\Ep$-ideal $\l_G \subseteq \c\{\kappa\}_\Ep$, where
$\kappa$ is the canonical coordinate function on $G^{\Ep}_{a}$.
Each $\Ep$-$\c$-subgroup $H \subseteq G$ is also the $\Ep$-zeros
of a linear $\Ep$-ideal $\l_H \subseteq \c\{y\}_\Ep$ such that
$\l_G \subseteq \l_H$. Recall, by the definition of a linear
$\Ep$-ideal $\l$, $\l=[\l_{1}]_\Ep$ where $\l_{1}$ is the subset
of elements of $\l$ of degree one. For each $H\subset G$, the
following proposition exhibits the subfield of $\g$ invariant
under the action of $H$ and, thus, specifies the Galois
correspondence, even if $\c=\f^\E$ is not constrainedly closed.

\begin{proposition}\label{Proposition exhibiting the Galois
correspondence for Ga}Let \,$\g$ be an \,$\Ep$-strongly normal
extension of \,$\f$ with Galois group \,$G(\g/\f)\subseteq
G_a^\Ep$. Assume that \,$\g=\f\langle b\rangle_{\Ep,\E}$ where
\,$b \in \u$ is a \,$\E$-primitive over \,$\f$. Then, their exists
a Galois correspondence which to each \,$\Ep$-$\c$-subgroup \,$H$
of \,$G(\g/\f)$ associates the \,\de subfield \,$\h =\f\langle
(L(b))_{L \in \l_{H}} \rangle_{\Ep,\E}\subseteq \g$, where \,$\l_H
\subseteq \c\{\kappa\}_\Ep$ is the
linear \,$\Ep$-ideal defining \,$H$ and \,$\kappa$ is the canonical coordinate function on \,$\k=G_a^\Ep$ .\\
\[
\begin{CD}
G(\g/\h)             @+>>>[80]     \f\\
    @AAA                      @VVV \\
    H          @>>>\h=\g^H=\f\langle (Lb)_{L \in \l_{H}}\rangle_{\Ep,\E}   \\
   @AAA   @VVV\\
1    @+>>>[80]     \g  \\
\end{CD}
\]\\

\end{proposition}
\proof{Since $\l_H=[\l_{H,1}]_\Ep$, it follows that $\h=\f\langle
(L(b))_{L \in \l_{H,1}} \rangle_{\Ep,\E}$. Let $\sigma \in
G(\g/\h)$. By Proposition \ref{Proposition on Ga extensions},
$\sigma(b)=b +c(\sigma)$ for $c(\sigma) \in \k$. For all $L \in
\l_{H,1}$,
$$L(b)=\sigma(L(b))=L(\sigma(b))=L(b+c(\sigma))=L(b)+L(c(\sigma)):$$ thus $L(c(\sigma))=0$.
Therefore, $c(\sigma)$ is an $\Ep$-zero of $\l_H$, $\sigma \in H$
and $H \supset G(\g/\h)$.  If $\sigma \in H$,
\begin{equation}\label{L(b)invariant}\sigma(L(b))=L(\sigma(b))=L(b +c(\sigma))=L(b)
+L(c(\sigma))=L(b) \end{equation} for $L \in \l_1$, and $H \subset
G(\g/\h)$.}\bigskip

 For simplicity, assume $\E=\{\delta\}$ throughout the remainder of
this section. In the next proposition, if $b$ is $\E$-primitive
over $\f$, the Galois group of $\g=\f\langle b \rangle_{\Ep,\E}$
over $\f$ is completely determined by $a=\delta b \in \f$.

\begin{proposition}\label{Proposition the Galois group is determined in F for Ga}
Let \,$b$ be a \,$\E$-primitive over \,$\f$, and let
\,$\g=\f\langle b \rangle_{\Ep,\E}$. Assume that \,$\g^\E=\f^\E$.
Let \,$a=\delta b$, let \,$\l_{a,1}=\{L(y) \in \c\{y\}_{\Ep,1}
\mid L(a) \in \delta \f\}$, and let \,$\l_a=[\l_{a,1}]_\Ep$. Let
\,$G=Gal(\g/\f)$, and let \,$c:G \rightarrow \v$ be the
\,$\Ep$-$\f$-homomorphism defined by \,$c(\sigma)=\sigma(b)-b$.
Then, the defining $\Ep$-ideal \,$\AA_{~c(G)} \subset \c\{y\}_\Ep$
of \,$c(G)$ is $\l_a$.
\end{proposition}\proof{By Proposition \ref{Proposition on Ga extensions}, $\g$ over $\f$ is $\Ep$-strongly
normal, and $G$ is an $\Ep$-$\c$-group. By \cite[page 151]{kol85},
the $\Ep$-$\c$-group $c(G)\subseteq G^\Ep_a$ is the set of
$\Ep$-zeros of a linear $\Ep$-ideal $\AA_{~c(G)}\subset
\c\{y\}_\Ep$.  Also, Proposition \ref{Proposition on Ga
extensions} shows that each $\sigma \in G$ is of the form
$\sigma(b)=b+c(\sigma)$ for an $\Ep$-zero $c(\sigma)$ of
$\AA_{c(G)}$.

For each linear $L(y) \in\AA_{~c(G)}$, Equation
\ref{L(b)invariant} above shows $L(b)$ is invariant under all
elements of $G$. Thus, $L(b) \in \f$, and $L(b)=f$ for some $f \in
\f$. Hence,$$L(a)=L(\delta b)=\delta(L(b))=\delta f.$$ Therefore,
$\AA_{~c(G)} \subseteq \l_a$.

On the other hand, let $L(y) \in \l_{a,1}$. Then $L(a)=\delta f$
for $f \in \f$, and $L(b)- f$ is a $\E$-constant because
$\delta(L(b)- f)=L(\delta b)-\delta f= L(a)-\delta f=0$.
Therefore, $L(b)-f \in \c \subseteq \f$, and $L(b) \in \f$. Hence,
for all $\sigma \in G$, $\sigma(L(b))=L(b)$, and the computation
$$L(c(\sigma))=L(\sigma(b)-b)=L(\sigma(b))-L(b)=\sigma(L(b))-L(b)=0$$ shows that $\AA_{~c(G)} \supseteq\l_a$. }\bigskip

The following is a simple example of an $\Ep$-strongly normal
extension $\g$ over $\f$ such that the transcendence degree of
$\g$ over $\f$ is infinite in the usual algebraic sense. Let $\f
\subset \u$ be an \de field containing an element $a$ that is
linearly $\Ep$-$\f^\E$-independent modulo $\delta\f$ (Definition
\ref{Definition of linearly E-independence}). For instance, any
$a\in \c\langle t \rangle_{\Ep,\E}$, $a\notin\f$, where $t$ is \de
independent over $\f$ satisfies this condition by Proposition
\ref{Exercise 1 of Kolchin}. Let $b \in \u$ be an \de zero of the
\de ideal $\{\delta y-a\}_{\Ep,\E} \subset \f\{y\}_{\Ep,\E}$. Let
$\g=\f\langle b \rangle_{\Ep,\E}$.  By Corollary \ref{Proposition
on no nontrivial epsilon-differential ideal in the differential
ring generated by logs}, $(\f\langle b
\rangle_{\Ep,\E})^\E=\f^\E$. Therefore, $\g$ is $\Ep$-strongly
normal over $\f$ by Proposition \ref{Proposition on Ga
extensions}. Since $b$ is $\Ep$-independent over $\f$,
$\g=\f\langle b \rangle_{\Ep,\E}$ has infinite transcendence
degree over $\f$. In fact $c(G(\g/\f))=G^\Ep_a$, because if a
nonzero $L(y) \in \c\{y\}_{\Epsilon,1}$ is in the defining
$\Ep$-ideal of $c(G(\g/\f))$ by the previous proposition, there
exist an $f \in \f$ such that $L(a)=\delta f$. This contradicts
the fact that $1,b,\epsilon b,\epsilon^2 b,\ldots$ are linearly
independent over $\f$ (Proposition \ref{Proposition on no
nontrivial differential ideal in the differential ring generated
by logs}).

\begin{corollary}
Assume \,$\Ep=\{\epsilon\}$ and \,$\Delta=\{\delta\}$. Let \,$\h$
be an algebraically closed \,\de field such that \,$\h^\E=\h$, let
\,$\f=\h\langle x \rangle_{\Ep,\E}$, where \,$x \in \u$,
\,$\epsilon x=0$ and \,$\delta x =1$, and as usual let
$\c=\f^\E=\h^\E$. Then, there is no \,$\E$-primitive
\,$\Ep$-strongly normal extension of \,$\f$ with Galois group
\,$\Ep$-$\c$-isomorphic to \,$G^\Ep_{a}$.
\end{corollary}\begin{remark}This remains true if the
hypothesis that \,$\h$ be an algebraically closed is omitted; the
following proof must be modified to take the structure of
irreducibles into account in the partial fraction decomposition.
\end{remark}\proof{
Assume that there exist an $\E$-primitive $\Ep$-strongly normal
extension $\g$ of $\f$ with Galois group G that is
$\Ep$-$\c$-isomorphic to $G^\Ep_{a}$. Let $b \in \u$ be a
$\E$-primitive over $\f$ such that $\delta b=a \in \f$ and
$\g=\f\langle b \rangle_{\Ep,\E}$. Let
$a=p(x)+\displaystyle{\Sigma}_{i,j} ~\frac{h_{i,j}}{(x-h_i)^j}$,
for $p(x) \in \h[x]$ and $h_i, h_{i,j} \in \h$, be the partial
fraction decomposition of $a$. If $h_{i,1}=0$ for all $i$,
$a=\delta f$ for $f \in \f$, and $b-f \in \g$ is a $\E$-constant
not in $\f$, which contracts the assumption that $\g$ over $\f$ is
$\Ep$-strongly normal (Proposition \ref{Proposition that there are
no new constants in an E strongly normal extension}). Therefore,
$h_{i,1}\neq 0$ for at least one $i$, and there exists a non-zero
$L(y)=$\[\begin{vmatrix} h_{1,1} & h_{2,1}&{\ldots} & h_{r,1} & y\\
\epsilon h_{1,1} &\epsilon h_{2,1}&{\ldots} &\epsilon h_{r,1} &
\epsilon y\\ :& :& :& :& : \\ \epsilon^r h_{1,1} &\epsilon^r
h_{2,1}&{\ldots} &\epsilon^r h_{r,1} & \epsilon^r y
\end{vmatrix}\] $\in \h\{y\}_{\Ep,1}$ such that the finitely
many $h_{i,1}$ span over $\h^{\Ep,\E}$ the linear space of
$\Ep$-zeros of $L(y)$. By Lemma \ref{Lemma on L kills a mod
deltaF} below, since $L(h_{i,1})=0$ for all $i$, $L(a)\in
\delta\f$. By Proposition \ref{Proposition the Galois group is
determined in F for Ga}, $L(y)$ is contained in the defining
$\Ep$-ideal of $c(G)$, which contradicts the assumption that $G$
is $\Ep$-$\c$-isomorphic to \,$G^\Ep_{a}$.}

\begin{lemma}\label{Lemma on L kills a mod deltaF}
Assume \,$\Ep=\{\epsilon\}$ and \,$\Delta=\{\delta\}$. Let \,$\h$
be an algebraically closed \,\de field such that \,$\h^\E=\h$, and
let \,$\f=\h\langle x \rangle_{\Ep,\E}$, where \,$x \in \u$,
\,$\epsilon x=0$ and \,$\delta x =1$. Let \,$M(y) \in
\h^\E\{y\}_{\Ep,1}$. For \,$\alpha \in \f$, let
\,$\alpha=p(x)+\displaystyle{\Sigma}_{i,j}
~\frac{h_{i,j}}{(x-h_i)^j}$ for \,$p(x) \in \h[x]$ and \,$h_i,
h_{i,j} \in \h$, be the partial fraction decomposition of
\,$\alpha$. Then, $M(\alpha)\in\delta \f$ if and only if
$M(h_{i,1})=0$ for all \nolinebreak$i$. \end{lemma} \proof{ The
only terms in the above representation of $\alpha$ not in $\delta
\f$ are those with $j=1$. Since $\delta M(y)=M(\delta y)$, if
$j>1$, $M(\frac{h_{i,j}}{(x-h_i)^j}) \in \delta \f$ because
$\frac{h_{i,j}}{(x-h_i)^j} \in \delta \f$. Therefore, the
condition $M(a) \in \delta \f$ is equivalent to
$M(\displaystyle{\Sigma}_{i} ~\frac{h_{i,1}}{(x-h_i)}) \in \delta
\f$. Since
$$\epsilon(\frac{h_{i,1}}{(x-h_i)})= \frac{\epsilon h_{i,1}}{(x-h_i)} - \frac{h_{i,1}
\epsilon h_i}{ (x-h_i)^2},$$ by induction $\frac{\epsilon^k
h_{i,1}}{(x-h_i)}=\frac{\epsilon^k h_{i,1}}{(x-h_i)} +{\rm an~
element ~of~ } \delta\f$. By the linearity of $M$, $M(\alpha) \in
\delta \f$ is equivalent to $\displaystyle{\Sigma}_{i}
~\frac{M(h_{i,1})}{(x-h_i)} \in \delta \f$.  This is true if and
only if $M(h_{i,1})=0$ for all $i$ since all the $\frac{1}{x-h_i}$
are linearly independent over $\h$ modulo $\delta \f$ (Corollary
\ref{Corollary on the sum of derivatives of logs that are not
exact}).}\bigskip

 The
next two results establish procedures for the construction of all
$G^\Ep_a$-extensions under the condition $\Ep=\{\epsilon\}$.

\begin{proposition}\label{Proposition on the construction
of a G(L) ext} Assume \,$E=\{\epsilon\}$ and \,$\Delta =
\{\delta\}$. Let \,$\h$ be an \, \de field, and let \,$h \in \h$.
Let $\u$ be \,\de universal over \,$\h$.
Let \,$L(y) \in \h^\E\{y\}_{\Ep,1}$ of positive order \,$n$ with
the coefficient of the highest order term equal to \nolinebreak
\,$1$.
\begin{enumerate} \item There exists \,$a \in \u$ be an
\,\de zero of \,$[L(y) -\delta h]_{\Del,\E} \subset \h\lbrace y
\rbrace_{\Del,\E} $ such that
 $ a,\epsilon a, \ldots, \epsilon^{n-1} a$ are linearly independent over
\,$(\h\langle a \rangle_{\Ep,\E})^\E$ modulo \linebreak
\,$\delta((\h\langle a \rangle_{\Ep,\E})$
  \item There exists \,$b \in \u$ be an
\,\de zero of \,$M=[\delta y - a, L(y)-h]_{E,\E} \subset
\linebreak \h\langle a \rangle_{\Ep,\E} \lbrace y \rbrace_{E,\E}$.
\end{enumerate} Put \,$\f =\h\langle a \rangle_{E,\E}$ and \,$\g
=\f \langle b\rangle_{E,\E}$. Then, \,$\g$ is an \,\dsn ~extension
of \,$\f$, and \,$\l=[L]_{\Ep}$ is the defining \,$\Ep$-ideal of
\,$c(G(\g/\f))\subseteq G^\Ep_a$.
\end{proposition}

\proof{ Let $a \in \u$ be an $\h$-generic $(\Ep,\E)$-zero of
$[L(y) -\delta h]_{\Del,\E}$. Clearly, $a$ satisfies 1. To show
there exists $b \in \u$ that satisfies 2, \cite[Lemma 5 and 6,
page 137]{kol73} will be applied to show that $M$ is a proper
prime\de ideal.  Since $\u$ is \de universal over $\h$, there
exists an \de zero $b \in \u$ as required.

To apply \cite[Lemma 5, page 137]{kol73}, $\{\delta y - a,
L(y)-h\}$  must be an coherent autoreduced set of $M$ relative to
some fixed ranking. It is clearly autoreduced. The coherence of
the follows by letting $L'(y) =L(y)-\epsilon^n y$ and computing
$$\delta( L(y)-h) -\epsilon^n(\delta y -a) =\delta L'(y)
+\epsilon^{n} a -\delta h$$ $$=\delta L'(y) +\epsilon^{n} a
-\delta h - (L( a) -\delta h) =\delta L'(y)- L'(a) =L'(\delta y-
a).$$

To show $M$ is a proper \de ideal, assume that it is not. Then $1
\in M$. Since $1$ is partially reduced with respect to $\{\delta y
- a, L(y)-h\}$, \cite[Lemma 5, page 137]{kol73} implies that $1
\in (\delta y - a, L(y)-h) \subset\h\langle a \rangle_{\Ep,\E}
\lbrace y \rbrace_{E,\E}$, which is impossible because $1$ is
reduced with respect to $\{\delta y - a, L(y)-h\}$ and is not
zero. To show $M$ is a prime \de ideal, since the separants and
initials of a coherent autoreduced set are $1$, it is sufficient
to observe that the ideal $(\delta y - a, L(y)-h)$ is prime by
\cite[Lemma 6, page 137]{kol73}. It is well known that an
inhomogeneous linear ideal is prime.

Next, $\g^\E=\f^\E$ follows from 1 and Proposition
\ref{Proposition on no nontrivial differential ideal in the
differential ring generated by logs} since $\delta(\epsilon^k
b)=\epsilon^k a$ for $k=0,\ldots, n-1$ and the set $b,\epsilon
b,\ldots,\epsilon^{n-1} b$ generate $\g$ as a field extension of
$\f$. Because $\g$ is \de generated by a $\E$-primitive element
over $\f$, Proposition \ref{Proposition on Ga extensions} implies
the $\g$ is $\Ep$-strongly normal over $\f$. Since $L( a)=\delta
h$, Proposition \ref{Proposition the Galois group is determined in
F for Ga} implies $L(y) \in \l$. Suppose there where a linear $M
\in \l$ of lower order than $L$. Then, again by Proposition
\ref{Proposition the Galois group is determined in F for Ga},
there exist $f\in \f$ such that $M(a)=\delta f$.
 Then, $M(b)-f$ is a $\E$-constant in
$\f$. Therefore, $M(b)=f_1 \in \f$. However, by Proposition
\ref{Proposition on no nontrivial differential ideal in the
differential ring generated by logs}, $1,b, \epsilon
b,\ldots,\epsilon^{n-1}b$ are linearly independent over $\f$. This
contradiction shows that $L$ is linear of minimal order in $\l$.
Therefore, $\l=[L]_\Ep$.}\bigskip

One may apply this proposition to the example of the introduction.
Let $\h=  \mathbb{C}(t,x,\cos t,\sin t)$ ($\epsilon t=1,\epsilon
x=0,\delta t=0,\delta x=0$), $h=0$, $\gamma = \cos t / \sin t $,
$a =\sin t /x \in \h$, and let $L(y)=\epsilon y-\gamma y \in
\h^\Ep\{y\}_{\Ep,1}$. Then $a$ is an \de zero of
$[L(y)]_{\Ep,\E}$, and $a$ is linearly independent over
$(\h\langle a \rangle_{\Ep,\E})^\E=(\h)^\E=\mathbb{C}(t,\cos
t,\sin t)$ modulo $\delta(\h\langle a
\rangle_{\Ep,\E})=\delta(\h)$ by Corollary \ref{Corollary on the
sum of derivatives of logs that are not exact}. Let $b=\log x \sin
t$. Then $b$ is an \de zero of $[ \delta y -a,L(y)]_{\Ep,\E}$. By
Proposition \ref{Proposition on the construction of a G(L) ext},
$\g=\h\langle b \rangle_{\Ep,\E}$ is an $\Ep$-strongly normal
extension of $\f=\h\langle a \rangle_{\Ep,\E}$, and $[L(y)]_{\Ep}$
is the defining ideal of the Galois group in $G^\Ep_a$.

The following corollary reformulates the previous proposition so
that other examples may be constructed easily.

\begin{corollary}Let \,$F$ be an \,\de field, let
\,$h \in F$, and let $d_1,\ldots,d_n \in F^\Ep\subset \u$. Let
\,$L(y)\in F^\E\{y\}_{\Ep,1}$ of positive order \,$n$ with the
coefficient of the highest order term equal to \nolinebreak \,$1$,
and, for \,$i=1,\ldots n$, let \,$e_i \in \u^\Ep$ be an
\,$\E$-zero of \,$\delta y - d_i \in F\{y\}_{\E}$. Assume
\begin{enumerate}\item \,$d_1,\ldots,d_n$ are linearly independent
over \,$F^\E$ modulo \,$\delta F$,  \item there exist
\,$\eta_1,\ldots,\eta_n \in F^\E$ such that
\,$\eta_1,\ldots,\eta_n$ are $\Ep$-zeros of \,$L( y)$ linearly
independent over \,$F^{\Ep,\E}$, and \item there exist an \,\de
zero \,$\eta \in F$ of \,$L(y)-h \in
F\{y\}_{\Ep,\E}$.\end{enumerate} Let \,$b=\eta+\Sigma_i \eta_i
e_i$. Then \,$F\langle b \rangle_{\Ep,\E}$ is \,$\Ep$-strongly
normal over \,$F$, and $c(G(F\langle
b\rangle_{\Ep,\E}/F))\subseteq G^\Ep_a$ is defined by the
\,$\Ep$-ideal \,$\l=[L]_{\Ep}$.
\end{corollary}
\proof{The assumptions of the proposition are satisfied by taking
$\h=F$, $\f=F$ and $a=\delta \eta+\Sigma ~\eta_i d_i \in F$.
Clearly
$$L(a)=L(\delta b)=\delta(L(b))=\delta(L(\eta+\Sigma_i \eta_i
e_i))=\delta(L(\eta)+\Sigma_i L(\eta_i e_i))$$
$$=\delta(L(\eta)+\Sigma_i L(\eta_i) e_i)=\delta(L(\eta))=\delta
h$$ shows that $a$ is an \de zero of $[L(y)-\delta h]_{\Ep,\E}$.
The computations $$\delta b=\delta(\eta+\Sigma_i \eta_i
e_i)=\delta\eta+\Sigma_i \eta_i \delta e_i= \delta\eta+\Sigma_i
\eta_i d_i=a$$ and $L(b)=L(\eta+\Sigma_i \eta_i
e_i)=L(\eta)+\Sigma_i e_iL(\eta_i)=L(\eta)=h$ demonstrate the $b$
is an \de zero of $[\delta y - a, L(y)-h]_{E,\E} \subset F \lbrace
y \rbrace_{E,\E}$.

It remains to be shown that $a,\epsilon a,\ldots,\epsilon^{n-1}a$
are linearly independent over $F^\E$ modulo $\delta F$. Since
$d_1,\ldots,d_n$ are linearly independent over $F^\E$ modulo
\,$\delta F$, Proposition \ref{Proposition on no nontrivial
differential ideal in the differential ring generated by logs}
implies that $(F\langle e_1,\ldots,e_n \rangle_{\Ep,\E})^\E=F^\E$
and $e_1,\ldots,e_n $ are algebraically independent over $F$.
Because $\eta_1,\ldots,\eta_n$ are linearly independent over
$F^{\Ep,\E}$, the matrix $(\epsilon^{i-1}
\eta_j)_{i=1,\ldots,n:j=1,\dots n}$ is invertible \cite[Theorem 1,
page 86]{kol73}, and, therefore, the map $\varphi$ of $F\langle
e_1,\ldots,e_n \rangle_{\Ep,\E}$ defined by $\varphi(e_i)=
\Sigma_j~ \epsilon^{i-1} \eta_j e_j$ is an automorphism of
$F\langle e_1,\ldots,e_n \rangle_{\Ep,\E}$ over $F$. The
composition $\rho$ of this automorphism with the translation that
sends $\varphi(e_i)$ to $\epsilon^{i-1}\eta+\varphi(e_i)$ is an
automorphism of $F\langle e_1,\ldots,e_n \rangle_{\Ep,\E}$ such
that $\rho(e_i)=\epsilon^{i-1}\eta+\Sigma_j~ \epsilon^{i-1} \eta_j
e_j$. Therefore, $\rho(e_1),\ldots,\rho(e_n)$ are also
algebraically independent over $F$, and $(F\langle
\rho(e_1),\ldots,\rho(e_n) \rangle_{\Ep,\E})^\E=F^\E$. Proposition
\ref{Proposition on no nontrivial differential ideal in the
differential ring generated by logs} implies that
$\delta(\rho(e_1)),\ldots, \delta(\rho(e_n))$ are linearly
independent over $F^\E$ modulo $\delta F$. The observation that
for each $i$
$$\delta(\rho(e_i))=\delta(\epsilon^{i-1}\eta +
\Sigma_j~ \epsilon^{i-1} \eta_j e_j)=\delta \epsilon^{i-1}\eta
+\Sigma_j~ \epsilon^{i-1} \eta_j \delta(e_j)$$
$$=\delta\epsilon^{i-1}
\eta +\Sigma_j~ \epsilon^{i-1} \eta_j d_j =\epsilon^{i-1}(\delta
\eta +\Sigma_j~  \eta_j d_j)=\epsilon^{i-1} a$$
 completes the proof. }\bigskip

A particularly simple example may be obtained by taking, in this
last corollary, $F=\mathbb{C}(t,x)$ ($\epsilon t=1,\epsilon
x=0,\delta t=0,\delta x=1$), $d_i=1/(x-i)$ for $i=0,\dots,n-1$,
$e_i=\ln(x-i)$, $L=\epsilon^n y$, $h=0$ and $\eta=0$. By Corollary
\ref{Corollary on the sum of derivatives of logs that are not
exact}, $1/(x-i)$ for $i=1,\dots,n$ are linearly independent over
$\mathbb{C}(t)$ modulo $\delta F$. Let
$\eta_1=1,\ldots,\eta_n=t^{n-1}$ be a fundamental system for
$\epsilon^{n} y$, and let
$$c= \ln x+t \ln (x-1)+\cdots +t^{n-1} \ln(x-(n-1)).$$
Then, $F\langle c\rangle_{\Ep,\E}=F(c,\ln x,\ldots,\ln (x-(n-1))$,
and $F\langle c\rangle_{\Ep,\E}$ is $\Ep$-strongly normal over
$\mathbb{C}(t,x)$. The operation of an element $g=\alpha_0 +t
\alpha_1+\ldots+t^{n-1} \alpha_{n-1}$ of Galois group
$\ZZ_{[\epsilon^n y]_\Ep}=\{v\in \v \mid \epsilon^n
v=0\}=\{\alpha_0 +t \alpha_1+\ldots+t^{n-1} \alpha_{n-1}  \mid
\alpha_i \in \mathbb{C}\}$ is defined by $gc=(\alpha_0+\ln
x)+t(\alpha_1+ \ln (x-1))+\cdots +t^{n-1}
(\alpha_{n-1}+\ln(x-(n-1))$. If $f=x$, $\eta$ may be taken to be
$t^{n} x/(n)!$. Then
$$c= t^{n} x/(n)!+\ln x+t \ln (x-1)+\cdots +t^{n-1} \ln(x-(n-1)),$$
and the Galois group is the same.

\subsection{$G^{\Ep}_m$-extensions}

Denote the multiplicative \de -$\mathbb{Q}$-group \cite[page
28]{kol85} (relative to $\u$) by the symbol $G^{\Ep,\E}_{m}$. The
elements of $G^{\Ep,\E}_{m}$ are those of $\u^*$, and its group
structure is that of the field $\u$ under multiplication.
Similarly, $G^{\Ep}_{m}$ will denote the multiplicative
$\Ep$-$\mathbb{Q}$-group (relative to $\k$) with elements those of
$\k^*$. Let $\kappa\in \FF(\u^*)$ be the canonical coordinate
function on $G^{\Ep,\E}_{m}$. Then, $\delta_i \kappa /\kappa \in
\FF(\u^*)$, and the $\Ep$-$\f$-mapping
$l\E=(\delta_1\kappa/\kappa,\dots,\delta_m\kappa/\kappa):G^{\Ep,\E}_{m}\rightarrow(G^{\Ep,\E}_{m})^m$
\cite[Proposition 6, page 129]{kol85} is the logarithmic
derivative on $G^{\Ep,\E}_{m}$ relative to $\E$ \cite[Example 2,
page 352]{kol73}. By \cite[Proposition 3, page 89]{kol85}, it is
an $(\Ep,\E)$-$\f$-homomorphism. The kernel of $l\E$ is the \de
$\f$-subgroup consisting of \de zeros of the \de ideal $[\delta_1
y,\ldots, \delta_m y] \subset \f\{y\}_{\Ep,\E}$ and can be
identified with $G^{\Ep}_{m}$ relative to the $\Ep$-universal
field $\k$.

\begin{definition}
An element \,$\alpha\in\u^{*}$ is \,$\Delta$-{\rm exponential}
over \,$\f$ if \linebreak$(\alpha^{-1}\delta_{1}\alpha ,
\ldots,\alpha^{-1}\delta_{m}\alpha)\in \f ^{m}$; that is, if for
suitable elements \,$a_{1},\ldots, a_{m}\in\f$, \,$\alpha$
satisfies the system of differential equations
$$ \delta_{i}\alpha=a_{i}\alpha ~~(1\leq i\leq m).$$
\end{definition}\bigskip

Let $\alpha$ be $\Delta$-exponential over $\f$, and suppose that
the field of $\E$-constants of $\f\langle\alpha\rangle_{\Del,\E}$
is $\c$ ($=\f^{\Delta}$). For any isomorphism $\sigma$ of
$\f\langle\alpha\rangle_{\Del,\E}$ over $\f$ and $\delta \in
\Delta$,
$$(\alpha^{-1}\sigma\alpha)^{-1} \delta( \alpha^{-1}\sigma\alpha)=
(\alpha^{-1}\sigma\alpha)^{-1}[\delta( \alpha^{-1})\sigma\alpha +
\alpha^{-1}\delta(\sigma\alpha)]$$ $$=
(\sigma\alpha)^{-1}\alpha[-\alpha^{-1} \delta \alpha~ \alpha^{-1}
\sigma\alpha + \alpha^{-1}\delta(\sigma\alpha)]=-\alpha^{-1}
\delta \alpha + (\sigma\alpha)^{-1}\delta(\sigma\alpha)$$
 $$=-\alpha^{-1} \delta \alpha +
\sigma(\alpha^{-1}\delta\alpha)=-\alpha^{-1} \delta \alpha
+\alpha^{-1}\delta\alpha=0.$$ Therefore,
$$l\E(\alpha^{-1}\sigma\alpha)=(( \alpha^{-1}\sigma\alpha)^{-1}
\delta_1(  \alpha^{-1}\sigma\alpha), \ldots,(
\alpha^{-1}\sigma\alpha)^{-1} \delta_m(
\alpha^{-1}\sigma\alpha))=0.$$

 Hence the element $c(\sigma)=\alpha^{-1}\sigma\alpha$ is in the
kernel of $l\E$ and is a $\E$-constant. Just as in the case of an
element $\Delta$-primitive over $\f$, $\f\langle\alpha\rangle$ is
$\Del$-strongly normal over $\f$ because  \[\f
\langle\alpha\rangle_{\Del,\E}\sigma(\f\langle\alpha\rangle_{\Del,\E})
= \f
\langle\alpha\rangle_{\Del,\E}\f\langle\sigma\alpha\rangle_{\Del,\E}
~~~~~~~\]
\[~~~~~~~=\f\langle\alpha\rangle_{\Del,\E}\f\langle\alpha\cdot
c(\sigma)\rangle_{\Del,\E}
=\f\langle\alpha\rangle_{\Del,\E}\c\langle
c(\sigma)\rangle_{\Del,\E}.\] The mapping $c:
G(\f\langle\alpha\rangle /\f)\to G^{\Ep}_{m}$ is clearly a group
homomorphism. It is injective because $1=c(\sigma)=\alpha^{-1}
\sigma \alpha$ implies $\alpha=\sigma \alpha$ and $\sigma={\rm
id}_{\f\langle\alpha\rangle_{\Del,\E}}$. This proves the first
part of the following proposition, and the remainder is a special
case of Proposition \ref{proposition given a primitive of G on the
injective homomorphism from the Galois group the G}

\begin{proposition}\label{Proposition on a Delta-exponential gives an E-strongly normal Gm extension}
 Let \,$\alpha$ be a \,$\Delta$-exponential over \,$\f$, and suppose that
\,$\c=(\f\langle\alpha\rangle_{\Del,\E})^\E$. Then, each \,\de
-$\f$-isomorphism \,$\sigma$ of \,$\f\langle \alpha
\rangle_{\Ep,\E}$ into \,$\u$ is of the form
\,$\sigma\alpha=\alpha \cdot c(\sigma)$ for \,$c(\sigma) \in
\k^*$. In addition, \,$\f\langle \alpha \rangle_{\Ep,\E}$ is
$\Ep$-strongly normal over \,$\f$, and the mapping \,$c:
G(\f\langle\alpha\rangle /\f)\to G^{\Ep}_{m}$ defined by
\,$c(\sigma)=\alpha^{-1}\sigma\alpha$ for \,$\sigma \in
G(\f\langle\alpha\rangle /\f)$
is an injective \,$\Del$-$%
\c$-homomorphism of \,$\Ep$-groups relative to the \,$\Ep$-field
\,$\k$. Consequently, \,$\f%
\langle\alpha\rangle_{\Del,\E}$ is a \,$G^{\Ep}_{m}$-extension of
\,$\f$.\end{proposition}

\begin{proposition}\label{Galois subgroups of Gm}
Let \,$G$ be a connected \,$\Del$-$\c$-subgroup of
\,$G^{\Ep}_{m}$. Let \,$\p$ the prime \,$\Del$-ideal in \,$\c
\left\{ y \right\} _{\Del }$ defining \,$G$. Let \,$b$ be a
generic zero in \,$\u$ of \,$\p_{\E,E} \subset \c \left\{ y
\right\} _{\Del,\E }$. Let \,$a = l\Delta b/b$. Put \,$\f=\c
\langle a \rangle_{E,\E}$, and \,$\g=\f \langle b \rangle_{E,\E}$.
Then \,$\g$ over \,$\f$ is an \,$\Ep$-strongly normal extension
with Galois group \,$G$.
\end{proposition}

\proof{This is a special case of Theorem \ref{Theorem on the
construction of an $E$-strongly normal extension from any
$E$-group}. }\bigskip

The $\Ep$-subgroups of $G^{\Ep}_{m}$ are the algebraic subgroups
$\mu_r=\{v \in G^{\Ep}_{m}\mid v^r=1\}$ for every positive integer
$r$ and $G_\l=\{v \in\v^* \mid L(l\Ep(v))=0 {\rm ~for ~} L(y) \in
\l\}$ where \,$\l \subset\f\{y\}_\Ep$ is a linear \,$\Ep$-ideal
\cite[Chapter 4]{cass72}. For $\mu_s \subseteq \mu_r$,  it is
necessary and sufficient for $s$ to be a divisor of $r$, and, for
$G_\l \subseteq G_{\l'}$, it is necessary and sufficient for $\l
\supseteq \l'$. Additionally, each subgroup of the form $G_\l$ is
connected and contains $\mu_r$ for each $r$ \cite[Chapter
4]{cass72}.

The following proposition exhibits the Galois correspondence even
if $\c=\f^\E$ is not constrainedly closed.

\begin{proposition}\label{Proposition exhibiting the Galois
correspondence for Gm} Assume that \,$\g$ is an \,$\Ep$-strongly
normal extension of \,$\f$ that is \,\de generated over \,$\f$ by
a \,$\E$-exponential \,$b$ over \,$\f$. Let \,$G={\rm
Gal}(\g/\f)\subseteq  \v^*$ be the Galois group.
\begin{enumerate}\item \,
If \,$G=\mu_r$, then each \,$\Ep$-$\c$-subgroup \,$H$ is
$\Ep$-$\c$-isomorphic to \,$\mu_s$ for some divisor \,$s$ of
\,$r$, and \,$\g^H=\f\langle b^s \rangle_{\Ep,\E}$.
 \item If \,$G=G_\l$, then each \,$\Ep$-$\c$-subgroup
\,$H$ is \,$\Ep$-$\c$-isomorphic to either \,$\mu_s$ for some
positive integer \,$s$ or \,$G_{\l'}$ such that \,$\l \subseteq
\l'$. If \,$H=\mu_s$, \,$\g^H=\f\langle b^s \rangle_{\Ep,\E}$. If
\,$H=G_{\l'}$, \,$\g^H =\f\langle (L(\epsilon b/b)_{L \in \l'}
\rangle_{\Ep,\E}$.\end{enumerate}
\end{proposition}
\proof{Let $\sigma \in G^{\Ep}_{m}$. Proposition \ref{Proposition
on a Delta-exponential gives an E-strongly normal Gm extension}
implies $\sigma_\zeta(b)= \zeta b$ for some $\zeta \in \k^*$. If
$\sigma$ leaves $\f\langle b^s \rangle_{\Ep,\E}$ invariant for
some positive integer $s$, then $\sigma (b^s)=(\sigma(b))^s=(\zeta
b)^s=\zeta^s b^s$ implies $\zeta^s =1$ and $\sigma^s=id$.
Therefore, $\sigma \in \mu_s$. If $\sigma$ leaves $\f\langle
(L(\epsilon b/b)_{L \in \l'} \rangle_{\Ep,\E}$ invariant, since
$\sigma(b)= \zeta b$, for each $L \in \l'$,
$$L(\epsilon \zeta/\zeta)=L( \epsilon(\zeta b)/(\zeta b)-\epsilon
b/ b)=L(\epsilon(\zeta b)/(\zeta b))-L(\epsilon b/ b)$$
$$=L(\epsilon(\sigma( b))/\sigma(b))-L(\epsilon b/
b)=\sigma(L(\epsilon b/b))-L(\epsilon b/ b)=0,$$ and $\sigma \in
G_{\l'}$.

If \,$G=\mu_r$, each $\Ep$-subgroup $H=\mu_s$, for $s$ a divisor
of $r$, clearly leaves invariant $\f\langle b^s \rangle_{\Ep,\E}$.
Conversely, by the above result, if an element of $G$ leaves
$\f\langle b^s \rangle_{\Ep,\E}$ invariant, it is in $H$.

If $G=G_\l$ and $H=\mu_s$, then the last paragraph shows that
$\g^H=\f\langle b^s \rangle_{\Ep,\E}$. If $G=G_\l$ and
$H=G_{\l'}$, then $H$ leaves invariant $\f\langle (L(\epsilon
b/b)_{L \in \l'} \rangle_{\Ep,\E}$ because, for $\sigma \in H$
$$\sigma( L'(\epsilon b/b))=L'(\epsilon (\sigma(b))/\sigma
b))=L'(\epsilon (\zeta b))/\zeta b)$$
$$=L'(\epsilon b/b+\epsilon \zeta/\zeta)= L'(\epsilon
b/b)+L'(\epsilon \zeta/\zeta)=L'(\epsilon b/b),$$ the \de field
$\f\langle (L'(\epsilon b/b))_{L' \in \l'} \rangle_{\Ep,\E}$ is
invariant under $G_{\l'}$. From the result in the first paragraph
of this proof, $\g^H =\f\langle (L(\epsilon b/b)_{L \in \l'}
\rangle_{\Ep,\E}$.}\bigskip

The following proposition characterizes certain $\Ep$-exponential
$G_m^\Ep$-extensions by the structure of $\f$.

\begin{proposition}\label{Proposition the characterization of Gm
extensions inside F} Let \,$\Delta=\{\delta\}$, let
\,$\Delta=\{\delta\}$, and let \,$\g$ be an \,$\Ep$-strongly
normal extension of \,$\f$ that is \,\de $\f$-generated by a
transcendental \,$\E$-exponential $c$ over \,$\f$. Let \,$a=\delta
c/c$, let $\l_{a,1}=\{L(y) \in \c\{y\}_{\Ep,1} \mid L(\epsilon a)
\in \delta \f\}$, and let \,$\l_a=[\l_{a,1}]_\Ep$. Then \,${\rm
Gal}(\g/\f)=G_{\l_a}$
\end{proposition}\proof{By Proposition \ref{Galois subgroups of Gm}, ${\rm Gal}(\g/\f) \subset
G^\Ep_m$, and, since $c$ is transcendental over $\f$, ${\rm
Gal}(\g/\f)=G_{\l}$ for some $\Ep$-ideal $\l \subset \c\{y\}_\Ep$.
Let $\sigma \in G^{\Ep}_{m}$. Proposition \ref{Proposition on a
Delta-exponential gives an E-strongly normal Gm extension} implies
$\sigma(b)= \zeta b$ for some $\zeta \in G_\l $ so that
$L(\epsilon \zeta/\zeta)=0$ for every $L(y) \in \l$.

Let $b=\epsilon c/c$. Clearly, $\delta b=\epsilon a$. Let $L(y)
\in \l$ of degree one. Then $L(\epsilon c/c)$ is invariant under
all elements of $G$ because $$\sigma(L(\epsilon c/c))=L((\epsilon
\sigma(c))/\sigma(c))=L(\epsilon(\zeta c)/(\zeta c))$$
$$=L(\epsilon c/c+\epsilon \zeta/\zeta)=L(\epsilon c/c)+L(\epsilon
\zeta/\zeta)=L(\epsilon c/c)$$
 Thus $L(\epsilon c/c) \in
\f$, and $L(\epsilon c/c)=f$ for some $f \in \g$. The computation
$$L(\epsilon a)=L(\delta b)=\delta(L(b))=\delta(L(\epsilon
c/c))=\delta f$$ shows $L \in \l_a$, and $\l\subseteq \l_a$ since
$\l$ is generated by elements of degree $1$.

On the other hand, let $L(y) \in \l_{a,1}$. Then $L(\epsilon
a)=\delta f$ for $f \in \f$, and $L(b)-f$ is a $\E$-constant
because $\delta(L(b)-f)=L(\delta b) -\delta f=L(\epsilon a)-
\delta f=0$. Therefore, $L(b)-f \in \c \subseteq \f$, and $L(b)
\in \f$. Hence, for all $\sigma \in G_\l$, $\sigma(L(b))=L(b)$,
and the computation $$L(\epsilon v/v)= L(\sigma(\epsilon
c/c)-\epsilon c/c)=L(\sigma(\epsilon c/c))-L(\epsilon c/c)$$
$$=\sigma(L(\epsilon c/c))-L(\epsilon
c/c)=\sigma(L(b))-L(b)=0$$ shows $L(y) \in \l$ and $\l\supseteq
\l_a$ since $\l_a$ is generated by elements of degree $1$.
\nolinebreak}

\begin{corollary}
Let \,$\h$ be an algebraically closed \,\de field such that
\,$\h^\E=\h$, and let \,$\f=\h\langle x \rangle_{\Ep,\E}$, where
\,$x \in \u$, \,$\epsilon x=0$ and \,$\delta x =1$.  Then, there
is no \,$\Ep$-strongly normal extension of \,$\f$ that is \,\de
generated by a \,$\E$-exponential over \,$\f$ and has Galois group
\,$\Ep$-$\h$-isomorphic to \,$G^\Ep_{m}$.
\end{corollary}\begin{remark}This remains true if the
hypothesis that \,$\h$ be an algebraically closed is omitted; the
following proof must be modified to take the structure of
irreducibles into account in the partial fraction decomposition.
\end{remark}\proof{
Assume that such an $\Ep$-strongly normal extension $\g$ of $\f$
exists. Let $b \in \u$ be a $\E$-exponential over $\f$ such that
$\delta b=a b $ for $a \in \f$, and $\g=\f\langle b
\rangle_{\Ep,\E}$. Let $\epsilon
a=p(x)+\displaystyle{\Sigma}_{i,j} ~\frac{h_{i,j}}{(x-h_i)^j}$ for
$p(x) \in \h[x]$ and $h_i, h_{i,j} \in \h$, be the partial
fraction decomposition of $\epsilon a$.

By Proposition \ref{Proposition the characterization of Gm
extensions inside F}, since the Galois group is $G^\Ep_m$, there
does not exist a non-zero $L(y) \in \h\{y\}_{\Ep,1}$ such that
$L(\epsilon a) \in \delta \f$. If all of the $h_{i,1}=0$, then
$\epsilon a \in \delta \f$, and $L(\epsilon a) \in \delta \f$ for
$L(y)=y$. If there exists a non-zero $h_{i,1}$, there exists a
non-zero
$L(y)=$\[\begin{vmatrix}  h_{1,1} &  h_{2,1}&{\ldots} &  h_{r,1} & y\\
\epsilon h_{1,1} &\epsilon h_{2,1}&{\ldots} &\epsilon h_{r,1} &
\epsilon y\\ :& :& :& :& : \\ \epsilon^r h_{1,1} &\epsilon^{r}
h_{2,1}&{\ldots} &\epsilon^{r} h_{r,1} & \epsilon^{r} y
\end{vmatrix}\] $\in \h\{y\}_{\Ep,1}$ such that the finitely
many $ h_{i,1}$ span over $\h^{\Ep,\E}$ the linear space of
$\Ep$-zeros of $L(y)$.
By Lemma \ref{Lemma on L kills a mod deltaF}, since $L(h_{i,1})=0$
for all $i$, $L(\epsilon a)\in \delta\f$. \nolinebreak }\bigskip

The following proposition shows how to construct an $\Ep$-strongly
normal extension for a given connected $\Ep$-subgroup of
$G^\Ep_m$.
\begin{proposition}\label{Proposition on the construction of a
GsubL ext} Assume \,$E=\{\epsilon\}$ and \,$\Delta = \{\delta\}$.
Let the \,\de field $\u$ be \,\de universal over the \,\de field
\,$\d$ of \,$\E$-constants.

\begin{enumerate} \item Let \,$G_\l=\{v\in G^\Ep_m\mid M(\epsilon v/v)=0, M(y)\in \l\}$ be
a connected \,$\Ep$-subgroup of \,$G^\Ep_m$ defined over an \,\de
subfield \,$\d \subset \u^\E$ where \,$L(y) \in \d\{y\}_{\Ep,1}$
of positive order \,$n$ with the coefficient of the highest order
term equal to \,$1$ and \,$\l=[L]_\Ep$.

\item Let the \,\de-field \,$\c \subset \u^\E$ be a strongly
normal extension of \,$\d$, considered as an \,$\Ep$-field, that
is \,$\Ep$-generated over \,$\d$ by a fundamental system
\,$1,\eta_1,\ldots,\eta_n$ of \,$\Ep$-zeros of \,$L(\epsilon y)$.

\item  Let the \,\de-field \,$\b \subset \u^\Ep$ be finitely
\,$\E$-generated over \,$\c^\Ep$, satisfy the condition
\,$\b^\E=\c^\Ep$, and contain the elements \,$f_1,\dots,f_n$ that
are assumed to be linearly independent over \,$\b^\E$ modulo
\,$\delta \b$.

\item Let \,$\f=\c\cdot\b$, and let \,$f \in \f$. Let \,$\eta \in
\f$ be an \,\de zero \,$\eta$ of \,$L(\epsilon y)-f \in
\f\{y\}_{\Ep,\E}$.

\item For each \,$i=1,\ldots,n$, let $g_i \in \u^\Ep$ be a
\,$\delta$-primitive of \,$f_i$, i.e. \,$\delta g_i =f_i$.

\item  Let \,$\h=\f\langle g_1,\ldots,g_n\rangle_{\Ep,\E}$, and
let \,$c$ be an \,$\h$-generic \,\de zero of \,$$N=[\delta
y-(\delta \eta+\Sigma ~\eta_i f_i)y,~\epsilon
y-(\epsilon\eta+\Sigma~\epsilon \eta_i g_i)y]_{\Ep,\E} \subset
\h\{y\}_{\Ep,\E}.$$ \end{enumerate}

\[
\begin{CD}
\eta_i \in\c             @+>>>    \f=\c\cdot\b  @+>>> \h=\f\langle g_1,\ldots,g_n \rangle_{\Ep,\E}\\
    @AAA                     @AAA  @AAA  \\
\d        @>>>  \d\cdot\b  @+>>> \d\b\langle g_1,\ldots,g_n \rangle_{\Ep,\E}\\
   @AAA     @AAA   @AAA
   \\ \c^\Ep =\b^\E  @+>>>   f_i \in \b  @+>>> \b\langle g_1,\ldots,g_n \rangle_{\Ep,\E}\\
\end{CD}\]
\\
Then \,$\f\langle c \rangle_{\Ep,\E}$ is \,$\Ep$-strongly normal
over \,$\f$ with Galois group \,$G_\l$.

\end{proposition}
\begin{remark}If the elements of the \,\de fields in the
proposition are of analytic functions of two variables, \,$c$ may
be taken to be \,$\exp(\eta+\Sigma ~\eta_i g_i)$.
\end{remark}

\proof{ Since $\b\langle g_1,\ldots,g_n \rangle_{\Ep,\E}$ and $\c$
are linearly disjoint over $\c^\Ep =\b^\E $ \cite[Corollary 1,
page 87]{kol73}, $\b\langle g_1,\ldots,g_n \rangle_{\Ep,\E}$ and
$\f$ are linearly disjoint over $\b$ \cite[Proposition 1, page
50]{langalggeom}. Since that $f_1,\dots,f_n$ are are assumed to be
linearly independent over $\b^\E$ modulo \,$\delta \b$,
Proposition \ref{Proposition on no nontrivial differential ideal
in the differential ring generated by logs} implies $1,
g_1,\ldots,g_n$ are linearly independent over $\b$ which, by the
linearly disjointness, are also linearly independent over $\f$.
Proposition \ref{Proposition on no nontrivial differential ideal
in the differential ring generated by logs} implies
$f_1,\dots,f_n$ are linearly independent over $\f^\E$ modulo
\,$\delta \f$, $ g_1,\ldots,g_n$ are algebraically independent
over $\f$ and $\h^\E=\f^\E$.

Let $a=\delta \eta+\Sigma ~\eta_i f_i$ ($\in \f$) and
$b=\epsilon\eta+\Sigma~\epsilon \eta_i g_i$ ($\in \h$). Clearly,
$\epsilon a=\delta b$. For any orderly ranking, the set $\{\delta
y - ay, \epsilon y -by\}$ is coherent and autoreduced because
$\epsilon(\delta y -ay)- \delta(\epsilon y -by) =0$. By
\cite[Lemma 5, page 137]{kol73}, $N$ is prime. No polynomial
non-zero $p(y) \in \h[y] \subset \h\{y\}_{\Ep,\E}$ is contained in
$N$ because if $p(y) \in N$ then because $p(y)$ is partially
reduced with respect to $\{\delta y - ay, \epsilon y -by\}$
\cite[Lemma 5, page 137]{kol73} implies $p(y) \in (\delta y - ay,
\epsilon y -by)$. This is impossible since $p(y)$ is reduced and
non-zero. By taking $p(y)=1$, the argument above shows $N$ is
proper. Therefore, there exist a nonzero \de zero $c \in \u$ that
is not algebraic over $\h$. This and the fact that
$\h[c]=\h\{c\}_{\Ep,\E}$ imply that $c$ is transcendental over
$\h$.

The Wronskian matrix $(\epsilon^j \eta_i)_{i=1,\ldots,n;j=1,n}$ is
invertible because $\epsilon \eta_1,\ldots, \epsilon \eta_n$ is a
fundamental system of zeros for $L(y)$. Therefore, the following
system of linear equations obtained by repeatedly differentiating
$b=\epsilon \eta +\Sigma_i \epsilon \eta_i g_i$ by $\epsilon$ may
be solved
for $g_1,\ldots,g_n$:
$$b=\epsilon \eta +\Sigma_i \epsilon \eta_i g_i$$
$$\epsilon b=\epsilon^2 \eta +\Sigma_i \epsilon^2 \eta_i g_i$$
$$\cdots $$
$$\epsilon^{n-1} b=\epsilon^n \eta +\Sigma_i \epsilon^n \eta_i g_i.$$
Because $\eta \in \f$, $\f(b,\epsilon b,\ldots,\epsilon^{n-1}
b)=\f(g_1,\ldots,g_n)$. From this and the fact that $b=\epsilon
c/c \in \f\langle c \rangle_{\Ep,\E}$, all $g_i \in \f\langle c
\rangle_{\Ep,\E}$. Since $g_1,\ldots,g_n$ are algebraically
independent over $\f$, so are $b,\epsilon b,\ldots,\epsilon^{n-1}
b$. Because $\delta(\epsilon^i b)=\epsilon^i (\delta b)=\epsilon^i
(\epsilon a)=\epsilon^{i+1} a$, Proposition \ref{Proposition on no
nontrivial differential ideal in the differential ring generated
by logs} implies $\epsilon a,\dots, \epsilon^n a$ are linearly
independent over $\f^\E$ modulo $\delta \f$.

To show $(\f\langle c \rangle_{\Ep,\E})^\E=\f^\E$, since
$\f^\E=\h^\E$, $(\h\langle c \rangle_{\Ep,\E})^\E=\h^\E$ must be
proved. Let $\alpha \in \h \langle c\rangle_{\Ep,\E}$ be a
non-zero $\E$-constant. First assume $\alpha \in \h \{ c\}_{\E,E}=
\h [c]$. Since $c$ is transcendental over $\h$. one may uniquely
write $\alpha= a_r c^r +a_{r-1}c^{r-1} + \ldots +a_0$ where
$a_r\neq 0$ and $a_i \in \h $ for $ i=0,\ldots,r$. Then $\delta
\alpha = A_r c^r + A_{r-1} c^{r-1} + \ldots + A_0$ where $A_i =
\delta a_i +i a a_i$ for $ i=0$ to $r$. Since $\delta \alpha =0$
and the powers of $c$ are linearly independent over $\h$, it
follows that $A_i=0$ for $i=0,\ldots,r$. By Corollary \ref{no
exponential solution}, $a_i \in \f$ for $ i=1$ to $r$. Therefore
$\epsilon a_r/a_r \in \f$, and $\delta (\epsilon
a_r/a_r)=\epsilon(\delta a_r/a_r)=\epsilon(-r a)=-r \epsilon a$
which unless $r=0$ contradicts the linear independence of the
family $\epsilon a,\dots, \epsilon^n a$ over $\f^\E$ modulo
$\delta \f$. Hence, $\alpha=a_0 \in\f^\E$. Similarly, if $1/\alpha
\in \h [c]$, then $1/\alpha \in\f^\E$.

 Second, if neither $\alpha$ nor $1/\alpha$ is in $\h
 [c]$, let $\alpha = A/B$ where $A$ and $B$ are in $\h
 [c]$ of positive degree such that $A$ has the minimal degree
 among all such choices of $A$ and $B$. It may be assumed that $\delta B \neq 0$ because
 otherwise $\delta A=0$ and $A \in \h^\E$. Since $\delta\alpha = 0$, $A/B =\delta A/ \delta B$.
 Write $A=a_r c^r + ,\ldots,+a_0$,
$a_i \in \h$ for $i=0$ to $r$, and $B=b_s c^s + ,\ldots,+b_0$,
$b_i \in \h$ for $i=0$ to $s$. Both $a_0$ and $b_0$ may not be $0$
because then the numerator and the denominator of $\alpha$ may be
divided by $c$ resulting in a fraction representing $\alpha$ with
a lower degree numerator. If $b_0 = 0$ and $a_0 \neq 0$, divide
the numerator and the denominator by $a_0$, then the derivatives
of both have no constant terms and may be divided by $c$ again to
produce an equivalent fraction with lower degree numerator. If
$b_0 \neq 0$ and $a_0 = 0$, apply the same reasoning. If $b_0 \neq
0$ and $a_0 \neq 0$, from $ \delta g f =g \delta f$, by comparing
zeroth degree terms in $c$, it follows that $ \delta b_0 a_0 =b_0
\delta a_0$. Therefore $\delta (a_0/b_0) =0$.  Divide the
numerator and the denominator both by $b_0$. The zeroth degree
terms in $c$ of both the numerator and the denominator are
$\E$-constants. Differentiate them and divide both by $c$ to
produce an equivalent fraction with lower degree numerator. So,
$\alpha \in \f^\E$.

Since $c$ is a $\E$-exponential over $\f$ and $(\f\langle c
\rangle_{\Ep,\E})^\E=\f^\E$, Proposition \ref{Proposition on a
Delta-exponential gives an E-strongly normal Gm extension} implies
$\f\langle c \rangle_{\Ep,\E}$ over $\f$ is $\Ep$-strongly normal.
It remains to show that the Galois group $G$ of $\f\langle c
\rangle_{\Ep,\E}$ over $\f$ is $G_\l$. Because $c$ is
transcendental over $\f$, $G$ is not finite, and $G=G_\m$ for some
linear $\Ep$-ideal $\m \subset \f^\E\{y\}_\Ep$.  Since it may be
verified that $L(\epsilon a)=\delta f$, Proposition
\ref{Proposition the characterization of Gm extensions inside F}
implies $L(y) \in \m$. For a linear $M(y) \in \m$, the same
proposition implies $M(\epsilon a)= \delta h$ for some $h \in \f$.
Then $M(b)-h \in \f^\E \subseteq \f$, and $M(b)=\overline{h}$ for
$\overline{h}\in \f$. Since $1, b,\epsilon
b,\ldots,\epsilon^{n-1}b$ are linearly independent over $\f$,
$M(y)$ has order greater than or equal to the order of $L(y)$.
Hence $\{L(y)\}_\Ep=\m =\l$, and $G=G_\l$.\nolinebreak}\bigskip

A particularly simple example may be obtained by taking, in
Proposition \ref{Proposition on the construction of a GsubL ext},
$\d=\mathbb{C}$, $L=\epsilon^n y$, $\c=\d(t)$ with $\epsilon t=1$
and $\delta t=0$, $\b=\mathbb{C}(x)$, $\f=\mathbb{C}(t,x)$,
$\eta_i=t^i$ for $i=1,\ldots,n$, $f_i=1/(x+i-1)$ for
$i=1,\ldots,n$, $g_i=\ln(x+i-1)$ for $i=1,\ldots,n$ and $\eta=0$.
A fundamental system of $\Ep$-zeros of $\epsilon^{n+1} y$ is
$1,t,t^2,\ldots,t^{n}$. By Corollary \ref{Corollary on the sum of
derivatives of logs that are not exact},
$1/(x),1/(x+1),\ldots,1/(x+n-1)$ are linearly independent over
$\mathbb{C}=\b^\E$ modulo $\delta \b$. Then,
$a=t/(x)+t^2/(x+1)+\cdots+t^n/(x+n-1)$, and
$b=\ln(x)+2t\ln(x+1)+\cdots+nt^{n-1}\ln(x+n-1)$.  One may take
$$c=\exp(t \ln x+t^2 \ln (x+1)+\cdots +t^n \ln(x+n-1)).$$
Then, $\f\langle c\rangle_{\Ep,\E}=\f(c,\ln x,\ldots,\ln
(x+n-1)))$, and $\f\langle c\rangle_{\Ep,\E}$ is $\Ep$-strongly
normal over $\f$. The operation of the Galois group
$G_{[\epsilon^n y]_\Ep}=\{v\in \v^* \mid \epsilon^n(\epsilon
v/v)=0\}=\{\exp(\alpha_0 +t \alpha_1+\ldots+t^n \alpha_n ) \mid
\alpha_i \in \mathbb{C}\}$ on $c$ is induced by addition in the
exponents. If $f=x$, $\eta$ may be taken to be $t^{n+1} x/(n+1)!$.
Then,
$$c=\exp(t^{n+1} x/(n+1)!+t \ln x+t^2 \ln (x+1)+\cdots +t^n \ln(x+n-1)),$$
and the Galois group is the same.

\section{Appendix}\label{subsection isom not strong}
Throughout this section, let $\Delta=\{\delta\}$, and write
$\delta w$ as $w'$ for some $\Delta$-ring element $w$. The
following proposition and its corollaries determine the
$\Delta$-zeros of $\delta y - \alpha$ from the factorization of
$\alpha$.

\begin{proposition} Let \,$R$ be a \,$\E$-ring that is a
factorial domain of characteristic zero. Extend \,$\delta$ to a
derivation of the quotient field \,$Q$ of \,$R$. For any \,$\alpha
\in Q$, write the reduced fraction \,$\alpha = \Pi p_i^{n_i}/\Pi
q_j^{m_j}$ where the \,$p_i$ and \,$q_j$ are non-associate
irreducible elements of \,$R$ and the \,$n_i$ and \,$m_j$ are
positive integers. If \,$ q'_j\notin (q_j)$, then \,$q_j$ is in
the denominator of the reduced fraction of \,$\alpha'$ with an
exponent of \,$m_j +1$.
\end{proposition}\proof{Examine the numerator of $ \alpha'$:
\[(\Pi p_i^{n_i})'\Pi q_j^{m_j}-\Pi
p_i^{n_i}(\Pi q_j^{m_j})'=(\Pi p_i^{n_i})'\Pi q_j^{m_j} - \Sigma_j
(\Pi p_i^{n_i})m_jq'_j q_j^{m_j -1} \Pi_{k\not=j} q_k^{m_k}.\] The
only term not divisible by $q_j^{m_j}$ is $(\Pi p_i^{n_i})m_j q'_j
q_j^{m_j -1} \Pi_{k\not=j} q_k^{m_k}$. So the power of $q_i$ in
the factorization of the numerator is ${m_i-1}$. Since
$q_i^{2m_i}$ is present in the denominator of the derivative
formula for $\alpha'$, in the reduced fraction of $ \alpha'$ the
irreducible element $q_j$ is present in the denominator with an
exponent of $m_j +1$.}

\begin{corollary}\label{Proposition on the image of delta}
Let \,$k$ be a field of characteristic zero, and let \,$k(x)$ be
the rational function field in one indeterminate \,$x$ such that
\,$ x'=1$ and \,$a'=0$ for every \,$a \in k$.
For any \,$\alpha \in k(x)$, write the reduced fraction \,$\alpha
= \Pi p_i^{n_i}/\Pi q_j^{m_j}$ where the \,$p_i$ and \,$q_j$ are
different irreducible elements of \,$k[x]$ and the \,$n_i$ and
\,$m_j$ are positive integers. If one \,$m_j=1$, \,$\alpha$ is not
a derivative of any element of \,$k(x)$.
\end{corollary}

\begin{corollary}\label{Corollary on the sum of derivatives of logs that are not exact}
Let \,$\u$ be \,$\Delta$-universal extension of the constant field
$\c$. Let \,$x \in \u$ be a \,$\E$-zero of \,$ y' -1 \in
\c\{y\}_\E$. For \,$i=1,\ldots,n$, let $p_i(x) \in \c[x]$ be
non-associate and irreducible.  Then the reciprocals of the
\,$p_i$ are linearly independent over $\c$ modulo \,$(\c(x))'$.
\end{corollary}\proof{ Express
 any linear combination $\Sigma_i c_i /p_i(x)$ ($c_i \in \c$ and all
$c_i \neq 0$) of the reciprocals of the $p_i(x)$ over $\c$ as a
rational fraction $\alpha$ in reduced form. Since the numerator is
not divisible by any $p_i(x)$, the denominator of $\alpha$ has
each $p_i(x)$ as a factor with exponent exactly $1$.  Now apply
Corollary \ref{Proposition on the image of delta}.
\vspace*{-.1in}\nolinebreak}
\bigskip

In the proof of the next proposition, the following order on
polynomials will be utilized. (See \cite[Lemma 3, page
58]{johnsonrr} for a similar argument.) Let $z_1,\ldots,z_n$ be
algebraic indeterminates over $\f$. Let $g \in
\f[z_1,\ldots,z_n]$, and let $d$ be the degree
 of $g$ in the indeteminates $z_1,\ldots,z_n$, with the convention
 ${\rm deg}~0 =-1$.  Write $g= \Sigma_M \alpha_M M$ where the
$M$ are monomials in $z_1,\ldots,z_n$ and $\alpha_M \in \f$.  Let
$c(g)$ denote the number of terms $\alpha_M M$ ( $\alpha_M \neq
0$) of degree d in $g$. Define the ${\rm level}(g)$ to be $({\rm
deg}~ g, c(g))$ in the lexicographical order on $\mathbb{N}\times
\mathbb{N}$.

Let  $a_i \in\f$ for $i=1,\ldots,n$, and define a \,$\E$-ring
structure $\f[z_1,\ldots,z_n]_\E$ on $\f[z_1,\ldots,z_n]$ by
\,$z'_i= a_i$ for \,$i=1,\ldots,n$.

\begin{lemma}\label{lemma on properties of the level}
Assume that \,$a_1,\ldots,a_n$ are linearly independent over
\,$\c$ modulo \,$\delta \f$. For each \,$g \in
\f[z_1,\ldots,z_n]_\E$ of degree $d$, \,${\rm deg}~ g' \geq d-1$.
If \,$g \neq 0$ and at least one of the non-zero coefficients of a
term of degree \,$d$ is in \,$\c$, then \,${\rm level}(g') <{\rm
level}(g)$.
\end{lemma}\proof{Write $g$ in the form
\[g = \Sigma_{{\rm deg}M=d}~ \alpha_M M+ \Sigma_{{\rm deg}N=d-1}~ ~\alpha_N N +P\]
where $\alpha_M,\alpha_N \in \f$, $P \in \f[z_1,\ldots,z_n]$ and
${\rm deg}~P<d-1$. Then, since $$\delta(\alpha_M M)
=\delta\alpha_M M+\Sigma_i~\Sigma_{z_i \mid M} ~~n_i\alpha_M  a_i
\frac{M}{z_i}$$        for a monomial M of positive degree and
integers $n_i$,
\[ g' =\Sigma_{{\rm deg}M=d}~
\alpha'_M M+\Sigma_{{\rm deg}~N=d-1}~ ~(\alpha'_N + \Sigma_{{\rm
deg}~L=d,~L=Nz_i}~~n_{L,N}\alpha_L  a_i
 ) N +Q \]
where $n_{L,N}$ are positive integers, $Q \in \f[z_1,\ldots,z_n]$
and ${\rm deg}~Q<d-1$.

Assume that $ \alpha'_M \neq 0$ for at least one monomial $M$ of
degree $d$ in $g$. Then ${\rm deg}~ g' ={\rm deg}~g > d-1$. If
also $ \alpha'_{M'}=0$ for at least one monomial $M'$ of degree
$d$ in $g$, then $c( g')< c(g)$. Therefore, ${\rm level}( g')
<{\rm level}(g)$.

Assume the negative of the assumption of the last paragraph:
 $ \alpha'_M = 0$ for all monomials $M$ of degree $d$ in $g$. If $g \neq 0$, then
${\rm deg}~ g'  < {\rm deg}~g $. Therefore, ${\rm level}(g') <{\rm
level}(g)$. To show ${\rm deg}~ g' \geq d-1$, first assume ${\rm
deg}~g \leq 0$. Then $g \in \c$, and ${\rm deg}~ g' = -1 \geq
d-1$. On the other hand, if ${\rm deg}~g
>0$, choose a monomial $N$ of degree $d-1$ such that, for some i,
$Nz_i$ is present in $g$, i.e., $a_{Nz_i}\neq 0$. In $ g'$, the
coefficient of $N$, $ a'_N +\Sigma_{L=Nz_i} \alpha_L a_i $, is not
equal to $0$ because $a_1,\ldots,a_n$ are assumed in 1 to be
linearly independent over $\c$ modulo $\f'$. This proves ${\rm
deg}~ g' = d-1$.}
\begin{lemma}\label{Lemma on Delta-simple rings}
Assume that \,$a_1,\ldots,a_n$ are linearly independent over
\,$\c$ modulo \,$\delta \f$. The \,$\E$-$\f$-ring
\,$\f[z_1,\ldots,z_n]_\E$ is \,$\E$-simple, i.e., has no proper
non-trivial \,$\E$-ideal.
\end{lemma}\proof{Let $\p \subset \f[z_1,\ldots,z_n]_\E$
be a proper $\E$-ideal. Assume there exists a nonzero element of
$\p$. Let $g \in \p$ have the lowest level of all nonzero elements
of $\p$.
 Since $\p$ is proper and, therefore,
has no non-zero elements of degree $0$, $d={\rm deg}~g>0$.
Multiply $g$ by a non-zero element of $\f$ to ensure that one of
the terms of degree $d$ has $1$ for a coefficient. This new
non-zero element, which again is denoted by $g$, is also in $\p$
and has level less than or equal to all of the non-zero elements
of $\p$. By Lemma \ref{lemma on properties of the level}, ${\rm
level}( g') <{\rm level}(g)$. Since $ g' \in \p$, $ g' =0$.
However, by the first part of the same lemma, $-1={\rm deg}~ g'
\geq d-1 \geq 0$ since $d> 0$.  This contradiction shows $\p$ is
the zero $\E$-ideal.}

\begin{proposition}\label{Proposition on no nontrivial differential ideal in the
differential ring generated by logs}Let \,$\u$ be
\,$\Delta$-universal extension of the $\E$-field $\f$, and let
$\c=\f^\E$. For \,$i=1,\ldots,n$, let \,$a_i \in \f$, and let
\,$b_i \in \u$ be such that \,$b'_i =a_i$.
The following four conditions are equivalent:
\begin{enumerate}\item \,$a_1,\ldots, a_n$ are
linearly independent over  \,$\c$ modulo \,$\delta \f$, \item
$b_1,\ldots,b_n$ are algebraically independent over \,$\f$, and
\,$\f\{b_1,\ldots,b_n\}_\E$ is \,$\E$-simple, \item
\,$1,b_1,\ldots,b_n$ are linearly independent over \,$\f$, and
$(\f\{b_1,\ldots,b_n\}_\E)^\E=\nolinebreak[4]\c$, \item
\,$1,b_1,\ldots,b_n$ are linearly independent over \,$\f$, and
$(\f \langle b_1,\ldots,b_n\rangle_\E)^\E=\nolinebreak\c $.
\end{enumerate}
\end{proposition}

\proof{1 $\Longrightarrow$ 2. Define a $\E$-ring structure
$\f[z_1,\ldots,z_n]_\E$ on $\f[z_1,\ldots,z_n]$ by \,$z'_i= a_i$
for \,$i=1,\ldots,n$. Clearly,
$\f\{z_1,\ldots,z_n\}_\E=\f[z_1,\ldots,z_n]_\E$. To show
$\f\{b_1,\ldots,b_n\}_\E=\f[b_1,\ldots,b_n]_\E$ is $\E$-simple,
define a surjective  $\E$-$\f$-homomorphism
$\rho:\f[z_1,\ldots,z_n]_\E \rightarrow \f[b_1,\ldots,b_n]_\E$
over $\f$ by $\rho(z_i)=b_i$ for $i=1,\ldots,n$. Then $\rho$ is a
$\E$-$\f$-isomorphism because the kernel of $\rho$, which is a
 $\E$-ideal, must be the zero ideal by Lemma \ref{Lemma on Delta-simple
 rings}.
Therefore, $\f\{b_1,\ldots,b_n\}_\E$, the codomain of $\rho$,
 also has no non-trivial $\E$-ideal, and $b_1,\ldots,b_n$
 are algebraically independent over $\f$
 because $z_1,\ldots,z_n$ are algebraically independent over $\f$
 and $\rho(z_i)=b_i$ for every $i$.

 2 $\Longrightarrow$ 3.  Let $g$ be a non-zero element of  $(\f\{b_1,\ldots,b_n\}_\E)^\E$.
 Because $g$ is a $\E$-constant, $(g) \subset \f\{b_1,\ldots,b_n\}_\E$ is a
 $\E$-ideal and must be the unit $\E$-ideal by 2. Because, by assumption, $\f\{b_1,\ldots,b_n\}_\E$ is a polynomial ring in the
 algebraically independent indeterminates $b_1,\ldots,b_n$, $g \in \f$ and $g \in \c=\f^\Delta$.
 That $b_1,\ldots,b_n$
are algebraically independent over $\f$ clearly implies that
$1,b_1,\ldots,b_n$ are linearly independent over $\f$.

 3 $\Longrightarrow$ 1. Assume $a_1,\ldots, a_n$ are
linearly dependent over $\c$ modulo $\delta \f$, i.e., $\Sigma_i
~\alpha_i a_i =\delta f$ for $\alpha_i \in \c$ and $f \in \f$.
Since $1,b_1,\ldots,b_n$ are assumed to be linearly independent
over $\f$, the element $\Sigma_i ~\alpha_i b_i -f$ ($\in
\f\{b_1,\ldots,b_n\}_\E$) is not in $\f$ and is a $\E$-constant of
$ \f\{b_1,\ldots,b_n\}_\E$. This contradicts 3. Therefore, the
$a_1,\ldots, a_n$ are linearly independent over $\c$ modulo
$\delta \f$. This proves 1.

3 $\Longleftrightarrow$ 4. For the non-obvious implication, assume
3.
Assume $g \in \f\langle b_1,\ldots,b_n \rangle^\Delta$. Then
\[\aa= \{a\in \f\{b_1,\ldots,b_n\}_\Delta \mid ag \in
\f\{b_1,\ldots,b_n\}_\Delta \}\] is a $\Delta$-ideal because $g$
is a $\Delta$-constant. Since it is non-zero and
$\f\{b_1,\ldots,b_n\}_\Delta$ is $\Delta$-simple, $1 \in \aa$,
which implies $g \in\f\{b_1,\ldots,b_n\}^\Delta$.
}\bigskip

\begin{corollary}[The Ostrowski Theorem] If \,$b_1,\ldots,b_n$
are algebraically dependent over \,$\f$ and \,$(\f \langle
b_1,\ldots,b_n\rangle_\E)^\E = \c$, then \,$1,b_1,\ldots,b_n$ are
linearly dependent over \,$\f$.
\end{corollary}\proof{(See \cite[Exercise 4, page 407]{kol73} or \cite[page 1155]{kol68}.)
The contrapositive of 4 $\Longrightarrow$ 2 is that, if
$\f\{b_1,\ldots,b_n\}_\E$ has a non-trivial $\E$-ideal or
$b_1,\ldots,b_n$ are algebraically dependent over $\f$, then $(\f
\langle b_1,\ldots,b_n\rangle_\E)^\E$ $ \neq
 \c$ or $1,b_1,\ldots,b_n$
are linearly dependent over $\f$.  Therefore, if $b_1,\ldots,b_n$
are algebraically dependent over $\f$ and $(\f \langle
b_1,\ldots,b_n\rangle_\E)^\E =
 \c$, then $1,b_1,\ldots,b_n$ are linearly dependent over $\f$. }

\begin{corollary}\label{Corollary on algebraic independence of
logs and no new constants} Let \,$\u$ be \,$\Delta$-universal
extension of the constant field $\c$. Let \,$x \in \u$ be a
\,$\E$-zero of \,$y' -1 \in \c\{y\}_\E$. For \,$i=1,\ldots,n$, let
\,$c_i \in \c$ such that \,$c_i \neq c_j$ if \,$i \neq j$, and let
\,$b_i \in \u$ be a \,$\E$-zero of the \,$\E$-polynomial \,$ y'
-\frac{1}{x+c_i} \in \c(x)\{y\}_\E$. Then \,$b_1,\ldots,b_n$ are
algebraically independent over \,$\c(x)$, and \,$(\c(x)\langle
b_1,\ldots,b_n \rangle_\E)^\E=\c$.
\end{corollary}\proof{By Corollary \ref{Corollary on the sum of derivatives of logs
that are not exact} with irreducible $p_i(x) =x +c_i$ for
$i=1,\ldots,n$,\linebreak
\,$\frac{1}{x+c_1},\ldots,\frac{1}{x+c_n}$ are linearly
independent over $\c$ modulo $\delta(\c(x))$.  Then apply
Proposition \ref{Proposition on no nontrivial differential ideal
in the differential ring generated by logs}.}

\begin{corollary}\label{no exponential solution} Assume
\,$\Delta=\{\delta\}$.  Let the conditions of the last corollary
be satisfied, let \,$a\in \f$, and let $\xi \in \f \langle
b_1,\ldots,b_n\rangle_{\E}$ be a \,$\Delta$-zero of \,$\delta y-ay
\in \f\{y\}_\E$. Then \,$\xi \in \f$.
\end{corollary}\proof{Let $\xi=A/B$ where $A,B \in  \f \{
b_1,\ldots,b_n\}_{\E}$ where $A$ and $B$ are relatively prime and
both $A$ and $B$ are not elements of $\f$. Then $\delta A B -A
\delta B -aAB=0$. If $A \notin \f$, then $A$ divides $\delta A$.
This is impossible because then the proper ideal $(A)\subset \f \{
b_1,\ldots,b_n\}_{\E}$ would be a $\E$-ideal, which is contrary to
2 of the proposition. If $B \notin \f$, the argument is similar.}

\begin{corollary}\label{no exponential solution} Assume
\,$\Delta=\{\delta\}$.  Let the conditions of the last corollary
be satisfied, let \,$a\in \f$, and let $\xi \in \f \langle
b_1,\ldots,b_n\rangle_{\E}$ be a \,$\Delta$-zero of \,$\delta y-ay
\in \f\{y\}_\E$. Then \,$\xi \in \f$.
\end{corollary}\proof{Let $\xi=A/B$ where $A,B \in  \f \{
b_1,\ldots,b_n\}_{\E}$ where $A$ and $B$ are relatively prime and
both $A$ and $B$ are not elements of $\f$. Then $\delta A B -A
\delta B -aAB=0$. If $A \notin \f$, then $A$ divides $\delta A$.
This is impossible because then the proper ideal $(A)\subset \f \{
b_1,\ldots,b_n\}_{\E}$ would be a $\E$-ideal, which is contrary to
2 of the proposition. If $B \notin \f$, the argument is similar.}

\begin{definition}\label{Definition of linearly E-independence}Let \,$\w$ be a \,$\E$-vector space over a
\,$\E$-field \,$\f$. Any set \,$\Sigma \subseteq \w$ is
\,$\E$-{\rm linearly} {\rm independent over} \,$\f$ if the family
\,$(\theta \alpha)_{\theta \in \Theta, \alpha \in \Sigma}$ is
linearly independent over \,$\f$. Let \,$\r$ be a \,$\E$-ring. A
family \,$(\alpha_i)_{i \in I}$ of elements of a \,$\E$-overring
of \,$\r$ is \,$\E$-{\rm algebraically}
 {\rm independent over} \,$\r$ or, more simply,
 \,$\E$-$\r$-{\rm algebraically independent}, or  \,$\E$-$\r$-{\rm
 independent},
if the family \,$(\theta \alpha)_{\theta \in \Theta, \alpha \in
\Sigma}$ is algebraically independent over \,$\r$.
\end{definition}

\begin{corollary}\label{Proposition on no nontrivial epsilon-differential ideal in the
differential ring generated by logs}Let \,$\u$ be
\,$(E,\Delta)$-universal extension of the $(E,\E)$-field $\f$, and
let \,$\c=\f^\E$. For \,$i=1,\ldots,n$, let \,$a_i \in \f$, and
let \,$b_i \in \u$ be such that \,$ b'_i = a_i$. The following
four conditions are equivalent:
\begin{enumerate}

\item \,$a_1,\ldots, a_n$ are $\Ep$-linearly independent over
\,$\c$ modulo \,$\delta \f$,

\item  \,$b_1,\ldots,b_n$ are \,$\Ep$-algebraically independent if
the family over \,$\f$, and \,$\f\{b_1,\ldots,b_n\}_{\Ep,\E}$ is
\,$\E$-simple,

\item \,$1,b_1,\ldots,b_n$ are \,$\Ep$-linearly independent over
\,$\f$, and \linebreak$(\f\{b_1,\ldots,b_n\}_{\Ep,\E})^\E=\c$,

\item \,$1,b_1,\ldots,b_n$ are \,$\Ep$-linearly independent over
\,$\f$, and \linebreak$(\f \langle
b_1,\ldots,b_n\rangle_{\Ep,\E})^\E =\c $.
\end{enumerate}
\end{corollary}
\proof{ For each positive integer $\nu$, let $\Psi(\nu)$ be the
set of monomials in $\Ep$ of order less than or equal to $\nu$.
Then for each $\nu$,$i$ and $\psi \in \Psi(\nu)$, $\psi b_i$ is a
$\E$-zero of the $\E$-polynomial $ y' -\psi a_i$. Since $\u$ is
clearly also a $\E$-universal extension of the $\E$-field $\f$,
Proposition \ref{Proposition on no nontrivial differential ideal
in the differential ring generated by logs} may be applied to the
families $(\psi b_i)_{\psi \in \Psi(\nu), i=1,\ldots,n}$ and
$(\psi a_i)_{\psi \in \Psi(\nu), i=1,\ldots,n}$ for each $\nu$.

 The equivalence of the four parts of the
proposition may be verified by the following four observations
which are true because each $\Ep$-algebraic relation only has a
finite number of $\Ep$-derivatives:
\begin{enumerate}
\item  $(\Psi_i a_i)_{\psi \in \Psi(\nu),1\leq i\leq n}$ are
linearly independent over $\c$ modulo $\delta \f$ for all $\nu$ if
and only if the $a_1,\ldots,a_n$ are $\Ep$-linearly independent
over $\c$ modulo \nolinebreak$\delta \f$, \item  $(\Psi_i
a_i)_{\psi \in \Psi(\nu),1\leq i\leq n}$ are algebraically
independent over $\c$ modulo $\delta \f$  for all $\nu$ if and
only if the $b_1,\ldots,b_n$ are $\Ep$-algebraically independent
over $\c$ modulo $\delta \f$,
 \item
 $\f\{(\psi b_i)_ {\psi \in \Psi(\nu),1\leq i\leq
n}\}_{\Ep,\E}$ is $\Delta$-simple  for all $\nu$ if and only if
$\f\{ b_i \}_{\Ep,\E}$ is $\Delta$-simple,
 \item
$(\f\{(\psi b_i)_ {\psi \in \Psi(\nu),1\leq i\leq
n}\}_{\Ep,\E})^\E =\c$  for all $\nu$ if and only if
 $(\f\{ b_i \}_{\Ep,\E})^\E=\c$.\end{enumerate}
}\bigskip

\begin{corollary}\label{no exponential solution1} Assume
\,$\Delta=\{\delta\}$.  Let the conditions of the last corollary
be satisfied, let \,$a\in \f$, and let $\xi \in \f \langle
b_1,\ldots,b_n\rangle_{\Ep,\E}$ be a \,$\Delta$-zero of \,$\delta
y-ay \in \f\{y\}_\E$. Then \,$\xi \in \f$.
\end{corollary} \proof{The proof is the same as \ref{no exponential
solution}.}\bigskip

 The main objective of
\cite{johnsonrr} by Johnson, Reinhart and Rubel is to construct a
prime \de ideal $\p \subset \f\{y\}_{E,\E}$ such that all \de
zeros $\zeta \in \u$ of $\p $ generate \de field extensions
$\f\langle \zeta \rangle_{E,\E}$ over $\f$ that have infinite
transcendence degree over $\f$. Using the techniques just
developed, the next proposition presents new simpler examples of
such prime ideals.  Recall $\E=\{\delta\}$.

\begin{lemma}\label{Lemma for the Johnson, Reinhart and
Rubel result}Let \,$z$ be an \,$(\Ep,\E)$-indeterminate over the
\,$(\Ep,\E)$-field \,$\h$. Let \,$a \in \h\langle z \rangle_\Ep$
and \,$a \notin \h$. Then
\begin{enumerate} \item $1$ and \,$a$ are \,$\Ep$-linearly
independent over \,$\h$, \item \,$a \notin \delta \h\langle z
\rangle_{\Ep,\E}$, i.e., \,$a$ has no primitive in \,$\h\langle z
\rangle_{\Ep,\E}$, \item $(\h\langle z\rangle_{\Ep,\E})^\E= \h^\E$
and \item \,$a$ is \,$\Ep$-linearly independent over \,$\h^\E$
modulo \,$\delta \h\langle z\rangle_{\Ep,\E}$.
\end{enumerate}
\end{lemma} \proof{ \begin{enumerate} \item Apply Proposition
\ref{Exercise 1 of Kolchin}.

\item Let $\Ep=\{\epsilon_1,\ldots,\epsilon_n\}$ and choose the
ranking on the $(\Ep,\E)$-indeterminate $z$ such that the rank of
$\delta^r \epsilon_1^{r_1},\cdots,\epsilon_n^{r_n}  z$ is
$(r,r_1,\ldots,r_n)$ in the lexicographical order on
$\mathbb{N}^{n+1}$. Extend this to a ranking of $\h\{ z
\}_{\Ep,\E}$. For an element $f \in \h\{ z \}_{\Ep,\E}$, let $S_f$
denote the separant of $f$.

Let $b\in \h\langle z \rangle_{E,\E}$ be represented as the
quotient $c/d$ with $c,d \in \h\{ z \}_{\Ep,\E}$ and $d\neq 0$
such that the maximum of the rank of $c$ and the rank of $d$ is
the least possible among all such representations. Let $w
$ be the highest ranking derivative of $z$ present in $c$ or $d$.

Suppose $a=b'$ where $b=c/d$ as above. If the rank of
$w=(0,\ldots,0)$, then $c \in \h$, $d \in \h$, $c/d \in \h$, and
$a=(c/d)' \in \h$. Assume the rank of $w$ is greater than
$(0,\ldots,0)$ and write $(c/d)'=$
\[\frac{ c'\cdot d -c \cdot
  d'}{d^2}= \frac{S_c d -c S_d}{d^2}  w' +\frac{{\rm terms~ of~ rank ~< ~rank}~
 w'}{d^2}.\] If $(S_c d -c S_d)\neq 0$, then $(c/d)' \notin \h\langle z
\rangle_{\Ep}$ because $ w' \notin\h\langle z \rangle_{\Ep}$. If
$(S_c d -c S_d)=0$, then, since $c\neq 0$ and $d\neq 0$, $S_d \neq
0$ because otherwise $S_c=0$. But $S_c/S_d =c/d=b$ is a
representation of $b$ such that $S_c$ and $S_d$ have lower rank
than $c$ and $d$, which is contrary to the assumptions on $c$ and
$d$.

\item For each positive integer $\nu$, let $\Psi(\nu)$ be the
monomials in $\Ep$ of order less than or equal to $\nu$.  Since
$(\psi z)_{\psi \in \Psi(\nu)}$ is a finite set of
$\E$-indeterminates over $\h$ and each $\E$-constant of $\h\langle
z\rangle_{E,\E}$ is in $\h \langle (\psi z)_{\psi \in \Psi(\nu)}
\rangle_\E$ for some $\nu$, Corollary \ref{Delta constants of a
transext} implies that $(\h\langle z\rangle_{E,\E})^\E= \h^\E$.

\item Every $\Ep$-linear combination of $a$ over $\h$ is not in
$\h$ because $a$ and $1$ are $\Ep$-linearly independent over $\h$
(part 1), is in $\h\langle z\rangle_\Ep$ by assumption, and not in
$\delta \h\langle z\rangle_{E,\E}$ by part 2. Therefore $a$ is
$\Ep$-independent over $\h$ modulo \,$\delta \h\langle
z\rangle_{E,\E}$.  A fortiori, $a$ is $\Ep$-linearly independent
over $\h^\E$ modulo \,$\delta \h\langle z\rangle_{E,\E}$, since
$\h^\E \subseteq \h$.
\end{enumerate} }

\begin{proposition}Let \,$\Ep$ be non-empty, and \,$\Delta=\{\delta\}$.
Let \,$z$ be an \,$(E,\E)$-indeterminate over the \,$(E,\E)$-field
\,$\h$. And, let \,$y$ be an \,$(E,\E)$-indeterminate over the
\,$(E,\E)$-field \,$\f=\h\langle z \rangle_{E,\E}$. Let \,$a \in
\h\langle z \rangle_\Ep$, and \,$a \notin \h$. Then for all
\,$(E,\E)$-zeros \,$b$ of the prime \,$(E,\E)$-ideal \,$[\delta y
-a]_{E,\E} \subset \f\{y\}_{E,\E}$, \,$b$ is \,$\Ep$-algebraically
independent over \,$\f$, and the algebraic transcendence degree of
\,$\f\langle b \rangle_{E,\E}$ over \,$\f$ is infinite.
\end{proposition}
\proof{ By part 4 of Lemma \ref{Lemma for the Johnson, Reinhart
and Rubel result}, $a$ is $\Ep$-linearly independent over
$\f^\E=\h^\E$ modulo $\delta(\f)$ (See \cite[Theorem 5, page
59]{johnsonrr}). This is the condition 1 of Corollary
\ref{Proposition on no nontrivial epsilon-differential ideal in
the differential ring generated by logs}. For any $b \in \u$ that
is an \de zero of the prime \de ideal $[\delta y -a]_{E,\E}
\subset \f\{y\}_{E,\E}$, condition 2 of Corollary \ref{Proposition
on no nontrivial epsilon-differential ideal in the differential
ring generated by logs} implies $b$ is $\Ep$-algebraically
independent over $\f$. Since $\Ep$ is non-empty, $\g= \f\langle b
\rangle_{E,\E}$ has $\Ep$-transcendence degree one over $\f$ and
has infinite algebraic transcendence degree over $\f$.}

\end{document}